\documentclass[letterpaper,11pt]{article}

\usepackage{ucs}
\usepackage[utf8x]{inputenc}
\usepackage{graphicx}
\usepackage{amsfonts}
\usepackage{dsfont}
\usepackage{amssymb}
\usepackage{amsmath}
\usepackage{amsthm}
\usepackage{enumerate}
\usepackage{stmaryrd}
\usepackage{fullpage}
\usepackage{ifthen}
\usepackage{subfigure}
\usepackage{epic}
\usepackage{authblk}
\usepackage{textcomp}
\usepackage{bm}
\usepackage[small]{caption}

\usepackage[hypertexnames=false,colorlinks=true,linkcolor=blue,citecolor=blue]{hyperref}
\usepackage[numbers,comma,square,sort&compress]{natbib}
\usepackage[letterpaper,text={7in,9in},centering]{geometry}

\usepackage{color}
\usepackage{titlesec}
\graphicspath{{Figures/}}

\newcommand{\be}{\begin{equation}}
\newcommand{\ee}{\end{equation}}
\newcommand{\bqs}{\begin{equation*}}
\newcommand{\eqs}{\end{equation*}}

\newcommand{\R}{\mathbb{R}}

\newcommand{\Ps}{\mathrm{P}^{ss}}
\newcommand{\yu}{y^u}
\newcommand{\yus}{y^{ss,u}}
\newcommand{\yw}{y^{ws}}
\newcommand{\yvs}{y^{ss,v}}
\newcommand{\Yv}{\yu,\yw,\yus,\yvs}
\newcommand{\zu}{z^u}
\newcommand{\zus}{z^{ss,u}}
\newcommand{\zw}{z^{ws}}
\newcommand{\zvs}{z^{ss,v}}
\newcommand{\Zs}{Z^s}
\newcommand{\bZs}{\bar{Z}^s}

\newcommand{\bzus}{\bar{z}^{ss,u}}
\newcommand{\bzw}{\bar{z}^{ws}}
\newcommand{\bzvs}{\bar{z}^{ss,v}}

\newcommand{\md}{\mathrm{d}}

\renewcommand{\O}{\mathcal{O}}

\numberwithin{equation}{section}

\newenvironment{Hypothesis}[1]%
  {\begin{trivlist}\item[]{\bf Hypothesis #1 }\em}{\end{trivlist}}

\theoremstyle{plain}
\newtheorem{theorem}{Theorem}

\newtheorem{lemma}[theorem]{Lemma}

\newtheorem{rmk}[theorem]{Remark}

\newcommand{\p}{\mathbf{p}}
\renewcommand{\P}{\mathbf{P}}

\newenvironment{Proof}[1][.]%
 {\begin{trivlist}\item[]\textbf{Proof#1 }}%
 {\hspace*{\fill}$\rule{0.3\baselineskip}{0.35\baselineskip}$\end{trivlist}}

\title{Bifurcation to  locked fronts in two component reaction-diffusion systems}
\author[1]{Gr\'egory Faye}
\author[2]{Matt Holzer}
\affil[1]{\small CNRS, UMR 5219, Institut de Math\'ematiques de Toulouse, 31062 Toulouse Cedex, France}
\affil[2]{\small Department of Mathematical Sciences, George Mason University, Fairfax, VA 22030, USA}

\begin{document}
\maketitle

\begin{abstract} We study invasion fronts and spreading speeds in two component reaction-diffusion systems.  Using a variation of Lin's method, we construct traveling front solutions and show the existence of a bifurcation to locked fronts where both components invade at the same speed.  Expansions of the wave speed as a function of the diffusion constant of one species are obtained.  The bifurcation can be sub or super-critical depending on whether the locked fronts exist for parameter values above or below the bifurcation value.  Interestingly, in the sub-critical case numerical simulations reveal that the  spreading speed of the PDE system does not depend continuously on the coefficient of diffusion.  
\noindent 
\end{abstract}


\section{Introduction}\label{sec:intro}
We study invasion fronts for general systems of reaction-diffusion equations, 
\begin{equation}
\begin{split}
u_t &= u_{xx}+F(u,v),\\
v_t&= \sigma v_{xx}+G(u,v),
\end{split}
\label{eq:main}%
\end{equation}
where $\sigma>0$ and $x\in\R$. More specifically, we are interested in traveling wave solutions of the form $(u(x-st),v(x-st))$ which satisfy 
\begin{align*}
-s u' &= u'' +F(u,v),\\
-s v'&= \sigma v''+G(u,v),
\end{align*}
where we have set $\xi=x-st$ and used the notation $u'$ for $\dfrac{\md u}{\md \xi}$ and $u''$ for $\dfrac{\md^2 u}{\md \xi^2}$. It will be more convenient to write this system as a first-order system
\begin{equation}
\begin{split}
u_1'&= u_2,  \\
u_2'&= -su_2 -F(u_1,v_1), \\
v_1'&= v_2,  \\
\sigma v_2'&= -sv_2 -G(u_1,v_1). 
\end{split}
\label{eq:TW}%
\end{equation}
Throughout this paper, the reaction terms are assumed to have the form,
\begin{equation}
F(u,v)=uf(u,v), \quad G(u,v)=vg(u,v), \quad \text{ with } f(0,0)>0 \text{ and } g(0,0)>0.
\label{eq:defFG}
\end{equation}

\begin{figure}[!t]
\centering
\includegraphics[width=0.65\textwidth]{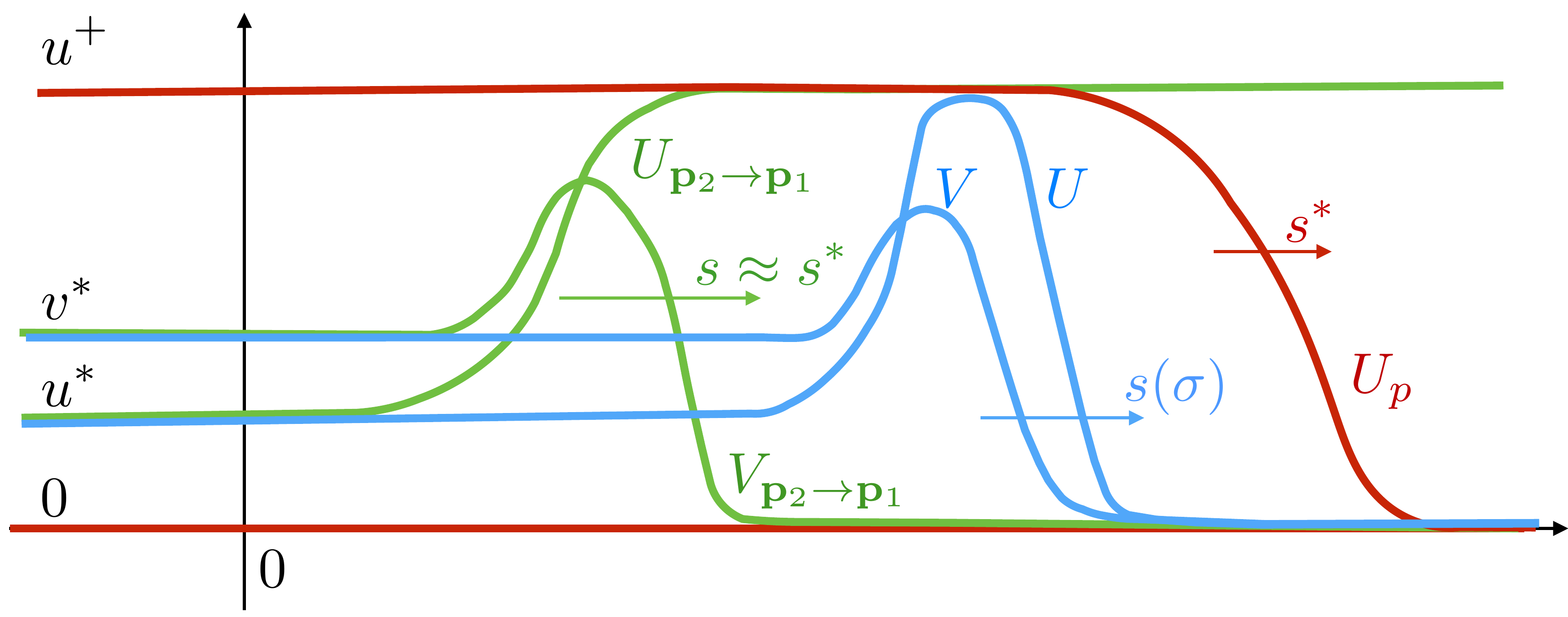}
\caption{Illustration of our assumptions leading to the existence of locked traveling front solutions (in blue) of \eqref{eq:main}. In red, we have represented the pushed front $(U_p(x-s^*t),0)$ connecting $\p_1=(u^+,0)$ to $\p_0=(0,0)$ that propagates to the right with speed $s^*$ given by assumption {\bf (H2)} below. In green, we have sketched one traveling front solution $(U_{\p_2\rightarrow\p_1}(x-st),V_{\p_2\rightarrow\p_1}(x-st))$ connecting $\p_2=(u^*,v^*)$ to $\p_1=(u^+,0)$ that propagates to the right with some speed $s\approx s^*$ given by assumption {\bf (H5)} below. Our main result demonstrates the existence of locked fronts $(U(x-s(\sigma)t),V(x-s(\sigma)t))$ connecting $\p_2=(u^*,v^*)$ to $\p_0=(0,0)$ that propagates to the right with speed $s(\sigma)$ for $\sigma \approx \sigma_*$, see {\bf (H3)} below for the definition of $\sigma^*$.}
\label{fig:frontsketch}
\end{figure}

Precise assumptions regarding the functions $F(u,v)$ and $G(u,v)$ are listed in Section~\ref{sec:setup}.  We sketch those assumptions now to better set the stage and we refer to Figure~\ref{fig:frontsketch} for an illustration.
\begin{enumerate}
\item[{\bf (H1)}] System (\ref{eq:main}) has three nonnegative homogeneous steady states: $\p_0=(0,0)$, $\p_1=(u^+,0)$ and $\p_2=(u^*,v^*)$ and the associated traveling wave equation \eqref{eq:TW} has three corresponding fixed points $\P_0=(0,0,0,0)$, $\P_1=(u^+,0,0,0)$ and $\P_2=(u^*,0,v^*,0)$.
\item[{\bf (H2)}] There exists a pushed front $(U_p(x-s^*t),0)$ connecting $\p_1$ to $\p_0$ that propagates to the right with speed $s^*$ and leaves the homogeneous state $\p_1$ in its wake.
\item[{\bf (H3)}] There exists a $\sigma^*>0$ such that the linearization of the $v$ component about the pushed front has marginally stable spectrum at $\sigma=\sigma^*$.  If  $\sigma<\sigma^*$, then small perturbations of the front $(U_p(x-s^*t),0)$ in the $v$ component propagate slower than $s^*$ whereas for $\sigma>\sigma^*$ these perturbations spread faster than $s^*$.  
\item[{\bf (H4)}] We assume an ordering of the eigenvalues for the linearization of the traveling wave equation \eqref{eq:TW} near $\P_0$ and $\P_1$ together with a condition on the ratio of the eigenvalues.
\item[{\bf (H5)}] There is a family of traveling front solutions connecting $\p_2$ to $\p_1$ for all wave speeds $s$ near $s^*$.  These fronts have weak exponential decay representing the fact that the invasion speed of $\p_2$ into $\p_1$ is slower than $s^*$.  
\end{enumerate}

\begin{figure}[!t]
\centering
 \subfigure[Staged invasion fronts.]{\includegraphics[width=0.475\textwidth]{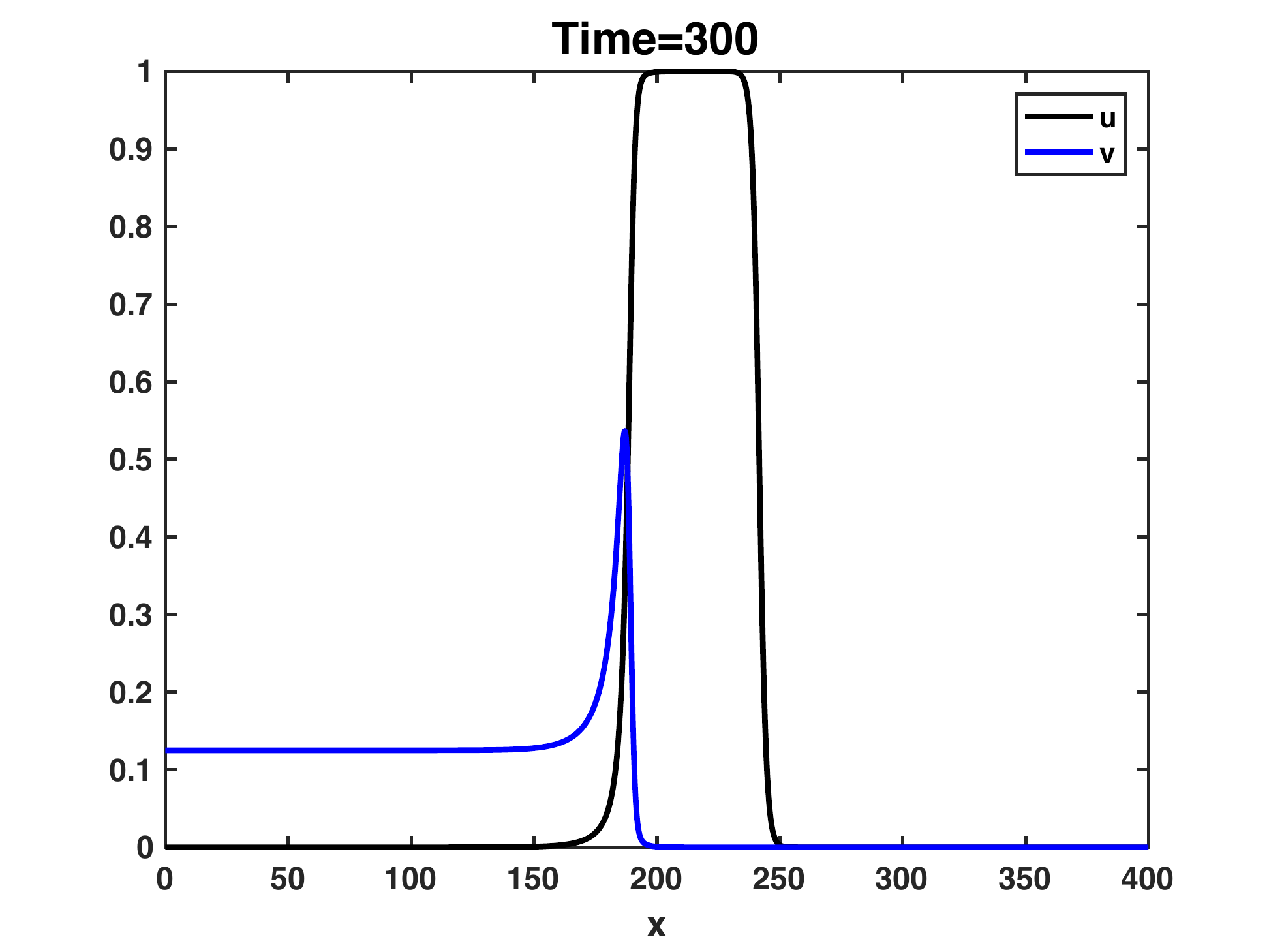}\label{fig:front1}}
 \subfigure[Locked fronts.]{\includegraphics[width=0.475\textwidth]{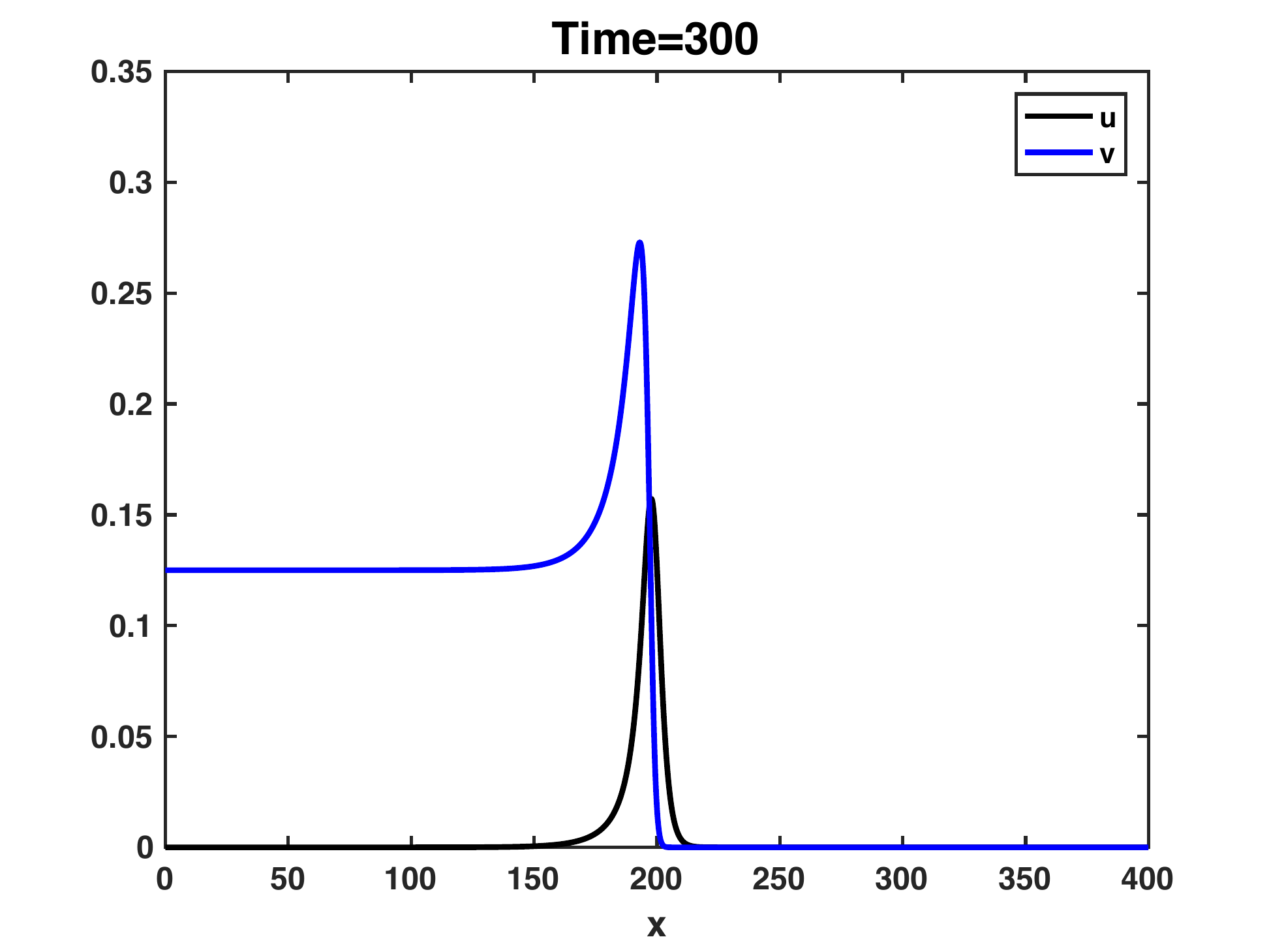}\label{fig:front2}}
\caption{Profiles of the solutions of (\ref{eq:main}), evaluated at time $t=300$, with nonlinear terms $f(u,v)=(1-u)(u+1/16)-v$ and $g(u,v)=2u(1-u)+1/8-v$ for different values of $\sigma$. (a) We observe a staged invasion process where the zero state is first invaded by the $u$ component, then at some later time is subsequently invaded by the $v$ component. Here we have set $\sigma=0.25$. (b) We observe locked fronts with both components traveling at the same wave speed. Here we have set $\sigma=0.3$. Note that $\p_0=(0,0)$, $\p_1=(1,0)$ and $\p_2=(0,1/8)$.}
\label{fig:fronts}
\end{figure}

One can think of $u$ and $v$ as representing independent species that diffuse through space and interact through the reaction terms $F(u,v)$ and $G(u,v)$.  When $\sigma$ is small, we expect the spreading speed of the $u$ component to exceed that of the $v$ component.  The dynamics in this regime is that of a staged invasion process: the zero state is first invaded by the $u$ component, then at some later time is subsequently invaded by the $v$ component, see Figure~\ref{fig:front1}.  As $\sigma$ is increased, the speed of this secondary front will increase until eventually the two fronts lock and form a coherent coexistence front where the unstable zero state $\p_0$ is invaded by the stable state $\p_2$, see Figures~\ref{fig:frontsketch} and~\ref{fig:front2}.  Broadly speaking, this transition to locking is the phenomena that we are concerned with in this article.  Our primary goal is to determine parameter values for which this onset to locking is to be expected and whether the speed of the combined front is faster or slower than the speed of the individual fronts.  

\begin{figure}[!t]
\centering
\includegraphics[width=0.75\textwidth]{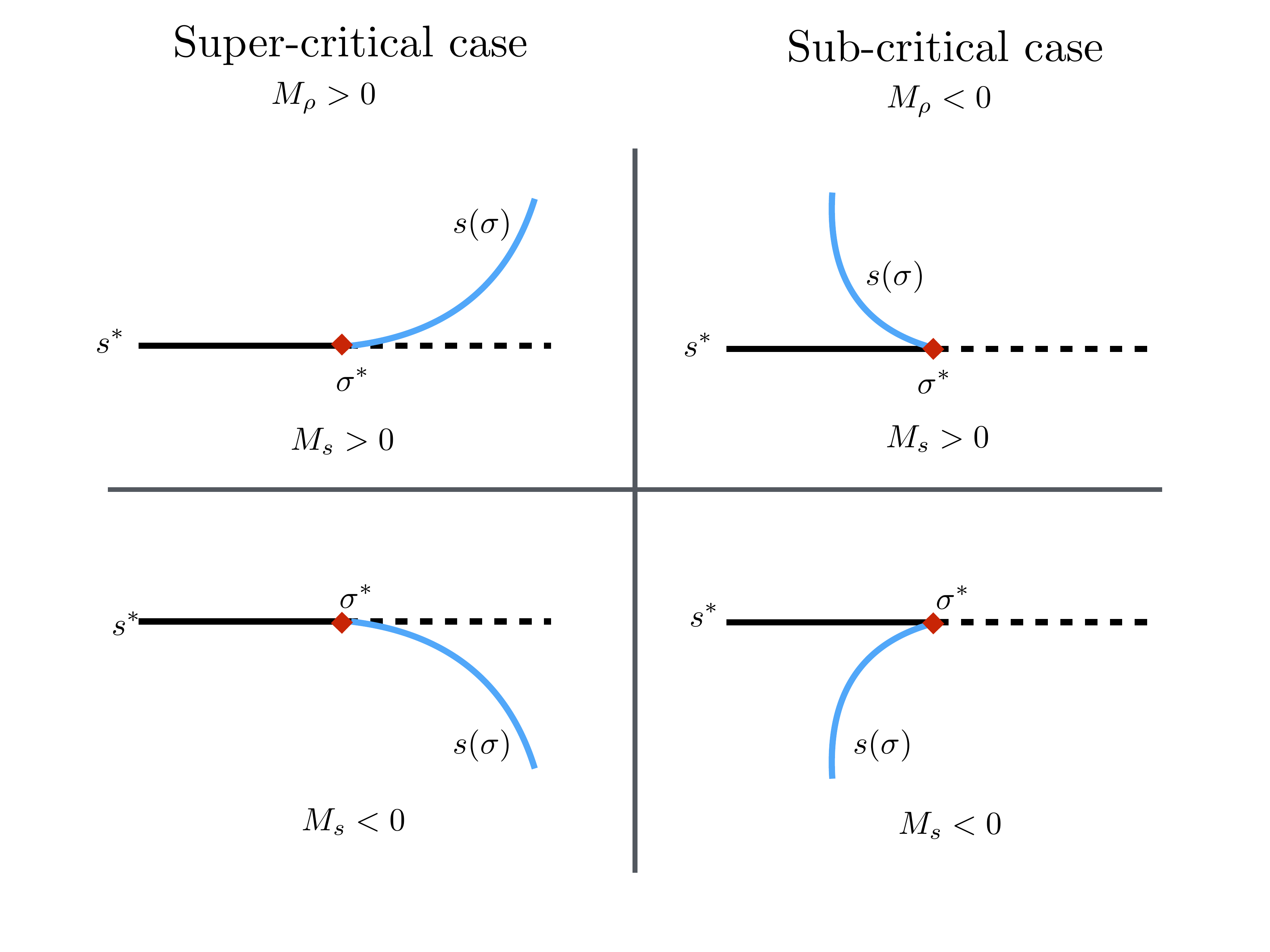}
\caption{Sketch of the different bifurcation scenarios covered by our main result. In each panel, the horizontal black line $s=s^*$ illustrates the marginal stability assumption {\bf (H3)} of the linearization of the $v$ component about the pushed front. The red diamond indicates the critical value $\sigma^*$ at which the pushed front has marginal stable spectrum. The solid part of the line indicates a negative principal eigenvalue of the corresponding linearized operator while the dashed part indicates a positive one. The bifurcating curve in blue illustrates the existence of locked front solutions with wave speed $s(\sigma)$ given by our main result. Two scenarios can happen: the bifurcation will occur either for $\sigma>\sigma^*$ (super-critical case) or for $\sigma<\sigma^*$ (sub-critical case), and in each case the direction of bifurcation can lead to larger wave speed (top panels) or slower wave speed (bottom panels). These different scenarios can be characterized by the signs of the constants $M_\rho$ and $M_s$ (see Theorem~\ref{thm:main}).}
\label{fig:bifdiag}
\end{figure}

Our main result is the existence of a bifurcation leading to locked fronts occurring at the parameter values $(s,\sigma)=(s^*, \sigma^*)$.  Depending on properties of the reaction terms the bifurcation will occur either for $\sigma>\sigma^*$ (super-critical) or for $\sigma<\sigma^*$ (sub-critical), see Figure~\ref{fig:bifdiag} for a sketch.  In the super-critical case, the coexistence front does not appear until after the bifurcation at $\sigma^*$ and the speed of the locked front changes continuously following the bifurcation -- varying quadratically in a neighborhood of the bifurcation point (see Figure~\ref{fig:speed_super} for an illustration on a specific example).  The dynamics of the system in the sub-critical case are much different.  In this scenario, the system transitions from a staged invasion process to locked fronts at a value of $\sigma$ strictly less than the critical value $\sigma^*$ and the spreading speed at this point is not continuous as a function of $\sigma$ and we refer to Figure~\ref{fig:speed_sub} for an illustration on a specific example.

We employ a dynamical systems approach and construct these traveling fronts as heteroclinic orbits of the corresponding traveling wave equation \eqref{eq:TW}, see Figure~\ref{fig:geomsketch}.  The traveling front solutions that we are interested in lie near a concatenation of traveling front solutions: the first  being the pushed front connecting $\P_1$ to $\P_0$ (see {\bf (H2)}) and the second connecting the stable coexistence state $\P_2$ to this intermediate state $\P_1$ (see {\bf (H5)}).  A powerful technique for constructing solutions near heteroclinic chains is Lin's method \cite{lin90,sandstedethesis,sandstede98}.  In this approach, perturbed solutions are obtained by variation of constants and these perturbed solutions are matched via Liapunov-Schmidt reduction  leading to a system of bifurcation equations.   Two common assumptions when using these techniques are a) that the dimensions of the stable and unstable manifolds of each fixed point in the chain are equal and b) the sum of tangent spaces of the intersecting unstable and stable manifolds have co-dimension one.  Neither of these assumptions hold in our case. As fixed points of the traveling wave equation the stable coexistence state $\P_2$ has two unstable eigenvalues and two stable eigenvalues, the intermediate saddle state $\P_1$ has three stable eigenvalues and one unstable eigenvalue and the unstable zero state $\P_0$ has four stable eigenvalues.  Restricting to fronts with strong exponential decay, the zero state can be thought of as having a two-two splitting of eigenvalues, but no such reduction is possible for the intermediate state.

One interesting phenomena that we observe is a discontinuity of the spreading speed as a function of $\sigma$ in the sub-critical regime.   The discontinuous nature of spreading speeds with respect to system parameters has been observed previously, see for example \cite{freidlin91,holzer14,criteria,faye17}.  However, the discontinuity in those cases is typically observed as a parameter is altered from zero to some non-zero value representing the onset of coupling of some previously uncoupled modes.  The mechanism here appears to be different.

There is a large literature pertaining to traveling fronts in systems of reaction-diffusion equations.  Directly related to the work here is \cite{accelerated}, where system (\ref{eq:main}) is studied under the assumption that the second component is decoupled from the first, i.e. that $g(u,v)=g(v)$.  Further assuming that the system obeys a comparison principle, precise statements regarding the evolution of compactly supported initial data can be made; see also \cite{berestycki08}.  Here, we do not assume monotonicity and therefore a dynamical system approach is required.  A similar approach is used in \cite{accelerated}, however, the decoupling of the $v$ component reduces the traveling wave equation to a three dimensional system.  

The present work is also partially motivated by recent studies of bacterial invasion fronts similar to \cite{korolev13}.  In this context, the $u$ component can be thought of as a bacterial population of cooperators while the $v$ component are defectors.  In a well mixed population the defectors out compete the cooperators.  However, in a spatially extended system the cooperators may persist via spatial movement by outrunning the defectors.  This depends on the relative diffusivities, where for $\sigma$ small the cooperators are able to escape.  However, for $\sigma$ sufficiently large the defector front is sufficiently fast to lock with the cooperator front and slow its invasion.  Our result characterizes how this locking may take place.  See also \cite{wakano06,wakano11} for similar systems of equations.

\paragraph{Discussion of methods: a dynamical systems viewpoint}

We have thus far focused primarily on properties of the PDE (\ref{eq:main}).  Mathematically, our main result regards the construction of traveling fronts in the associated traveling wave ODE, (\ref{eq:TW}).  We include a short discussion now to connect these two perspectives; see also \cite{vansaarloos03} for a longer discussion.  To keep this discussion as straightforward as possible we restrict ourselves only to the simplest case of constant coefficient reaction-diffusion systems giving rise to fixed form traveling front solutions connecting homogeneous steady states and ignore complications that can arise for pattern forming systems, inhomogeneous problems, or systems including advective terms to name a few.  

The notion of spreading speeds for a PDE typically refers to the asymptotic speed of invasion of compactly supported perturbations of an unstable state; see for example \cite{aronson75}.  For scalar equations having a comparison principle or for monotone systems of equations, it is often possible to rigorously establish spreading speeds. In doing so, it is often the case that the compactly supported initial conditions eventually converge to a traveling front. Thus, the system identifies a unique {\em selected} front propagating at the selected spreading speed and the proof implies stability (in an appropriate sense) of this front with respect to a large class of initial conditions.

Many systems, including the ones considered here, lack a comparison structure and consequently it becomes extremely difficult to rigorously establish PDE spreading speeds in the traditional sense.  In these cases, one approach is to consider the speed selection problem as a front selection problem and identify fronts which are consistent with selection from compactly supported initial data.   In doing this, one weakens the "global" stability requirement of the selected front to a local stability criterion.   This local stability criterion is referred to as {\em marginal stability}; see \cite{dee83,vansaarloos03}.

Marginal stability requires that the selected front be pointwise marginally stable with respect to compactly supported perturbations.  As fronts propagating into unstable states, the essential spectrum of any invasion front is unstable (in $L^2(\mathbb{R})$ for example).  A common technique to stabilize the essential spectrum is to work in exponentially weighted spaces.  Weights shift the essential spectrum and there is typically an optimal weight that pushes the essential spectrum as far to the left as possible; see \cite{sandstede00a} for an introduction to the absolute spectrum and its role in this regard.  Marginal stability can then be defined in terms of stability properties in this optimally weighted space.  Generally speaking, there are two possibilities.   For a pushed front, the essential spectrum is stabilized  while the point spectrum is stable with the exception of a translational eigenvalue on the imaginary axis.  For a pulled front, the essential spectrum is itself marginally stable and there are no unstable eigenvalues.

Invasion fronts typically come in families parameterized by their speed of propagation.  With the previous discussion in mind, given this family of fronts we seek to identify the unique marginally stable front.  The speed of this marginally stable front then provides a prediction for the spreading speed of compactly supported initial conditions for the original PDE (\ref{eq:main}).

We are interested in constructing candidate pushed fronts for (\ref{eq:main}) by constructing heteroclinic orbits for (\ref{eq:TW}).  The fronts of interest must possess two qualitative features that are indicative of the existence of a pushed front.  First, it must be possible to stabilize the essential spectrum using exponential weights.  Secondly, the decay of the front must be sufficiently steep so that the derivative of the front profile remains as an eigenvalue in the weighted space.  

For the problem considered in this paper, the second property is key and we focus on constructing traveling front solutions with sufficiently steep exponential decay rates.  These are candidate solutions for the selected front and their speed then gives a prediction for the spreading speeds of the original PDE system (\ref{eq:main}).  We do not pursue a full stability analysis of the fronts that we construct, although such an analysis is conceivably possible through similar means as those used in the existence proof.  In fact, we do not necessarily believe these fronts to always be marginally stable.  For example, in the sub-critical regime depicted in Figure~\ref{fig:bifdiag} we expect the bifurcating fronts to be pointwise unstable and this feature is essential to the jump in spreading speed observed numerically in this regime.

We now proceed to outline our assumptions in more detail and state our main result.

\section{Set up and statement of main results}\label{sec:setup}
In this section, we specify the precise assumptions required of (\ref{eq:main}) and state our main result.   We first make some assumptions on the reaction terms $F(u,v)$ and $G(u,v)$ that have the specific form defined in \eqref{eq:defFG}.

\begin{Hypothesis}{(H1)}  Assume that that homogeneous system
\begin{eqnarray*}
u_t &=& F(u,v),  \\
v_t&=& G(u,v),
\end{eqnarray*}
with $F(u,v)=uf(u,v)$ and $G(u,v)=vg(u,v)$, has three non-negative equilibrium points which we denote by $\p_0=(0,0)$, $\p_1=(u^+,0)$ and $\p_2=(u^*,v^*)$ for some $u^*\geq 0$ and $v^*>0$.  We assume that $f(\p_0)>0$ and $g(\p_0)>0$ so that  $\p_0$ is an unstable node for the homogeneous system.  We assume that $F_u(\p_1)<0$ and $g(\p_1)>0$ so that  $\p_1$ is a saddle with one stable direction in the $v=0$ coordinate axis and an unstable direction transverse to this axis.  Finally, we assume that $\p_2$ is a stable node.  
\end{Hypothesis}

The traveling wave equation \eqref{eq:TW} naturally inherits equilibrium points from the homogeneous equation which we denote as $\P_0=(0,0,0,0)$, $\P_1=(u^+,0,0,0)$ and $\P_2=(u^*,0,v^*,0)$.  At either the fixed point $\P_0$ or $\P_1$, the linearization is block triangular and eigenvalues can be computed explicitly.  At $\P_0$, the four eigenvalues are
\begin{eqnarray*}
\mu_u^\pm(s) &=& -\frac{s}{2}\pm\frac{1}{2}\sqrt{s^2-4f(\p_0)}, \\
 \mu_v^\pm(s,\sigma) &=& -\frac{s}{2\sigma}\pm\frac{1}{2\sigma}\sqrt{s^2-4\sigma g(\p_0)},
\end{eqnarray*}
where we used the fact that $F_u(\p_0)=f(\p_0)$ and $G_v(\p_0)=g(\p_0)$. Similarly, at $\P_1$, the linearization has eigenvalues
\begin{eqnarray*}
\nu_u^\pm(s) &=& -\frac{s}{2}\pm\frac{1}{2}\sqrt{s^2-4F_u(\p_1)}, \\
 \nu_v^\pm(s,\sigma) &=& -\frac{s}{2\sigma}\pm\frac{1}{2\sigma}\sqrt{s^2-4\sigma g(\p_1)},
\end{eqnarray*}
where once again we used the fact that $G_v(\p_1)=g(\p_1)$.

When the $v$ component is identically zero, system (\ref{eq:main}) reduces to a scalar reaction-diffusion equation
\be u_t=u_{xx}+F(u,0),\label{eq:scalaru}\ee
and the traveling wave equation \eqref{eq:TW} reduces to the planar system
\begin{eqnarray*}
u_1'&=& u_2,  \\
u_2'&=& -su_2 -F(u_1,0).
\end{eqnarray*}
We now list assumptions related to traveling front  solutions of (\ref{eq:scalaru}). 
\begin{Hypothesis}{(H2)} We assume that there exists  $s^*>2\sqrt{f(\p_0)}$ for which (\ref{eq:scalaru}) has a pushed front solution $U_p(x-s^*t)$ moving to the right with speed $s^*$.  By pushed front, we mean that the solution has steep exponential decay $U_p(\xi)\sim Ce^{\mu_u^-(s^*)\xi}$ as $\xi\to\infty$ and has stable spectrum in the weighted space $L^2_\alpha(\mathbb{R})$, for some $\alpha>0$, with the exception of an eigenvalue at zero due to translational invariance.   There is, in fact,  a one parameter family of translates of these fronts and we therefore impose that   $U_p''(0)=0$ and restrict to one element of the family.  
\end{Hypothesis}

To reiterate the connection to the PDE (\ref{eq:main}), we are interested in reaction terms for which non-negative and compactly supported initial data for (\ref{eq:main}) of the form $(u_0(x),0)$ would spread with speed $s^*>2\sqrt{f(\p_0)}$.  Note that the quantity $2\sqrt{f(\p_0)}$ is the linear spreading speed of the $u$ component near $\p_0$ and so we require faster than linear invasion speeds.  For the traveling wave ODE, this translates to the existence of a marginally stable pushed front -- which is exactly what is laid out by assumption ${\bf (H2)}$.

Now consider the linearization  of the $v$ component of (\ref{eq:main}) around the traveling front solution $(U_p(x-s^*t),0)$,
\[ \mathcal{L}_v :=\sigma \partial_{\xi\xi}+s^*\partial_\xi+g(U_p(\xi),0).\]
The spectrum of this operator posed on $L^2(\mathbb{R})$ is unstable due to the instability of the asymptotic rest states.  However, this spectrum may be stable when  $\mathcal{L}_v$ is  viewed as an operator on the exponentially weighted space
\[ L^2_d(\mathbb{R})=\left\{ \phi(\xi)\in L^2(\mathbb{R})\ | \ \phi(\xi)e^{d \xi}\in L^2(\mathbb{R})\right\}.\]
Let $d=\frac{s^*}{2\sigma}$. Then the operator $\mathcal{L}_v=\sigma \partial_{\xi\xi}+s^*\partial_\xi+g(U_p(\xi),0)$ restricted to $L^2_d$ is isomorphic to the operator $H_\sigma:L^2(\mathbb{R})\to L^2(\mathbb{R})$, where
\[ H_\sigma:=\sigma \partial_{\xi\xi}+\left( -\frac{(s^*)^2}{4\sigma} + g(U_p(\xi),0)\right).\]
We now state our assumptions on the spectrum of $H_\sigma$.  
\begin{Hypothesis}{(H3)} We suppose that the most unstable spectra of $H_\sigma$ is point spectra and define 
\[ \lambda(\sigma)=\sup_{\omega\in\mathrm{spec}(H_\sigma)} \omega.\]
Let $\sigma^*$ be defined such that $\lambda(\sigma^*)=0$.  Associated to this eigenvalue is a bounded eigenfunction which we denote $\tilde{\phi}(\xi)$.  In the unweighted space, this eigenfunction becomes $\phi(\xi)=e^{-\frac{s^*}{2\sigma^*}\xi}\tilde{\phi}(\xi)$ which is unbounded as $\xi\to-\infty$.  We further assume that $G_v(u,0)=g(u,0)>0$ for all $u\in[0,u^+]$ such that $\phi'(\xi)<0$ for all $\xi$.  
\end{Hypothesis}

We will require some properties of the eigenvalues of the linearization of $\P_0$ and $\P_1$ in a neighborhood of the critical parameter values $(s^*,\sigma^*)$.  These are outlined next.

\begin{Hypothesis}{(H4)}
The eigenvalues of the linearization of (\ref{eq:TW}) at $\P_0$ has four unstable eigenvalues.  We assume for some open neighborhood of parameter space including $(s^*,\sigma^*)$ that there exists an   $\alpha>0$ such that 
\be \mu_u^-(s)<-\alpha< \mu_u^+(s), \quad \mu_v^-(s,\sigma)<-\alpha<\mu_v^+(s,\sigma). \label{eq:eigsnearp0}\ee
The fixed point $\P_1$ is a saddle point of (\ref{eq:TW}) with a $3:1$ splitting of the eigenvalues.  We assume that the eigenvalues of the linearization at $\P_1$ can be ordered
\be \nu_v^-(s,\sigma)<\nu_u^-(s)<\nu_v^+(s,\sigma)<0<\nu_u^+(s), \label{eq:eigsnearp1} \ee
again for for some open set of parameters including $(s^*,\sigma^*)$.  In addition, we assume the following condition on the ratio of the eigenvalues:
\be \nu_u^-(s)<2\nu_v^+(s,\sigma).\label{nonresonance}\ee 
\end{Hypothesis}
The eigenvalue splitting (\ref{eq:eigsnearp0}) in Hypothesis {\bf (H4)} guarantees the existence of a two dimensional strong stable manifold which we denote $W^{ss}(\P_0)$.   Initial conditions in $W^{ss}(\P_0)$ correspond to solutions of (\ref{eq:TW}) that decay to $\P_0$ with exponential rate greater than $e^{-\alpha \xi}$ at $\xi=+\infty$.

The final set of assumptions pertain to the existence and character of traveling front solutions connecting $\P_2$ to $\P_1$.  
\begin{Hypothesis}{(H5)} We assume a transverse intersection of the manifolds  $W^u(\P_2)$ and $W^s(\P_1)$ for all $(s,\sigma)$ in a neighborhood of $(s^*,\sigma^*)$.  For $(s^*,\sigma^*)$ we assume the existence of a heteroclinic connection between $\P_2$ and $\P_1$ that approaches $\P_1$ tangent to the weak-stable eigenspace corresponding to the eigenvalue $\nu_v^+(s^*,\sigma*)$, see (\ref{eq:eigsnearp1}). Thus, the two dimensional tangent space of $W^u(\P_2)$ enters a neighborhood of $\P_1$  approximately tangent to the unstable/weak-stable manifold of $\P_1$.

\end{Hypothesis}

In terms of PDE assumptions, ${\bf (H5)}$ is consistent with a staged invasion process where compactly supported perturbations of the steady state $\p_1$ form a traveling front propagating with speed $s<s^*$ replacing the unstable state $\p_1$ with the stable state $\p_2$.  Since the selected invasion speed of fronts propagating into the state $\p_1$ is slower than $s^*$, any traveling front solution with speed $s^*$ should be {\em pointwise stable} which requires that they converge to $\p_1$ with weak exponential decay  precluding the existence of a marginally stable translational eigenvalue.

\paragraph{Remarks on assumptions ${\bf (H1)-(H5)}$.}

We remark that ${\bf (H1)}$ and $\bf{(H4)}$ are straightforward to verify for a specific choice of $F(u,v)$ and $G(u,v)$.  Assumption ${\bf (H2)}$ is more challenging, but due to the planar nature of the traveling wave equation it is plausible that such a condition could be checked in practice.  We refer the reader to \cite{lucia04} for a general variational method suited to such problems.  Assumption ${\bf (H3)}$ is yet more challenging to verify, however as a Sturm-Liouville operator there are many results in the literature pertaining to qualitative features of the spectrum of these operators.  Finally, assumption ${\bf (H5)}$ is the most difficult to verify in practice, as it requires a rather complete analysis of a fully four dimensional system of differential equations (\ref{eq:TW}).  Nonetheless, our assumptions there simply state that the traveling front solutions have the most generic behavior possible as heteroclinic orbits between $\P_2$ and $\P_1$.  In this sense, we argue that assumption ${\bf (H5)}$ is not so extreme, in spite of the challenge presented in actually verifying that it would hold in specific examples.    

We also remark that the precise ordering of the eigenvalues assumed in ${\bf (H4)}$ are technical assumptions and could likely be relaxed in some cases.    

\paragraph{Main Result.}  We can now state our main result.  

\begin{theorem}\label{thm:main} Consider (\ref{eq:main}) and assume that Hypotheses {\bf (H1)-(H5)} hold.  Then there exists a constant $M_\rho$ such that:
\begin{itemize}
\item (sub-critical) if $M_\rho<0$ then there exists $\delta>0$ such that there exists positive traveling front solutions $\left(U(x-s(\sigma)t),V(x-s(\sigma)t)\right)$ for any $\sigma^*-\delta<\sigma<\sigma^*$ with speed 
\[ s(\sigma)=s^*+M_s(\sigma-\sigma^*)^2+\O(3); \]
\item (super-critical) if $M_\rho>0$ then there exists $\delta>0$ such that there exists positive traveling front solutions $\left(U(x-s(\sigma)t),V(x-s(\sigma)t)\right)$ for any $\sigma^*<\sigma<\sigma^*+\delta$ with speed 
\[ s(\sigma)=s^*+M_s(\sigma-\sigma^*)^2+\O(3). \]
\end{itemize}
These traveling fronts belong to the intersection of the unstable manifold $W^u(\P_2)$ and the strong stable manifold $W^{ss}(\P_0)$.
\end{theorem}

We make several remarks. 

\begin{rmk} As part of the proof of Theorem~\ref{thm:main} we obtain expressions for $M_\rho$ and $M_s$.  In particular,
\begin{eqnarray*} \mathrm{sign}(M_\rho)&=& \mathrm{sign}\left( -r_2 \int_{\xi_0}^\infty e^{\frac{s^*}{\sigma^*}\xi}\left( \frac{G_{uv}(U_p(\xi),0)}{\sigma^*} a_1(\xi) \phi(\xi)^2 +\frac{G_{vv}(U_p(\xi),0)}{2\sigma^*} \phi^3(\xi) \right)\md\xi \right. \\
&-&\left. r_1\left(\tilde\phi''(\xi_0)\tilde\phi(\xi_0)-(\tilde\phi'(\xi_0))^2\right)
+ \frac{1}{r_2} e^{\frac{s^*}{\sigma^*}\xi_0}\gamma^{(2)}(s^*,\sigma^*)\left(\nu_v^-(s^*,\sigma^*)\phi(\xi_0)-\phi'(\xi_0) \right)\right),  \end{eqnarray*}
where $r_{1,2}$, $a_1(\xi)$ and $\gamma^{(2)}(s^*,\sigma^*)$ are all defined below.  A similar  expression holds for $M_s$, but is quite complicated.  
\end{rmk}

\begin{figure}[!t]
\centering
\includegraphics[width=0.65\textwidth]{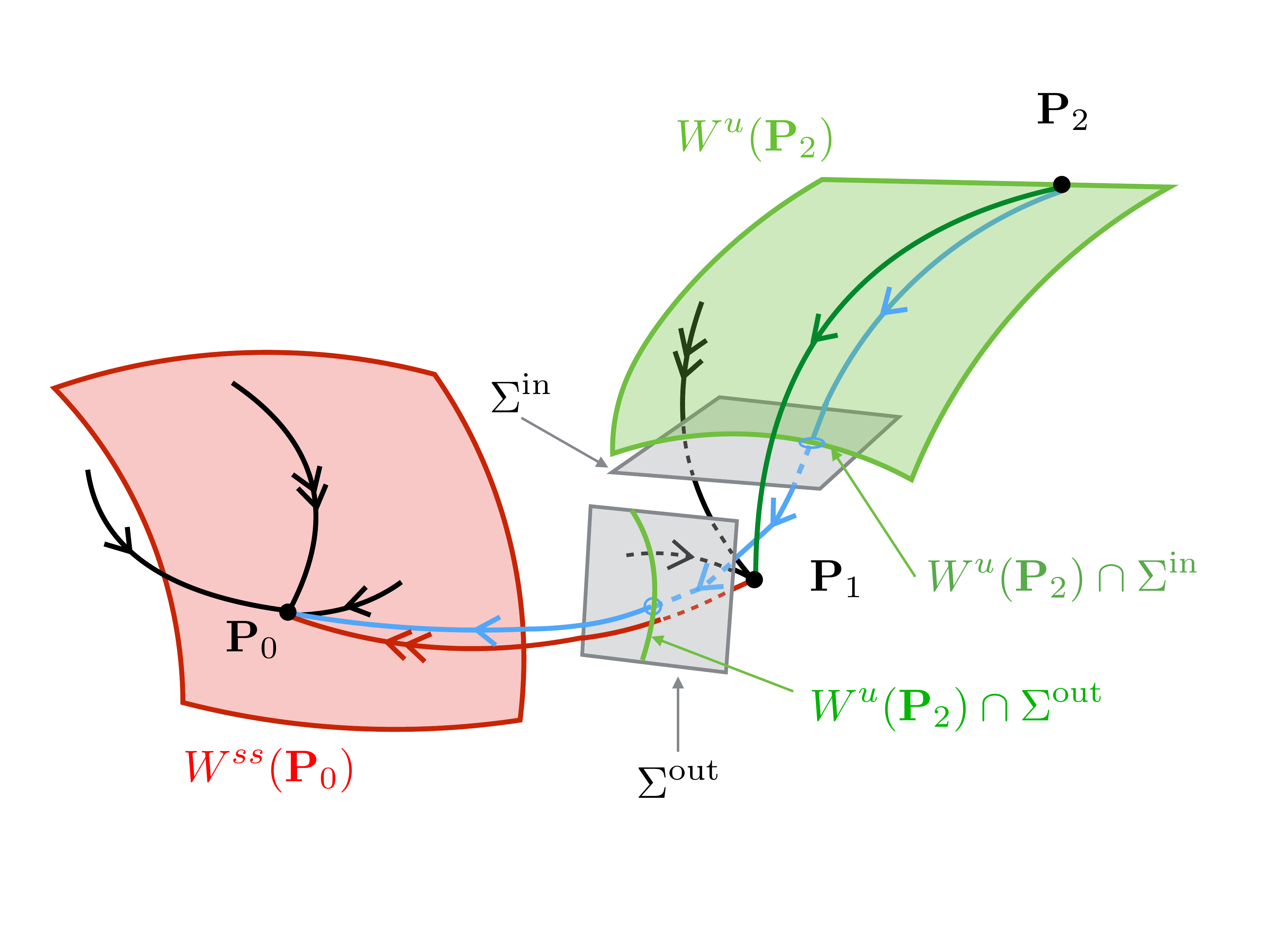}
\caption{Geometrical illustration in $\R^4$ of the construction of locked fronts. Locked fronts are heteroclinic orbits connecting $\P_2$ to $\P_0$ that lie at the intersection of the unstable manifold $W^u(\P_2)$ and the strong stable manifold $W^{ss}(\P_0)$. We track $W^{ss}(\P_0)$ backwards along the pushed front heteroclinic $(U_p(\xi),U_p'(\xi),0,0)^T$, represented by the dark red heteroclinic orbit on the figure, to a neighborhood of $\P_1$ and track $W^{u}(\P_2)$ forwards past the fixed point $\P_1$ from $\Sigma^{in}$ to $\Sigma^{out}$ to compare the two manifolds near a common point on the heteroclinic $(U_p(\xi),U_p'(\xi),0,0)^T$ in $\Sigma^{out}$. In the figure, we represented in dark green one heteroclinic orbit connecting $\P_2$ to $\P_1$ within $W^u(\P_2)$. Schematically, the locked front, represented by the blue heteroclinic orbit on the figure, is found to be close to the concatenation of the two heteroclinic orbits connecting first $\P_2$ to $\P_1$ (dark green) and then $\P_1$ to $\P_0$ (dark red). In that respect, our strategy of proof is a variation of Lin's method.}
\label{fig:geomsketch}
\end{figure}

\begin{figure}[!t]
\centering
 \subfigure[$\epsilon=-1$.]{\includegraphics[width=0.475\textwidth]{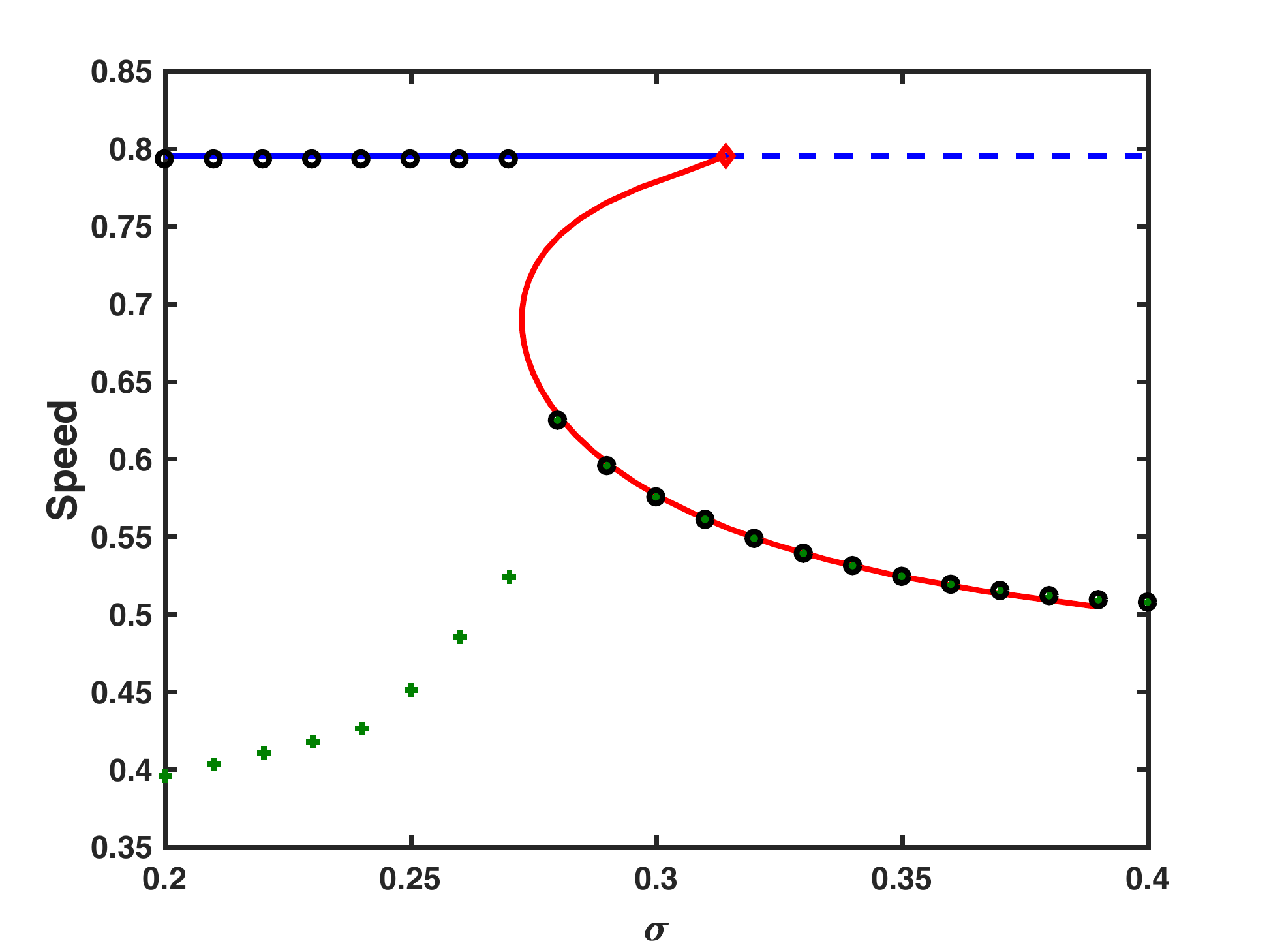}}
 \subfigure[Zoom of Figure (a) near $\sigma \sim 0.27$.]{\includegraphics[width=0.475\textwidth]{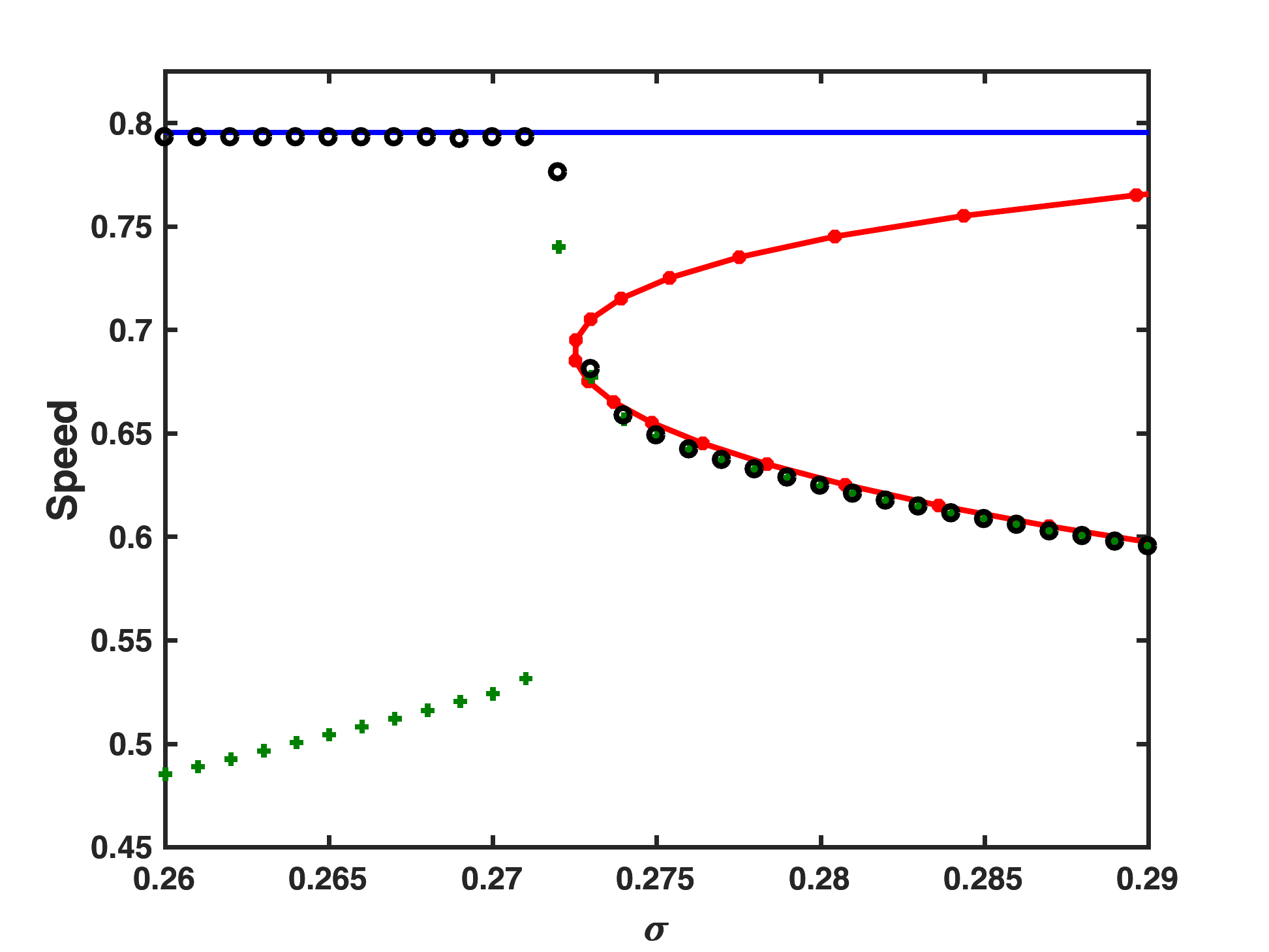}} 
\caption{ (a) Numerically computed wave speeds of the $u$-component, black circles, and of the $v$-component, green plus sign for $\epsilon=-1$ in \eqref{eq:example1}. We observe a discontinuity in the value of the measured wave speed as $\sigma$ is varied indicating a sub-critical bifurcation of the locked fronts. The horizontal blue line $s=s^*=\sqrt{2}(a+1/2)$ represents the sign of the associated principal eigenvalue of the operator $H_\sigma$ and the red diamond indicates the critical value $\sigma^*$ at which this principal eigenvalue vanishes. The solid part of the line indicates a negative principal eigenvalue while the dashed part indicates a positive one. Here, $\sigma^*\simeq0.314$. The red curve is a continuation of the wave speed of locked fronts up to the bifurcation point  $\sigma=\sigma^*$. (b) Refinement of Figure (a) near the fold point. Here, the red dots are wave speeds obtained by numerical continuation. For all numerical simulations we have set $a=1/16$.}
\label{fig:speed_sub}
\end{figure}

\begin{rmk} We comment on the sub-critical case.  Our analysis holds only in a neighborhood of the bifurcation point.  However, we expect that this curve could be followed in $(s,\sigma)$ parameter space to a saddle-node bifurcation where the curve would subsequently reverse direction with respect to $\sigma$.  This curve can be found numerically using numerical continuation methods, see Figure~\ref{fig:speed_sub}.  These numerics reveal two branches of fronts that appear via a saddle node bifurcation.   It is the lower branch of solutions that  appear to be marginally stable and reflect the invasion speed of the system.  

For systems of equations without a comparison principle, the selected front is classified as the marginal stable front, see  \cite{dee83,vansaarloos03} and the discussion at the end of Section~\ref{sec:intro}.  It is interesting to note that in these examples there appear to be two marginally (spectrally) stable fronts -- the original front $(U_p(x-s^*t),0)$ and the coexistence front -- and the full system selects the slower of these two fronts.  
\end{rmk}

We now comment on the strategy of the proof that employs a variation of Lin's method; see Figure~\ref{fig:geomsketch} for a geometrical illustration of our dynamical systems approach.  The traveling fronts that we seek are heteroclinic orbits in the traveling wave equations connecting $\P_2$ to $\P_0$.  We further require that these fronts have strong exponential decay in a neighborhood of $\P_0$.  As such, these traveling fronts belong to the intersection of the unstable manifold $W^u(\P_2)$ and the strong stable manifold $W^{ss}(\P_0)$.  Therefore, the goal is to track $W^{ss}(\P_0)$ backwards along the pushed front heteroclinic $(U_p(\xi),U_p'(\xi),0,0)^T$ to a neighborhood of $\P_1$.  The dependence of this manifold on the parameters $s$ and $\sigma$ can be characterized using Melnikov type integrals and the manifold can be expressed as a graph over the strong stable tangent space.  To track $W^{u}(\P_2)$ forwards we use {\bf (H5)} to get an expression for this manifold as it enters a neighborhood of $\P_1$.  To track this manifold past the fixed point requires a Shilnikov type analysis near $\P_1$.  Finally, we compare the two manifolds near a common point on the heteroclinic $(U_p(\xi),U_p'(\xi),0,0)^T$  and following a Liapunov-Schmidt reduction we obtain the required expansions of $s$ as a function of $\sigma$.  

\paragraph{Numerical illustration of the main result.}

Before proceeding to the proof of Theorem~\ref{thm:main}, we illustrate the result on an example.  We consider the following nonlinear functions $f_\epsilon(u,v)$ and $g(u,v)$  that lead to a supercritical bifurcation when $\epsilon=1$ and exhibit a sub-critical bifurcation for $\epsilon=-1$:
\begin{equation}
 f_\epsilon(u,v)=(1-u)(u+a)+\epsilon v, \text{ and } \quad g(u,v)=2u(1-u)+2a-v,
\label{eq:example1}
\end{equation}
where $\epsilon\in\{\pm1\}$. In both cases, when $v$ is set to zero the system reduces to the scalar Nagumo's equation
\be u_t=u_{xx}+u(1-u)(u+a).\label{eq:nagumo} \ee

\begin{figure}[!t]
\centering
\includegraphics[width=0.475\textwidth]{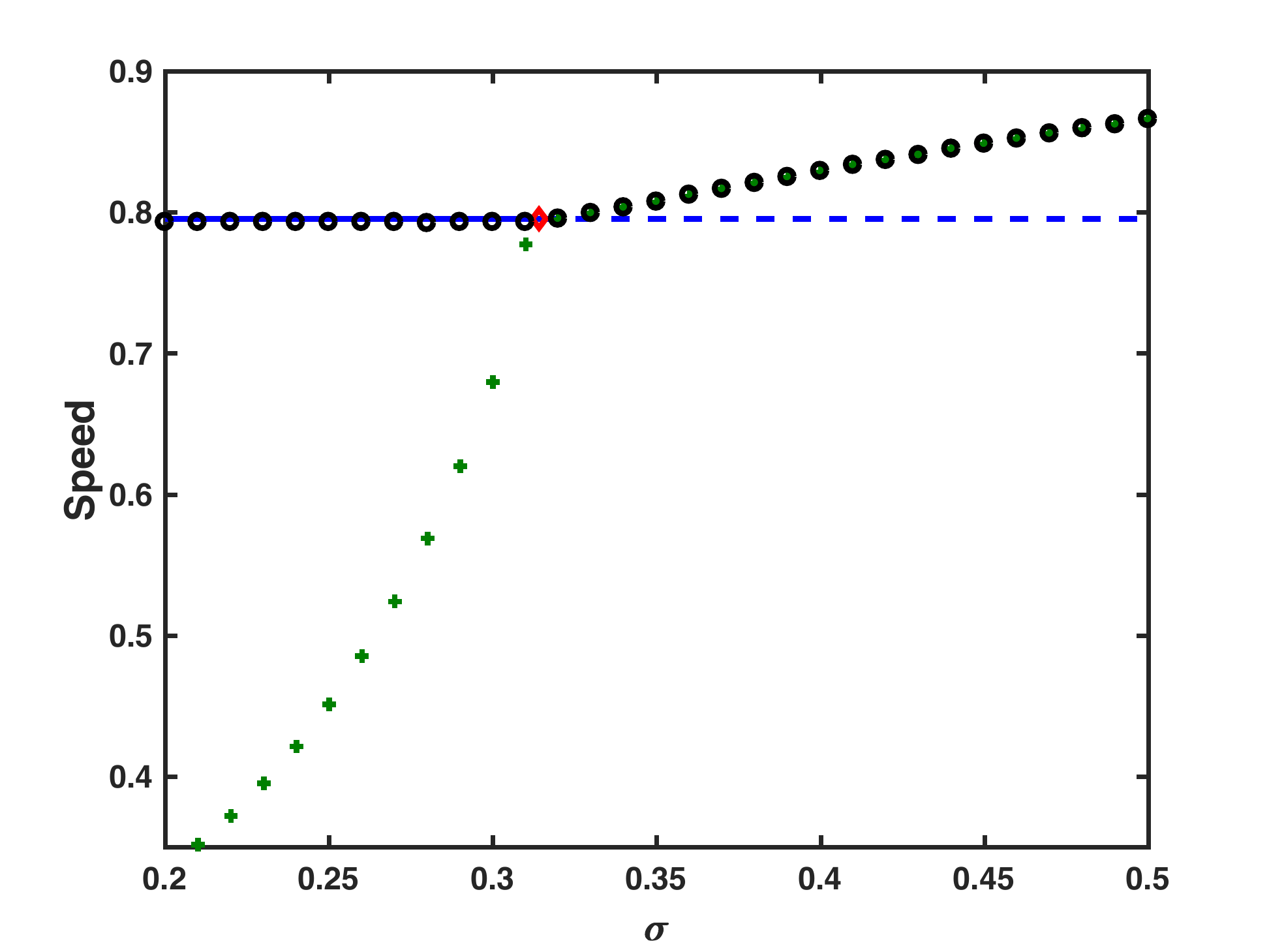}
\caption{Numerically computed wave speeds of the $u$-component, black circles, and of the $v$-component, green plus sign for $\epsilon=1$ in \eqref{eq:example1}.  The horizontal blue line $s=s^*=\sqrt{2}(a+1/2)$ represents the sign of the associated principal eigenvalue of the operator $H_\sigma$ and the red diamond indicates the critical value $\sigma^*$ at which this principal eigenvalue vanishes. The solid part of the line indicates a negative principal eigenvalue while the dashed part indicates a positive one. Here, $\sigma^*\simeq0.314$. For all numerical simulations we have set $a=1/16$.}
\label{fig:speed_super}
\end{figure}

The dynamics of (\ref{eq:nagumo}) are well understood, see for example \cite{fife77}.  For $a<1/2$, the system forms a pushed front propagating with speed $s^*=\sqrt{2}\left(\frac{1}{2}+a\right)$. For the numerical computations presented in both Figures~\ref{fig:speed_super} and \ref{fig:speed_sub}, we have discretized \eqref{eq:main} by the method of finite differences and used a semi-implicit scheme with time step $\delta t= 0.05$ and space discretization $\delta x=0.05$ with $x\in[0,400]$ and imposed Neumann boundary conditions. All simulations are done from compactly initial data and the speed of each component was calculated by computing how much time elapsed between the solution surpassing a threshold at two separate points in the spatial domain. In Figure~\ref{fig:speed_super}, we present the case of a super-critical bifurcation where locked fronts are shown to exist past the bifurcation point $\sigma=\sigma^*$. In Figure~\ref{fig:speed_sub}, we illustrate the case of a sub-critical bifurcation where locked fronts are shown to exist before the bifurcation point $\sigma=\sigma^*$. We observe a discontinuity of the wave speed as $\sigma$ is increased. We then implemented a numerical continuation scheme to continue the  wave speed of these locked fronts back to the bifurcation point $\sigma=\sigma^*$.  In the process, we see a turning point for some value of $\sigma$ near $0.273$. We expect that locked fronts on this branch to be unstable as solutions of (\ref{eq:main}) which explains why one observes the lower branch of the bifurcation curve. It is interesting to note the relative good agreement between the wave speed obtained by numerical continuation and the wave speed obtained by direct numerical simulation of the system \eqref{eq:main}.

\paragraph{Outline of the paper.} In Section~\ref{sec:ssb}, we track the strong stable manifold $W^{ss}(\P_0)$  backwards and derive expansions. In the following Section~\ref{sec:uf}, we track the unstable manifold $W^u(\P_2)$ forwards using the  Shilnikov Theorem to obtain precise asymptotics past the saddle point $\P_1$. Finally, in the last Section~\ref{sec:bif} we prove our main Theorem~\ref{thm:main} by resolving the bifurcation equation when matching the strong stable manifold $W^{ss}(\P_0)$ with the unstable manifold $W^u(\P_2)$ in a neighborhood of $\P_1$.  Some proofs and calculations are provided in the Appendix.  

\section{Tracking the strong stable manifold  $W^{ss}(\P_0)$  backwards}\label{sec:ssb}
In this section, we derive an expression for the strong stable manifold of the fixed point $\P_0$ near the fixed point $\P_1$.  Recall that for $(s,\sigma)=(s^*,\sigma^*)$, there exists a heteroclinic orbit given by $(U_p(\xi),U_p'(\xi),0,0)^T$ that connects $\P_1$ to $\P_0$.  By assumption ${\bf (H2)}$, this orbit lies in the strong stable manifold.  We will use this orbit to track the strong stable manifold back to a neighborhood of $\P_1$.  Before proceeding, we remark that $\bf{(H2)}$ and $\bf{(H3)}$ combine to provide a description of the tangent space to $W^{ss}(\P_0)$ for $(s,\sigma)=(s^*,\sigma^*)$ and at any point along the heteroclinic $(U_p(\xi),U_p'(\xi),0,0)^T$.  Importantly, we will see that the criticality of the principle eigenvalue in $\bf{(H3)}$ implies that the tangent space of $W^{ss}(\P_0)$ at $(s^*,\sigma^*)$ aligns with the unstable/weak-stable eigenspace near $\P_1$; see also $\bf{(H4)}$.  Looking ahead to Section~\ref{sec:uf}, we recall that the tracked manifold $W^u(\P_2)$ also enters a neighborhood of $\P_1$ tangent to the unstable/weak-stable manifold; see $\bf{(H5)}$.  Thus, on a linear level we anticipate intersections of these two manifolds for parameter values near $(s^*,\sigma^*)$ with a precise description involving how these individual manifolds vary with respect to $s$, $\sigma$ and their nonlinear characteristics.  

We first prove the existence of the manifold $W^{ss}(\P_0)$ and derive expansions of the manifold near $\P_1$.  To begin, change variables via
\begin{eqnarray*}
u_1&=& U_p(\xi)+p_1 \\
u_2&=&  U_p'(\xi)+p_2 \\
v_1&=& q_1  \\
v_2&=& q_2.
\end{eqnarray*}
Writing $z=(p_1,p_2,q_1,q_2)^T$, then we can express (\ref{eq:TW}) as the non-autonomous system in compact form,
\be z'=A(\xi,s^*,\sigma^*)z+n(\xi,s)+N(\xi,z,s,\sigma), \label{eq:z} \ee
where
\be A(\xi,s^*,\sigma^*)=\left(\begin{array}{cccc}  0 & 1 &0 & 0 \\ -F_u(U_p(\xi),0) & -s^* & -F_v(U_p(\xi),0) & 0 \\ 0 & 0 &0 & 1 \\ 0 & 0 &-\frac{g(U_p(\xi),0)}{\sigma^*} &-\frac{s^*}{\sigma^*} \end{array}\right), \  \ee
and 
\be 
n(\xi,s)=\left(\begin{array}{c} 0 \\ -(s-s^*)U_p'(\xi) \\0 \\ 0 \end{array}\right), \
N(z,\xi,s,\sigma)=\left(\begin{array}{c} 0 \\ N_p(z,\xi,s,\sigma) \\ 0 \\ N_q(z,\xi,s,\sigma) \end{array}\right) \ee
with
\begin{align*}
N_p(z,\xi,s,\sigma) =& -(s-s^*)p_2-\frac{F_{uu}(U_p(\xi),0)}{2}p_1^2-F_{uv}(U_p(\xi),0)p_1q_1-\frac{F_{vv}(U_p(\xi),0)}{2}q_1^2 +\mathcal{O}(3)\\
N_q(z,\xi,s,\sigma) =& -\frac{1}{\sigma^*}(s-s^*)q_2+\frac{s^*}{(\sigma^*)^2}(\sigma-\sigma^*)q_2 -\frac{G_{uv}(U_p(\xi),0)}{\sigma^*}p_1q_1 \\
& -  \frac{G_{vv}(U_p(\xi),0)}{2\sigma^*}q_1^2
+\frac{g(U_p(\xi),0)}{(\sigma^*)^2}q_1(\sigma-\sigma^*)+\mathcal{O}(3).
\end{align*}
These expressions have been simplified by noting that $G_{u}(U_p(\xi),0)=0$ and $G_{uu}(U_p(\xi),0)=0$, together with $G_{v}(U_p(\xi),0)=g(U_p(\xi),0)$.  

\begin{lemma}\label{lem:expdich} Recall $\alpha$ as defined in ${\bf (H4)}$.  Let $\Phi(\xi,\tilde{\xi})$ be the fundamental matrix solution of 
\be z'=A(\xi,s^*,\sigma^*)z.\label{eq:A} \ee
Then (\ref{eq:A}) has a generalized exponential dichotomy on $[\xi_0,\infty)$ with strong stable projection $\Ps(\xi)$ satisfying $\mathrm{dim}(\mathbf{Rg}(\Ps(\xi)))=2$, and there exists a $K>0$ and $0<\gamma<\alpha$  for which 
\begin{eqnarray*}
\left\|\Phi(\xi,\tilde{\xi})\Ps(\tilde{\xi})\right\|&\leq& K e^{-\alpha (\xi-\tilde{\xi})} \ \text{for} \ \xi>\tilde{\xi}, \\
\left\|\Phi(\xi,\tilde{\xi})\left(\mathrm{Id}-\Ps(\tilde{\xi})\right)\right\|&\leq& K e^{\gamma (\tilde{\xi}-\xi)} \ \text{for} \ \xi<\tilde{\xi}.
\end{eqnarray*}
\end{lemma}
\begin{Proof} This is a standard result on exponential dichotomies, see for example \cite{coppel78}.  Define $A_\infty(s^*,\sigma^*)=\lim_{\xi\to\infty}A(\xi,s^*,\sigma^*)$.  Since the convergence is exponential and there is a gap between the strong stable and weak stable eigenvalues, see {\bf (H4)}, the constant-coefficient asymptotic system has an exponential dichotomy and the non-autonomous system inherits one with the same decay rates . 
\end{Proof}

With the existence of an exponential dichotomy, we can express the strong stable manifold in the usual way as the fixed point of a variation-of-constants formula.  In the following, we use the the notation
\[ \Phi^{ss}(\xi,\xi_0)= \Phi(\xi,\xi_0)\Ps(\xi_0), \quad  \Phi^{ws}(\xi,\xi_0)= \Phi(\xi,\xi_0)\left(\mathrm{Id}-\Ps(\xi_0)\right).\]

\begin{lemma}\label{lem:SScontraction} Let $\xi_0<0$ be arbitrary and let $\alpha$ and $\gamma$ be as in Lemma~\ref{lem:expdich}.  Define
\[ S=\left\{ \phi\in C^0_{\beta}([\xi_0,\infty),\mathbb{R}^4)\right\}, \]
with norm $\|\phi\|_S=\sup_{\xi\in[\xi_0,\infty)} e^{\beta (\xi-\xi_0)}\|\phi(\xi)\|$ for $\gamma<\beta<\alpha$.  Given $Y\in\mathrm{Rg}\left(\Ps(\xi_0\right)) $, consider the operator $T$ defined for all $\xi\geq\xi_0$ as
\begin{equation}\label{eq:T}
\begin{split}
TQ(\xi) :=&~ \Phi^{ss}(\xi,\xi_0)Y+\int_{\xi_0}^\xi  \Phi^{ss}(\xi,\tau)\left(n(\tau,s)+N(Q(\tau),\tau,s,\sigma)\right) \md\tau  \\
 &- \int_{\xi}^\infty \Phi^{ws}(\xi,\tau)\left(n(\tau,s)+N(Q(\tau),\tau,s,\sigma)\right) \md\tau. 
\end{split}
\end{equation}
There exists an $r>0$ and a $c>0$ such that for any small $Y\in\mathrm{Rg}\left(\Ps(\xi_0\right))$ and all $(|s-s^*|+|\sigma-\sigma^*|)<c$ the operator $T$ is a contraction mapping on $\mathrm{B}_r(0)\subset S$, where $\mathrm{B}_r(0)$ stands for the ball of radius $r$ centered at $Q=0$ in $S$.
\end{lemma}
\begin{Proof} The proof is standard, but we include it since we will require some information regarding the value of the contraction constant.  Note first that $\|n(\tau,s,\sigma)\|<C|s-s^*|e^{-\alpha \tau}$.  Also, for $r$ sufficiently small there exists positive constants $l(r), l_s$ and $l_\sigma$ such that for any $\tau\in[\xi_0,\infty)$, 
\[ \|N(Q_1(\tau),\tau,s,\sigma)-N(Q_2(\tau),\tau,s,\sigma)\|\leq \left(l(r)+l_s|s-s^*|+l_\sigma |\sigma-\sigma^*|\right)\|Q_1(\tau)-Q_2(\tau)\|. \]
Note that $l(r)\to 0$ as $r\to 0$.  For brevity, let 
\[ L(r,s,\sigma)= l(r)+l_s|s-s^*|+l_\sigma |\sigma-\sigma^*|. \]
Then
\begin{align*} 
e^{\beta (\xi-\xi_0)}\|TQ(\xi)\| \leq&~ Ke^{(\beta-\alpha)(\xi-\xi_0)}\|Y\|+e^{(\beta-\alpha)\xi} \int_{\xi_0}^\xi K e^{-\alpha\tau-\beta\xi_0}\left(C|s-s^*|e^{-\alpha\tau}+L(r,s,\sigma)\|Q(\tau)\|\right)\md \tau \\
&+ e^{(\beta-\gamma)\xi} \int_\xi^\infty K e^{\gamma\tau-\beta\xi_0} \left(C|s-s^*|e^{-\alpha\tau}+L(r,s,\sigma)\|Q(\tau)\|\right)\md\tau.
\end{align*} 
Since $\beta-\alpha<0$ we obtain constants $C_g$, $C_N$ such that
\be \|TQ\|_S \leq  K\|Y\| +C_g|s-s^*|+L(r,s,\sigma)C_N\|Q\|_S,\label{eq:TQest}\ee
and we observe that for $|s-s^*|$, $|\sigma-\sigma^*|$ and $\| Y \|$ sufficiently small  the operator maps $T:\mathrm{B}_r(0)\to \mathrm{B}_r(0)$.  For any fixed  $Y$, we have
\begin{align*}
e^{\beta (\xi-\xi_0)}\|TQ_1(\xi)-TQ_2(\xi)\| \leq &~ e^{\beta(\xi-\xi_0)}
\int_{\xi_0}^\xi  
\Phi^{ss}(\xi,\tau)\|N(Q_1(\tau),\tau,s,\sigma)-N(Q_2(\tau),\tau,s,\sigma)\| \md\tau  \\
&+
 e^{\beta (\xi-\xi_0)}  \int_{\xi}^\infty \Phi^{ws}(\xi,\tau)\|N(Q_1(\tau),\tau,s,\sigma)-N(Q_2(\tau),\tau,s,\sigma)\| \md\tau \\
\leq&~ e^{\beta(\xi-\xi_0)}e^{-\alpha\xi}KL(r,s,\sigma)||Q_1-Q_2||_S\int_{\xi_0}^\xi e^{(\alpha-\beta)\tau}\md\tau
  \\
&+
 e^{\beta (\xi-\xi_0)}e^{-\gamma\xi} KL(r,s,\sigma)||Q_1-Q_2||_S \int_{\xi}^\infty e^{(\gamma-\beta)\tau} \md\tau.
\end{align*}
Since $\gamma<\beta<\alpha$ the integrals converge and we obtain that $T$ is a contraction for $L$ sufficiently small, or equivalently for $r>0$ and $c>0$ sufficiently small. And for future reference, we denote by $\kappa(r,s,\sigma)$ the associated contraction constant so that
\[
\| TQ_1-TQ_2 \|_S \leq  \kappa(r,s,\sigma) ||Q_1-Q_2||_S.
\]
\end{Proof}

The strong stable manifold is therefore given as the fixed point of (\ref{eq:T}) and at $\xi_0$ this manifold can be expressed as a graph from $\mathrm{Rg}(\Ps(\xi_0))$ to $\mathrm{Rg}(\mathrm{Id}-\Ps(\xi_0))$. We now select coordinates.  The range of the strong stable projection is spanned by the vectors
\begin{equation} \theta_1=\left(\begin{array}{c} U_p'(\xi_0) \\ U_p''(\xi_0) \\0 \\ 0 \end{array}\right), \quad \theta_2=\left(\begin{array}{c} a_1(\xi_0) \\ a_2(\xi_0) \\ \phi(\xi_0) \\ \phi'(\xi_0) \end{array}\right),
\label{eq:theta12}\end{equation}
where $\phi(\xi)$ is defined in ${\bf (H3)}$ and $a_1(\xi)$ and $a_2(\xi)$ are solutions of 
\begin{eqnarray*}
a_1'(\xi)&=& a_2(\xi) \\
a_2'(\xi)&=& - F_u(U_p(\xi),0)a_1(\xi)-s^*a_2(\xi)-F_v(U_p(\xi),0)\phi(\xi).
\end{eqnarray*}
The homogeneous equation has a pair of linearly independent solutions,
\be A_1(\xi)=U_p'(\xi), \quad A_2(\xi)=U_p'(\xi)\int_{\xi_0}^\xi \frac{e^{-s^*\tau}}{(U_p(\tau)')^2}\md\tau.\label{eq:A1A2} \ee
Note that $A_1(\xi)<0$ and $A_2(\xi)< 0$ for $\xi>\xi_0$.  A family of  solutions with strong exponential decay as $\xi\to\infty$ is given by
\begin{align} a_1(\xi)=&~ c_1A_1(\xi)+A_1(\xi)\int_{\xi_0}^\xi e^{s^*\tau}A_2(\tau)F_v(U_p(\tau),0)\phi(\tau)\md\tau  \nonumber \\
&+A_2(\xi)\int_{\xi}^\infty e^{s^*\tau}A_1(\tau)F_v(U_p(\tau),0)\phi(\tau)\md\tau. \label{eq:a1}
\end{align}
Then
\begin{align*} a_2(\xi)=&~c_1A_1'(\xi)+A_1'(\xi)\int_{\xi_0}^\xi e^{s^*\tau}A_2(\tau)F_v(U_p(\tau),0)\phi(\tau)\md\tau \\
&+A_2'(\xi)\int_{\xi}^\infty e^{s^*\tau}A_1(\tau)F_v(U_p(\tau),0)\phi(\tau)\md\tau.
\end{align*}
We select $c_1$ so that $\theta_1$ and $\theta_2$ are orthogonal at $\xi_0$.  This implies
\[ c_1\left( (U_p'(\xi_0))^2+(U_p''(\xi_0)^2)\right)=-\frac{U_p''(\xi_0)}{U_p'(\xi_0)}e^{-s^*\xi_0}\int_{\xi_0}^\infty e^{s^*\tau}A_1(\tau)F_v(U_p(\tau),0)\phi(\tau)\md\tau.\] 
We make several observations here that will be of importance later.  First, the sign of $c_1$ depends on the value of $F_v(U_p(\xi),0)$.  If $F_v(U_p(\xi),0)$ has one sign, then $c_1$ shares that sign. Second, if we set $\xi=\xi_0$ we observe that the integrand $e^{s^*\tau}A_1(\tau)F_v(U_p(\tau),0)\phi(\tau)$ converges exponentially as $\tau\to -\infty$.  Finally, we note that  $a_1(\xi)$ and $a_2(\xi)$ share the same decay rate as $\phi(\xi)$ as $\xi\to -\infty$ while their decay rate exceeds that of $\phi(\xi)$ as $\xi\to\infty$. 

The range of $(\mathrm{Id}-\Ps(\xi_0))$ can be expressed in terms of solutions to the adjoint equation, 
\be \psi'=-A(\xi,s^*,\sigma^*)^T\psi.\label{eq:adj} \ee
Note that the adjoint equation also admits a generalized exponential dichotomy with fundamental matrix solution $\tilde{\Phi}(\xi,\xi_0)=\left(\Phi(\xi,\xi_0)^{-1}\right)^T$.  The generalized exponential dichotomy distinguishes between solutions with weak and strong unstable dynamics.   The weak unstable projection for the adjoint equation has two dimensional range  spanned by,  
\begin{equation}
 \psi_1=\left(\begin{array}{c} -e^{s^*\xi_0}U_p''(\xi_0) \\ e^{s^*\xi_0}U_p'(\xi_0) \\b_1(\xi_0) \\ b_2(\xi) \end{array}\right), \quad \psi_2=e^{\frac{s^*}{\sigma^*}\xi_0}\left(\begin{array}{c} 0 \\ 0 \\ -\phi'(\xi_0) \\ \phi(\xi_0) \end{array}\right),\label{eq:psi12}\end{equation}
where $b_1(\xi)$ and $b_2(\xi)$ satisfy
\begin{eqnarray*}
b_1'(\xi)&=&\frac{g(U_p(\xi),0)}{\sigma^*} b_2(\xi)+F_v(U_p(\xi),0) e^{s^*\xi}U_p'(\xi) \\
b_2'(\xi)&=& -b_1(\xi)+\frac{s^*}{\sigma^*}b_2(\xi).
\end{eqnarray*}
This system can be re-expressed as the second order equation,
\be \sigma^* b_2''(\xi)-s^*b_2'(\xi)+g(U_p(\xi),0)b_2(\xi)=-\sigma^*F_v(U_p(\xi),0)e^{s^*\xi}U_p'(\xi).\label{eq:binhom}\ee
The homogeneous system has a pair of linearly independent solutions,
\[ B_1(\xi)=e^{\frac{s^*}{\sigma^*}\xi} \phi(\xi), \quad B_2(\xi)= e^{\frac{s^*}{\sigma^*}\xi} \phi(\xi)\int_{\xi_0}^\xi \frac{e^{-\frac{s^*}{\sigma^*}\tau}}{\phi^2(\tau)}\md\tau.  \]
Note that $B_1(\xi)$ possesses weak unstable growth as $\xi\to\infty$ and $B_2(\xi)$ has strong unstable growth.  For $\xi$ tending to $-\infty$, we have that $B_1(\xi)$ and $B_2(\xi)$ both converge exponentially to zero.

Variation of parameters yields a solution to the inhomogeneous equation  (\ref{eq:binhom}) with weak-unstable growth as $\xi\to\infty$,
\begin{align} b_2(\xi)=&~ \tilde{c}_1B_1(\xi)+B_1(\xi)\int_{\xi_0}^\xi e^{-\frac{s^*}{\sigma^*}\tau}B_2(\tau)F_v(U_p(\tau),0)e^{s^*\tau}U_p'(\tau)\md\tau  \nonumber \\
&+ B_2(\xi)\int_{\xi}^\infty e^{-\frac{s^*}{\sigma^*}\tau}B_1(\tau)F_v(U_p(\tau),0)e^{s^*\tau}U_p'(\tau)\md\tau, \label{eq:b2}
\end{align}
where we note that the integrand converges exponentially as $\tau\to\infty$ and, hence, the integral converges.  Finally, we select $\tilde{c}_1$ so that $\psi_1$ and $\psi_2$ are orthogonal.  Orthogonality requires that 
\[ -\phi'(\xi_0)\left(\frac{s^*}{\sigma^*} b_2(\xi_0)-b_2'(\xi_0)\right)+b_2(\xi_0)\phi(\xi_0)=0,\]
from which 
\[ \tilde{c}_1\left((\phi(\xi_0))^2+(\phi'(\xi_0))^2\right) = - \frac{\phi'(\xi_0)}{\phi(\xi_0)}\int_{\xi_0}^\infty e^{-\frac{s^*}{\sigma^*}\tau}B_1(\tau)F_v(U_p(\tau),0)e^{s^*\tau}U_p'(\tau)\md\tau.\]
We introduce the notation,
\be \Omega_1={\langle \psi_1(\xi_0),\psi_1(\xi_0)\rangle}, \quad \Omega_2={\langle \psi_2(\xi_0),\psi_2(\xi_0)\rangle}. \label{eq:omega} \ee
\begin{lemma}\label{lem:h1h2} There exists functions $h_1$ and $h_2$ such that the manifold $W^{ss}(\P_0)$ can be expressed as 
\be \left(\begin{array}{c} U_p(\xi_0) \\ U_p'(\xi_0) \\0 \\ 0 \end{array}\right)+\eta_1\theta_1+\eta_2\theta_2+(s-s^*)\Gamma_0 \psi_1+h_1(\eta_1,\eta_2,s,\sigma)\psi_1+h_2(\eta_1,\eta_2,s,\sigma)\psi_2, \label{eq:wssexp} \ee
where $h_{1,2}$ are quadratic or higher order in all their arguments.  Expansions of $h_{1,2}$ are obtained in Appendix~\ref{sec:hexp}.
\end{lemma}
\begin{Proof}
Given $Y=\eta_1\theta_1+\eta_2\theta_2 \in\mathrm{Rg}\left(\Ps(\xi_0\right))$ and $|\eta_1|+|\eta_2|+|s-s^*|+|\sigma-\sigma^*|$ small enough, let $Q^*(\cdot,\eta_1,\eta_2,s,\sigma) \in S$ be the unique fixed point solution to $TQ^*=Q^*$ in $\mathrm{B}_r(0)$  from which evaluating (\ref{eq:T}) at $\xi=\xi_0$ we obtain
\begin{equation} Q^*(\xi_0,\eta_1,\eta_2,s,\sigma)=\eta_1\theta_1+\eta_2\theta_2- \int_{\xi_0}^\infty \Phi^{ws}(\xi_0,\tau)\left(n(\tau,s)+N(Q^*(\tau,\eta_1,\eta_2,s,\sigma),\tau,s,\sigma)\right) \md\tau.
\label{eq:defQs}
\end{equation}
Using the fact that $\mathrm{P}^{ss}(\xi)\Phi(\xi,\tau)=\Phi(\xi,\tau)\mathrm{P}^{ss}(\tau)$  we have that 
\[\Phi^{ws}(\xi_0,\tau)=\Phi(\xi_0,\tau)\left(\mathrm{Id}-\mathrm{P}^{ss}(\tau)\right)= \left(\mathrm{Id}-\mathrm{P}^{ss}(\xi_0)\right)\Phi(\xi_0,\tau),\]
which shows that the second term in \eqref{eq:defQs} belongs to $\mathrm{Rg}\left(\mathrm{Id}-\Ps(\xi_0)\right)$ and thus
\begin{eqnarray*}
\Ps(\xi_0)Q^*(\xi_0,\eta_1,\eta_2,s,\sigma)&=&\eta_1\theta_1+\eta_2\theta_2,\\
\left(\mathrm{Id}-\Ps(\xi_0)\right)Q^*(\xi_0,\eta_1,\eta_2,s,\sigma)&=&- \int_{\xi_0}^\infty \Phi^{ws}(\xi_0,\tau)\left(n(\tau,s)+N(Q^*(\tau,\eta_1,\eta_2,s,\sigma),\tau,s,\sigma)\right) \md\tau.
\end{eqnarray*}
It is first easy to check, using the specific form of $n(\tau,s)$ that
\begin{align*}
\int_{\xi_0}^\infty \left\langle \psi_1(\xi_0),\Phi^{ws}(\xi_0,\tau)n(\tau,s)\right\rangle \md\tau = \int_{\xi_0}^\infty \big\langle \psi_1(\tau),n(\tau,s) \big\rangle \md\tau = -(s-s^*) \int_{\xi_0}^\infty e^{s^*\tau}\left(U_p'(\tau)\right)^2\md\tau,
\end{align*}
together with 
\[
\int_{\xi_0}^\infty \left\langle \psi_2(\xi_0),\Phi^{ws}(\xi_0,\tau)n(\tau,s)\md\tau \right\rangle =\int_{\xi_0}^\infty \big\langle \psi_2(\tau),n(\tau,s) \big\rangle \md\tau =0.
\]
As a consequence, there exists $h_{1,2}(\eta_1,\eta_2,s,\sigma)$ so that equation \eqref{eq:defQs} can be written as
\[
Q^*(\xi_0,\eta_1,\eta_2,s,\sigma)=\eta_1\theta_1+\eta_2\theta_2+(s-s^*)\Gamma_0\psi_1+h_1(\eta_1,\eta_2,s,\sigma)\psi_1+h_2(\eta_1,\eta_2,s,\sigma)\psi_2,
\]
where
\begin{align*}
\Gamma_0 &:= \frac{1}{\Omega_1} \int_{\xi_0}^\infty e^{s^*\tau}\left(U_p'(\tau)\right)^2\md\tau, \\
h_{1,2}(\eta_1,\eta_2,s,\sigma) &:= -\frac{1}{\Omega_{1,2}} \int_{\xi_0}^\infty \left\langle \psi_{1,2}(\xi_0),\Phi^{ws}(\xi_0,\tau)N(Q^*(\tau,\eta_1,\eta_2,s,\sigma),\tau,s,\sigma)\right\rangle \md\tau  ,
\end{align*}
and $\Omega_{1,2}$ have been introduced in \eqref{eq:omega}. 

In the remaining of the proof, we show that the maps $h_{1,2}(\eta_1,\eta_2,s,\sigma)$ are at least quadratic in their arguments and present  a procedure which allows one to compute the leading order terms in their expansions, the explicit formulae being provided in Appendix~\ref{sec:hexp}. Let $Q^0(\xi):=\eta_1\theta_1(\xi)+\eta_2\theta_2(\xi)+(s-s^*)\theta_s(\xi)$ where 
\[  (s-s^*)\theta_s(\xi)= \int_{\xi_0}^\xi  \Phi^{ss}(\xi,\tau)n(\tau,s) \md\tau - \int_{\xi}^\infty \Phi^{ws}(\xi,\tau)n(\tau,s)\md\tau, \]
with  $\theta_s(\xi_0)=\Gamma_0\psi_1$. Define now $Q^1:=TQ^0$, that is
\begin{equation}\label{eq:Q1}
\begin{split}
Q^1(\xi) =&~ \Phi^{ss}(\xi,\xi_0)Y+\int_{\xi_0}^\xi  \Phi^{ss}(\xi,\tau)\left(n(\tau,s)+N(Q^0(\tau),\tau,s,\sigma)\right) \md\tau  \\
 &- \int_{\xi}^\infty \Phi^{ws}(\xi,\tau)\left(n(\tau,s)+N(Q^0(\tau),\tau,s,\sigma)\right) \md\tau. 
\end{split}
\end{equation}
Let us remark that $\Phi^{ss}(\xi,\xi_0)Y=\Phi^{ss}(\xi,\xi_0)\left(\eta_1\theta_1+\eta_2\theta_2 \right)=\eta_1\theta_1(\xi)+\eta_2\theta_2(\xi)$ for any $\xi\geq\xi_0$ such that \eqref{eq:Q1} can be written in a condensed form
\begin{equation*}
Q^1(\xi)=Q^0(\xi)+\int_{\xi_0}^\xi  \Phi^{ss}(\xi,\tau)N(Q^0(\tau),\tau,s,\sigma) \md\tau- \int_{\xi}^\infty \Phi^{ws}(\xi,\tau)N(Q^0(\tau),\tau,s,\sigma) \md\tau.
\end{equation*}
From the contraction mapping theorem, we find that $\|Q^1-Q^*\|_S<\frac{\kappa}{1-\kappa}\|Q^1-Q^0\|_S$ where $\kappa(r,s,\sigma)$ is the contraction constant from Lemma~\ref{lem:SScontraction}.  Essentially repeating the estimate in (\ref{eq:TQest}), we also find that there exists a constant $C_L>0$ for which 
\be \|Q^1-Q^0\|_S\leq C_L L(r,s,\sigma) \|Q^0\|_S.\label{eq:contractionest} \ee
Let $\xi=\xi_0$ in \eqref{eq:Q1} to obtain
\[ Q^1(\xi_0)= \eta_1\theta_1+\eta_2\theta_2+(s-s^*)\Gamma_0 \psi_1- \int_{\xi_0}^\infty \Phi^{ws}(\xi_0,\tau)N(\eta_1\theta_1(\tau)+\eta_2\theta_2(\tau)+(s-s^*)\theta_s(\tau),\tau,s,\sigma) \md\tau.\]
The inequality (\ref{eq:contractionest}) implies that $h_{1,2}$ are at least quadratic in their arguments and that we can compute terms up to quadratic order in $h_{1,2}$ by projecting $Q^1(\xi_0)$ onto $\psi_{1,2}$.  We now refer to Appendix~\ref{sec:hexp} for the quadratic expansions of the maps $h_{1,2}$.
\end{Proof}

\begin{rmk}\label{rmk:thetas} An explicit expression for $\theta_s$ can be obtained in a fashion analogous to that of the terms $a_{1,2}(\xi)$.  Namely, we find that $\theta_s(\xi)=(\theta_s^1(\xi),\theta_s^2(\xi),0,0)^T$ solves
\begin{eqnarray*}
 \frac{\md \theta_s^1}{\md \xi} &=& \theta_s^2  \\
 \frac{\md \theta_s^1}{\md \xi}&=& -s^* \theta_s^2 -F_u(U_p(\xi),0)\theta_s^1-U_p'(\xi).
\end{eqnarray*}
Then a solution with strong exponential decay as $\xi\to\infty$ is given by
\be \label{eq:thetas} \theta_s^1(\xi)= \hat{c}_1A_1(\xi)+A_1(\xi)\int_{\xi_0}^\xi e^{s^*\tau}A_2(\tau)U_p'(\tau) \md\tau+A_2(\xi)\int_{\xi}^\infty e^{s^*\tau}A_1(\tau)U_p'(\tau) \md\tau, \ee
where $A_{1,2}(\xi)$ are defined in (\ref{eq:A1A2}) and $\hat{c}_1$ is chosen so that $\theta_s(\xi_0)$ is orthogonal to $\theta_1$.  
\end{rmk}

\subsection{The tangent space of $W^{ss}(\P_0)$}\label{sec:tangent}
Before proceeding to a local analysis of the dynamics near $\P_1$, we pause to comment on the behavior of the tangent space of $W^{ss}(\P_0)$ in the limit as $\xi\to -\infty$.  
This is most easily accomplished in the coordinates of (\ref{eq:z}), where we focus on the system $z'=A(\xi,s^*,\sigma^*)z$ with $z=(p_1,p_2,q_1,q_2)^T$.  

We will be interested in tracking the tangent space of $W^{ss}(\P_0)$ backwards along $(U_p(\xi),U_p'(\xi),0,0)$ until it reaches a neighborhood of $\P_1$.  We will first show that for $s=s^*$ and $\sigma=\sigma^*$ that this tangent space will align with the weak-unstable eigenspace of $\P_1$.  Here the weak-unstable eigenspace includes both the unstable eigendirection as well as the weak-stable eigendirection corresponding to the eigenvalue $\nu_v^+(s,\sigma)$;  recall assumption $\bf{(H4})$.  The fact that this alignment occurs at $s=s^*$ and $\sigma=\sigma^*$ is to be expected.  First, due to the existence of the pushed front we know that $W^u(\P_1)\cap W^{ss}(\P_0) \neq \emptyset$ so that their tangent spaces must also intersect.  Second, assumption ${\bf (H3)}$ gives the existence of second, linearly independent vector (see $\theta_2(\xi)$ in (\ref{eq:theta12})) that converges to the weak-stable eigendirection  associated to $\nu_v^+(s^*,\sigma^*)$.  After verifying this, we turn our attention to computing how this tangent space perturbs with $s$ and $\sigma$.  This is more involved and complicated by the unboundedness of individual vectors near the weak-stable eigenspace as $\xi\to -\infty$.  To deal with this, we use a generalization of projective coordinates that was used in \cite{sandstede00}.  

To begin, we track two dimensional subspaces using the coordinates, 
\[ \left(\begin{array}{c} p_2 \\ q_2 \end{array}\right)= \left(\begin{array}{cc} z_{11} &  z_{12} \\ 0 & z_{22} \end{array}\right)\left(\begin{array}{c} p_1 \\ q_1 \end{array}\right)
\]
wherein,
\begin{eqnarray}
z_{11}' &=& -s^*z_{11}-F_u(U_p(\xi),0)-z_{11}^2 \nonumber \\
z_{12}'&=& -s^*z_{12} -F_v(U_p(\xi),0) -z_{12}(z_{11}+z_{22}) \label{eq:proj} \\
z_{22}' &=& -\frac{s^*}{\sigma^*}z_{22}-\frac{1}{\sigma^*}g(U_p(\xi),0)-z_{22}^2. \nonumber
\end{eqnarray}
Using the expressions for the vectors $\theta_1(\xi)$ and $\theta_2(\xi)$, we find corresponding solutions 
\[ Z_{11}(\xi)=\frac{U_p''(\xi)}{U_p'(\xi)}, \ Z_{22}(\xi)=\frac{\phi'(\xi)}{\phi(\xi)}
, \ Z_{12}(\xi)= \frac{a_2(\xi)}{\phi(\xi)}-\frac{U_p''(\xi)}{U_p'(\xi)}\frac{a_1(\xi)}{\phi(\xi)}.\]
The tangent space of the manifold $W^{ss}(\P_0)$ is then expressed as a graph over $p_1$ and $q_1$ coordinates
\[ \left(\begin{array}{c} p_2 \\ q_2 \end{array}\right)= \left(\begin{array}{cc} Z_{11}(\xi) &  Z_{12}(\xi) \\ 0 & Z_{22}(\xi) \end{array}\right)\left(\begin{array}{c} p_1 \\ q_1 \end{array}\right).
\]
A calculation reveals that the expression for $Z_{12}$ can be simplified to 
\[ Z_{12}(\xi)=\frac{1}{\phi(\xi)}\left( \frac{e^{-s^*\xi}}{U_p'(\xi)}\int_\xi^\infty e^{s^*\tau} U_p'(\tau) F_v(U_p(\tau),0)\phi(\tau)\md \tau\right) .\]
It follows from Hypothesis ${\bf (H3)}$ that as $\xi\to -\infty$, 
\[ Z_{11}(\xi)\to \nu_u^+(s^*), \ Z_{22}(\xi)\to \nu_v^+(s^*,\sigma^*), \ Z_{12}(\xi)\to \frac{-F_v(\p_1)}{s+\nu_u^+(s^*)+\nu_v^+(s^*,\sigma^*)},\]
which we verify to be fixed points of the system (\ref{eq:proj}).  These fixed points correspond to the  unstable and weak stable eigenvectors for $\P_1$ (see (\ref{eq:evec}) below) and we have shown that $\mathrm{span}\{\theta_1(\xi),\theta_2(\xi)\}$ coincides with the weak-unstable eigenspace of $\P_1$ in the limit as $\xi\to -\infty$.  

To understand how this heteroclinic perturbs with $s$ and $\sigma$, we let 
\[ \left(\begin{array}{c} z_{11} \\ z_{12} \\ z_{22}\end{array}\right)=\left(\begin{array}{c} Z_{11}(\xi)+\zeta_{11} \\ Z_{12}(\xi)+\zeta_{12} \\ Z_{22}(\xi)+\zeta_{22} \end{array}\right).\]
Let $\Xi=(\zeta_{11},\zeta_{12},\zeta_{22})^T$, then we obtain
\be \Xi'=A(\xi,s^*,\sigma^*)\Xi+(s-s^*)n(\xi)+(\sigma-\sigma^*) m(\xi)+N(\Xi,s,\sigma) ,\label{eq:Xi} \ee
where we have momentarily re-purposed the notations $A$, $n$ and $m$ with, 
\[ A(\xi,s^*,\sigma^*):=\left(\begin{array}{ccc} 
 -s^*-2Z_{11}(\xi) & 0 & 0 \\
 -Z_{12}(\xi) & -s^*-Z_{11}(\xi)+Z_{22}(\xi) & -Z_{12}(\xi)\\
 0 & 0 & -\frac{s^*}{\sigma^*}-2Z_{22}(\xi)
\end{array}\right) \]
and 
\[ n(\xi):=\left(\begin{array}{c} 
 -Z_{11}(\xi) \\
 -Z_{12}(\xi)\\
-\frac{1}{\sigma^*}Z_{22}(\xi)
\end{array}\right) , \quad m(\xi) :=\left(\begin{array}{c} 
0 \\
 0\\
\frac{1}{(\sigma^*)^2}g(U_p(\xi),0)+\frac{s^*}{(\sigma^*)^2} Z_{22}(\xi)
\end{array}\right).\]
Since we are only interested in the linear dependence on $s$ and $\sigma$, we henceforth ignore the nonlinear terms $N(\Xi,s,\sigma)$.  We will also require  linearly independent solutions to the associated adjoint equation,
\[ \psi'=-A^T(\xi,s^*,\sigma^*)\psi.\]
The adjoint equations form a system
\begin{eqnarray*}
\psi_{11}'&=& (s^*+2Z_{11}(\xi))\psi_{11}+Z_{12}\psi_{12}, \\
\psi_{12}'&=& (s^*+Z_{11}(\xi)+Z_{22}(\xi))\psi_{12},\\
\psi_{22}'&=& Z_{12}(\xi)\psi_{12}+\left(\frac{s^*}{\sigma^*}+2Z_{22}(\xi)\right)\psi_{22},
\end{eqnarray*}
We have solutions
\[ \psi_3(\xi)=\left(\begin{array}{c} C_3(\xi) \\ 0 \\ 0 \end{array}\right),
\psi_4(\xi)=\left(\begin{array}{c} C_4(\xi) \\ D_4(\xi) \\ E_4(\xi) \end{array}\right), \psi_5(\xi)=\left(\begin{array}{c}  0 \\ 0 \\ E_5(\xi) \end{array}\right),\]
with 
\[ C_3(\xi)=\left(U_p'(\xi)\right)^2 e^{s^*\xi}, \quad  D_4(\xi)= U_p'(\xi) \phi(\xi)e^{s^*\xi}, \quad E_5(\xi)=\phi(\xi)^2e^{\frac{s^*}{\sigma^*}\xi}.\]
Requiring orthogonality of the three vectors at $\xi=\xi_0$ implies that $C_4(\xi_0)=E_4(\xi_0)=0$.  Let $\Phi(\xi,\xi_0)$ be the fundamental matrix solution to $\Xi'=A(\xi,s^*,\sigma^*)\Xi$.  Bounded solutions of (\ref{eq:Xi}) can be expressed in integral form as 
\[ \Xi(\xi)= \int_{-\infty}^\xi \Phi(\xi,\tau) \left((s-s^*)n(\tau)+(\sigma-\sigma)^* m(\tau) +N(\Xi(\tau),s,\sigma)\right)\md\tau.\]

We focus on the leading order dependence on $\sigma$.  At $\xi=\xi_0$, we write 
\[ \Xi(\xi_0)= h_3(\sigma) \psi_3(\xi_0)+ h_4(\sigma) \psi_4(\xi_0)+ h_5(\sigma)\psi_5(\xi_0).\]
Observe that $h_3(\sigma)=0$ due to the block structure of $\Phi(\xi,\xi_0)$ and the specific form of $m(\xi)$.  We focus first on the projection onto $\psi_5$
\begin{align}h_5(\sigma)= \frac{ \langle \psi_5(\xi_0),\Xi(\xi_0)\rangle}{E_5^2(\xi_0)}& = \frac{(\sigma-\sigma^*)}{E_5^2(\xi_0)}\int_{-\infty}^{\xi_0}\langle \psi_5(\xi_0), \Phi(\xi_0,\tau) m(\tau)\rangle \md\tau \nonumber\\
& =  \frac{(\sigma-\sigma^*)}{E_5^2(\xi_0)}\int_{-\infty}^{\xi_0}E_5(\tau)  \left( \frac{1}{(\sigma^*)^2}g(U_p(\tau),0)+\frac{s^*}{(\sigma^*)^2}  Z_{22}(\tau)\right)\md\tau \nonumber\\
& =  \frac{(\sigma-\sigma^*)}{E_5^2(\xi_0)}\int_{-\infty}^{\xi_0} e^{\frac{s^*}{\sigma^*}\tau}\left(\frac{1}{(\sigma^*)^2}g(U_p(\tau),0) \phi(\tau)^2+\frac{s^*}{(\sigma^*)^2} \phi(\tau)\phi'(\tau) \right)\md\tau.\label{eq:h5sigma}
\end{align}
In a similar fashion we compute,
\begin{align}h_4(\sigma)= \frac{ \langle \psi_4(\xi_0),\Xi(\xi_0)\rangle}{D_4^2(\xi_0)}& =\frac{(\sigma-\sigma^*)}{D_4^2(\xi_0)}\int_{-\infty}^{\xi_0}\langle \psi_4(\xi_0),  \Phi(\xi_0,\tau) m(\tau)\rangle \md\tau \nonumber \\
&= \frac{(\sigma-\sigma^*)}{D_4^2(\xi_0)}\int_{-\infty}^{\xi_0} E_4(\tau) \left( \frac{1}{(\sigma^*)^2}g(U_p(\tau),0)+\frac{s^*}{(\sigma^*)^2}  Z_{22}(\tau)\right)\md\tau .
\label{eq:h4sigma}
\end{align}

Returning now to the original change of coordinates, we find 
\[ \left(\begin{array}{c} p_2 \\ q_2 \end{array}\right)= \left(\begin{array}{cc} Z_{11}(\xi_0) &  Z_{12}(\xi_0)+ h_4(\sigma)D_4(\xi_0) \\ 0 & Z_{22}(\xi_0)+ h_5(\sigma)E_5(\xi_0)\end{array} \right)\left(\begin{array}{c} p_1 \\ q_1 \end{array}\right).
\]
This describes a two dimensional subspace of the form,
\be \mathcal{R}(p_1,q_1,\sigma)= \left(\begin{array}{c} 
p_1 \\
Z_{11}(\xi_0)p_1+Z_{12}(\xi_0)q_1 + h_4(\sigma)D_4(\xi_0)q_1 \\
q_1 \\
\left(Z_{22}(\xi_0)+ h_5(\sigma)E_5(\xi_0)\right)q_1 
\end{array}\right).
\label{eq:R} \ee
We now decompose this subspace into the basis $\{ \theta_1,\theta_2,\psi_1,\psi_2\}$.  To recover $\theta_1$, we require $\sigma=\sigma^*$,  $p_1=U_p'(\xi_0)$ and $q_1=0$. To recover $\theta_2$, we require $\sigma=\sigma^*$,  $p_1=a_1(\xi_0)$ and $q_1=\phi(\xi_0)$. Projecting onto $\psi_2$, we find
\be \langle \psi_2, \mathcal{R}(p_1,q_1,\sigma)\rangle = \frac{q_1 (\sigma-\sigma^*)}{\phi(\xi_0)}\int_{-\infty}^{\xi_0} e^{\frac{s^*}{\sigma^*}\tau}\left(\frac{1}{(\sigma^*)^2}g(U_p(\tau),0) \phi(\tau)^2+\frac{s^*}{(\sigma^*)^2} \phi(\tau)\phi'(\tau) \right)\md\tau,
\label{eq:psi2Rpqs}\ee
and projecting onto $\psi_1$, we find 
\be \langle \psi_1, \mathcal{R}(p_1,q_1,\sigma)\rangle = \frac{q_1 (\sigma-\sigma^*)}{\phi(\xi_0)} \int_{-\infty}^{\xi_0}b_2(\tau)\left(\frac{1}{(\sigma^*)^2}g(U_p(\tau),0)\phi(\tau)+\frac{s^*}{(\sigma^*)^2}\phi'(\tau)\right)\md\tau.
\label{eq:psi1Rpqs}\ee
We refer to Lemma~\ref{lemexppsiR} of the Appendix for the details of the computations.

\section{Tracking the unstable manifold $W^u(\P_2)$ forwards}\label{sec:uf}
We now derive an expression for $W^u(\P_2)$ in a neighborhood of the fixed point $\P_1$.  Hypothesis {\bf (H5)} will be key here.  We delay a precise description of this assumption and its consequences until Section~\ref{sec:shilnikov} and instead begin with a required normal form transformation for the traveling wave equation in a neighborhood of $\P_1$.  

\subsection{A normal form in a neighborhood of $\P_1$}
We begin with a local analysis of the dynamics of (\ref{eq:TW}) near the fixed point $\P_1=(u^+,0,0,0)^T$.  The Jacobian evaluated at this fixed point is
\[ Df(\P_1)=\left(\begin{array}{cccc} 0 & 1 &0 & 0 \\ -F_u(\p_1) & -s & -F_v(\p_1) & 0 \\ 0 & 0 &0 & \frac{1}{\sigma} \\ 0 & 0 &-\frac{g(\p_1)}\sigma &-\frac{s}{\sigma} \end{array}\right),\]
where we note that $G_u(\p_1)=0$ and hence the linearization is block triangular and the eigenvalues and eigenvectors can be computed explicitly.  The characteristic polynomial is $d(\nu)=d_u(\nu)d_v(\nu)=(\nu^2+s\nu+F_u(\p_1))(\sigma\nu^2+s\nu+g(\p_1))$. 
The eigenvalues are
\begin{eqnarray*}
\nu_u^\pm(s) &=& -\frac{s}{2}\pm\frac{1}{2}\sqrt{s^2-4F_u(\p_1)} \\
 \nu_v^\pm(s,\sigma) &=& -\frac{s}{2\sigma}\pm\frac{1}{2\sigma}\sqrt{s^2-4\sigma g(\p_1)}.
\end{eqnarray*}
Recall Hypothesis {\bf (H4)} and the assumed ordering $\nu_v^-(s,\sigma)<\nu_u^-(s)<\nu_v^+(s,\sigma)<0<\nu_u^+(s)$.  The corresponding eigenvectors are 
\be e_u^\pm(s) =\left(\begin{array}{c} 1 \\ \nu_u^\pm(s) \\0 \\ 0\end{array}\right), \quad e_v^\pm(s,\sigma) =\left(\begin{array}{c} -\frac{F_v(\p_1)}{d_u(\nu_v^\pm(s,\sigma))} \\ -\frac{F_v(\p_1) \nu_v^\pm(s,\sigma)}{d_u(\nu_v^\pm(s,\sigma))} \\1 \\ \nu_v^\pm(s,\sigma) \end{array}\right).\label{eq:evec} \ee
We introduce new coordinates, first by shifting the fixed point $\P_1$ to the origin and then  diagonalizing the linearization via
\[ \left(\begin{array}{c} u_1-u^+ \\ u_2 \\ v_1 \\ v_2 \end{array}\right)=T(s,\sigma)\left(\begin{array}{c} \yu \\ \yus \\ \yw \\ \yvs \end{array}\right),\]
where 
\begin{equation} T(s,\sigma) := \left(\begin{array}{cccc} 1 &  1 & -\frac{F_v(\p_1)}{d_u(\nu_v^+(s,\sigma))}  & -\frac{F_v(\p_1)}{d_u(\nu_v^-(s,\sigma))} \\ \nu_u^+(s)  & \nu_u^-(s) & -\frac{F_v(\p_1)\nu_v^+(s,\sigma)}{d_u(\nu_v^+(s,\sigma))}& -\frac{F_v(\p_1)\nu_v^-(s,\sigma)}{d_u(\nu_v^-(s,\sigma))} \\ 0& 0 & 1 & 1 \\ 0& 0 & \nu_v^+(s,\sigma) &  \nu_v^- (s,\sigma)\end{array}\right).
\label{matrixT}\end{equation}
In these new coordinates, the vector field assumes the form,
\begin{equation}
\begin{split}
\frac{d\yu}{d\xi} &= \nu_u^+(s) \yu+ \mathcal{N}_u(\Yv,s,\sigma),  \\
\frac{d\yw}{d\xi} &= \nu_v^+(s,\sigma) \yw +\mathcal{N}_{ws}(\Yv,s,\sigma), \\
\frac{d\yus}{d\xi} &= \nu_u^-(s) \yus +\mathcal{N}_{ss,u}(\Yv,s,\sigma), \\
\frac{d\yvs}{d\xi} &= \nu_v^-(s,\sigma) \yvs +\mathcal{N}_{ss,v}(\Yv,s,\sigma) . 
\end{split}
\label{eq:diagonalized}
\end{equation}
Invariance of the $v_1=v_2=0$ subspace implies that $\mathcal{N}_{ws}(\yu,0,\yus,0,s,\sigma)=0$ and $ \mathcal{N}_{ss,v}(\yu,0,\yus,0,s,\sigma)=0$.  We expand the nonlinear terms as follows to isolate the quadratic terms,
\begin{equation}
\mathcal{N}_u(\Yv,s,\sigma)=\sum_{i+j+k+l=2} (\yu)^i (\yw)^j (\yus)^k (\yvs)^l \mathbf{N}_u^{(i,j,k,l)}(s,\sigma)+\O(3), 
\label{nonlinearN}
\end{equation}
with the natural analogs for $\mathcal{N}_{ws}$ $\mathcal{N}_{ss,u}$ and $\mathcal{N}_{ss,v}$.

The goal is to perform a Shilnikov type analysis of the origin in (\ref{eq:diagonalized}) and obtain asymptotic expansions for solutions that enter a neighborhood of the origin near the weak-stable eigendirection and exit near the unstable manifold.  To do this a sequence of near-identity coordinate changes are required to place (\ref{eq:diagonalized}) into a suitable normal form.  These changes of coordinates are outlined in \cite{homburg}, but we include them in detail here because they will be relevant for deriving the bifurcation equations later.  

\paragraph{Straightening of the stable and unstable manifolds}
The origin is a hyperbolic equilibrium for  (\ref{eq:diagonalized}) with corresponding stable and unstable manifolds.  The following result transforms (\ref{eq:diagonalized}) into new coordinates where these stable and unstable manifolds have been straightened.

\begin{lemma}\label{lem:cov1} There exists a smooth change of coordinates,
\begin{equation}
\begin{split}
\zu &= \yu-\mathcal{H}_s(\yw,\yus,\yvs,s,\sigma) \\
\zw &=  \yw  \\
\zus &= \yus -\mathcal{H}_u(\yu,s)   \\
\zvs &=  \yvs, 
\end{split}
\label{eq:cov1}
\end{equation}
defined on a neighborhood of the origin that transforms (\ref{eq:diagonalized}) to the system
\begin{eqnarray}
\frac{d\zu}{d\xi} &=& \nu_u^+(s) \zu+ \mathcal{M}_u(\zu,\zw,\Zs,s,\sigma) \nonumber \\
\frac{d\zw}{d\xi} &=& \nu_v^+(s,\sigma) \zw +\gamma_{11}(\zu,\zw,\Zs,s,\sigma)\zw+\gamma_{12}(\zu,\zw,\Zs,s,\sigma)\Zs \label{eq:maindiag2} \\
\frac{d\Zs}{d\xi} &=& \Lambda_{ss}(s,\sigma) \Zs+\gamma_{21}(\zu,\zw,\Zs ,s,\sigma)\zw+\gamma_{22}(\zu,\zw,\Zs,s,\sigma)\Zs, \nonumber
\end{eqnarray}
where we have let $\Zs=(\zus,\zvs)^T$ and $\Lambda_{ss}(s,\sigma)=\mathrm{diag}(\nu_u^-(s),\nu_v^-(s,\sigma))$ and we have that $\mathcal{M}_u(0,\zw,\Zs,s,\sigma)=0$.   

\end{lemma}

\begin{Proof} The origin in (\ref{eq:diagonalized}) is hyperbolic with smooth stable and unstable manifolds.  The unstable manifold is contained within the invariant sub-space $\yw=\yvs=0$ and can be expressed as the graph
\[ \yus=\mathcal{H}_u(\yu,s),\]
which admits the expansion,
\[ \mathcal{H}_u(\yu,s)=\frac{\mathbf{N}_{ss,u}^{(2,0,0,0)}(s,\sigma)}{2\nu_u^+(s)-\nu_u^-(s)} (\yu)^2 +\O(3).\]
Let us remark here that $\mathbf{N}_{ss,u}^{(2,0,0,0)}(s,\sigma)$ does not depend on $\sigma$ and can be expressed as
\bqs
\mathbf{N}_{ss,u}^{(2,0,0,0)}(s,\sigma)=-\frac{F_{uu}(\p_1)}{2(\nu_u^-(s)-\nu_u^+(s))}.
\eqs
The proof of this statement is left to the Appendix (see Lemma~\ref{lemexpN}). The stable manifold has a similar expansion,
\[ \mathcal{H}_s(\yw,\yus,\yvs)=\sum_{j+k+l=2} \mathbf{n}_{(i,j,k)}(s,\sigma)(\yw)^j (\yus)^k (\yvs)^l+\O(3),\]
where 
\[ \mathbf{n}_{(j,k,l)}(s,\sigma)=\frac{\mathbf{N}_u^{(0,j,k,l)}(s,\sigma)}{j\nu_v^+(s,\sigma)+k\nu_u^-(s)+l\nu_v^-(s,\sigma)-\nu_u^-(s)}. \]
Following these changes of coordinates, we have transformed system (\ref{eq:diagonalized}) into (\ref{eq:maindiag2}) as required.  
\end{Proof}

\paragraph{Removal of terms $\gamma_{j1}(\zu,0,0,s,\sigma)$}
We will eventually employ a Shilnikov type analysis where solutions of (\ref{eq:maindiag2}) are obtained as solutions of a boundary value problem on the interval $\xi\in [0,T]$ with $T\gg 1$.  This boundary value problem imposes conditions on the unstable coordinate at  $\xi=T$ and thereby the instability is controlled by evolving that coordinate backwards.  One would then hope that the linear behavior would dominate in (\ref{eq:maindiag2}).  This is not the case due to the presence of the terms $\gamma_{j1}(\zu,0,0,s,\sigma)$.  To obtain useful asymptotics, we require a further change of coordinates that removes those terms. This is accomplished in the following lemma.

\begin{lemma}\label{lem:cov2} There exists functions $p(\zu,s,\sigma)$ and $q(\zu,s,\sigma)$, with $p:\mathbb{R}^3\to\mathbb{R}$, and $q:\mathbb{R}^3\to\mathbb{R}^2$,  valid for $|\zu|+|s-s^*|+|\sigma-\sigma^*|$ sufficiently small such that the change of coordinates,
\begin{equation} 
\begin{split}
\bzw &= \zw (1-p(\zu,s,\sigma))  \\
\bZs &= \Zs-q(\zu,s,\sigma)\zw, 
\end{split}
\label{eq:bartrans} 
\end{equation}
transforms (\ref{eq:maindiag2}) to the normal form
\begin{equation}
\begin{split}
\frac{d\zu}{d\xi} &= \nu_u^+(s) \zu+ \overline{\mathcal{M}}_u(\zu,\bzw,\bZs,s,\sigma)  \\
\frac{d\bzw}{d\xi} &= \nu_v^+(s,\sigma) \bzw +\bar{\gamma}_{11}(\zu,\bzw,\bZs,s,\sigma)\bzw+\bar{\gamma}_{12}(\zu,\bzw,\bZs,s,\sigma)\bZs  \\
\frac{d\bZs}{d\xi} &= \Lambda_{ss}(s,\sigma) \bZs+\bar{\gamma}_{21}(\zu,\bzw,\bZs,s,\sigma)\bzw+\bar{\gamma}_{22}(\zu,\bzw,\bZs,s,\sigma)\bZs, 
\end{split}
\label{eq:maindiag3}
\end{equation}
where $\bar\gamma_{11}(\zu,0,0,s,\sigma)=0$ and $\bar \gamma_{21}(\zu,0,0,s,\sigma)=0$.  
\end{lemma}

\begin{Proof}
We use a change of coordinates outlined in \cite{deng95,shilnikov98}. In a first step, we let 
\begin{eqnarray*}
\bzw &=& \zw (1-g^{ws})  \nonumber \\
\bZs &=& \Zs-G^s\zw,  
\end{eqnarray*}
for two smooth functions $g^{ws}:\R\rightarrow\R$ and $G^s:\R\rightarrow\R^2$. We substitute this change of coordinates into \eqref{eq:maindiag2} and obtain
\begin{equation}
\begin{split} \frac{\md\bzw}{\md\xi}=&~ \nu_v^+(s,\sigma)\bzw+\bzw\left( \gamma_{11}+\gamma_{12}G^s-\frac{1}{1-g^{ws}}\frac{\md g^{ws}}{\md\xi} \right)+\gamma_{12}\bZs (1-g^{ws}),  \\
 \frac{\md\bZs}{\md\xi}=& ~\Lambda_{ss}\bZs +\frac{\bzw}{1-g^{ws}} \left( \Lambda_{ss}G^s-\nu_v^+(s,\sigma)G^s+\gamma_{21} +\gamma_{22}G^s-\frac{\md G^s}{\md\xi} - \gamma_{11}G^s \right)  \\
&+\gamma_{22}\bZs-G^s\gamma_{12}G^s, 
\end{split}
\label{eq:covprelim}
\end{equation}
where we have suppressed the functional dependence of $\gamma_{ij}$ for convenience.  
Recall our original intention -- to remove those terms $\gamma_{j1}(\zu,0,0,s,\sigma)$ from (\ref{eq:maindiag2}).  To accomplish this, we set the terms multiplying $\bzw$ in (\ref{eq:covprelim}) to zero and find differential equations for $g^{ws}$ and $G^s$.  Since we are interested in these changes of coordinates along the unstable manifold, we augment these equations with the one for $\zu$ and obtain 
\begin{equation}\label{eq:covfinal}
\begin{split}
 \frac{\md g^{ws}}{\md\xi} &= (1-g^{ws})\left( \gamma_{11}(\zu,0,0,s,\sigma)+ \gamma_{12}(\zu,0,0,s,\sigma)G^s\right),  \\
\frac{\md G^s}{\md\xi}&= \left(\Lambda_{ss}(s,\sigma)-\nu_v^+(s,\sigma)\mathrm{I}+\gamma_{22}(\zu,0,0,s,\sigma)- \gamma_{11}(\zu,0,0,s,\sigma) \right)G^s+\gamma_{21}(\zu,0,0,s,\sigma),   \\
\frac{\md\zu}{\md\xi} &= \nu_u^+(s) \zu+ \mathcal{M}_u(\zu,0,0,s,\sigma). 
\end{split}
\end{equation}
The origin is a fixed point for (\ref{eq:covfinal}) with one unstable eigenvalue ($\nu_u^+(s)$), one zero eigenvalue and two stable eigenvalues $(\nu_u^-(s)-\nu_v^+(s,\sigma),\nu_v^{-}(s,\sigma)-\nu_v^+(s,\sigma))$.  Thus, there exists a one dimensional unstable manifold given as graphs over the $\zu$ coordinate.  These graphs provide the requisite change of variables, namely we have
\begin{align*}
g^{ws}&:=p(z^u,s,\sigma),\\
G^s&:=q(z^u,s,\sigma).
\end{align*}
 We also obtain expansions,
\begin{subequations}
\begin{align}
p(\zu,s,\sigma)&= \frac{\gamma_{11}^{(1)}(s,\sigma)}{\nu_u^+(s)}\zu +\O\left((\zu)^2\right),  \\
q(\zu,s,\sigma) &= - \left( \Lambda_{ss}(s,\sigma)-(\nu_v^+(s,\sigma)+\nu_u^+(s))\mathrm{I}\right)^{-1}  \gamma_{21}^{(1)}(s,\sigma) \zu +\O\left((\zu)^2\right),\end{align}
\label{eq:pqzu}
\end{subequations}
where we have employed the notations
\[ \gamma_{i1}(\zu,0,0,s,\sigma)=\gamma_{i1}^{(1)}(s,\sigma)\zu+\gamma_{i1}^{(2)}(s,\sigma)(\zu)^2+\O\left((\zu)^3\right),\quad i \in\{1,2\}. \]
Quadratic expansions of $p(\zu,s,\sigma)$ and $q(\zu,s,\sigma)$ can be found in Lemma~\ref{lem:expqzu} in the Appendix.
\end{Proof}

\paragraph{The Shilnikov Theorem}

\begin{theorem}\label{thm:shilnikov} 
Consider the boundary value problem consisting of  (\ref{eq:maindiag3}) with boundary conditions
\[ \bzw(0)=\kappa, \ \bZs(0)=\mathcal{Z}_0, \ \zu(T)=-\kappa,\]
for some $T>0$.  Then there exists a $\delta>0$ such that for any $|2\kappa+|\mathcal{Z}_0||<\delta$ and any $T>1/\delta$ then the boundary value problem has a unique solution and the following asymptotic expansions hold for large $T$,
\begin{equation}
\begin{split} 
\zu(0)&=-\kappa e^{-\nu_u^+(s)T}+\O(e^{(-\nu_u^+(s)+\omega)T})  \\
\bzw(T)&=\kappa e^{\nu_v^+(s,\sigma)T}+\O(e^{(\nu_v^+(s,\sigma)-\omega)T})  \\
\bZs(T)&=\gamma(s,\sigma) \kappa^2 e^{2\nu_v^+(s,\sigma)T}+\O(e^{(2\nu_v^+(s,\sigma)-\omega)T}),
\end{split}
\label{eq:asy} 
\end{equation}
for some $\omega>0$ where 
\[ \gamma(s,\sigma)=\frac{\partial \bar{\gamma}_{21}}{\partial \bzw}(0,0,0,s,\sigma)=\left(\begin{array}{c} \frac{ \mathbf{N}_{ss,u}^{(0,2,0,0)}(s,\sigma)}{2\nu_v^+(s,\sigma)-\nu_u^-(s)}\\\ \frac{\mathbf{N}_{ss,v}^{(0,2,0,0)}(s,\sigma)}{2\nu_v^+(s,\sigma)-\nu_v^-(s,\sigma)}\end{array}\right).\]

\end{theorem}
\begin{Proof}
A full proof of this result is detailed elsewhere and we refer the reader to \cite{shilnikov68} for example.  We sketch the ideas here.  Transform the system of differential equations (\ref{eq:maindiag3}) into a system of integral equations using variation of constants,
\begin{eqnarray*}
\zu(\xi)&=& e^{\nu_u^+(s) (\xi-T)} \zu(T) -e^{\nu_u^+(s)\xi}\int_\xi^T e^{-\nu_u^+(s) \tau} \overline{\mathcal{M}}_u(\zu(\tau),\bzw(\tau),\bZs(\tau))\md \tau \nonumber \\
\bzw(\xi) &=& e^{\nu_v^+(s)\xi}\bzw(0)+e^{\nu_v^+(s,\sigma)\xi}\int_0^\xi e^{-\nu_v^+(s,\sigma)\tau}\left(
\bar{\gamma}_{11}(\zu(\tau),\bzw(\tau),\bZs(\tau))\bzw(\tau)\right. \nonumber \\
& & \left. + \bar{\gamma}_{12}(\zu(\tau),\bzw(\tau),\bZs(\tau))\bZs(\tau)\right)\md\tau  \nonumber \\
\bZs(\xi) &=& e^{\Lambda_{ss}(s,\sigma)\xi}\mathcal{Z}_0
+e^{\Lambda_{ss}(s,\sigma)\xi}\int_0^\xi e^{-\Lambda_{ss}(s,\sigma)\tau}
\left(\bar{\gamma}_{21}(\zu(\tau),\bzw(\tau),\bZs(\tau))\bzw(\tau)\right. \nonumber \\
& & \left. +\bar{\gamma}_{22}(\zu(\tau),\bzw(\tau),\bZs(\tau))\bZs(\tau)\right)\md\tau.
\end{eqnarray*}
The solution is obtained as a fixed point of the mapping defined by the right hand side of the above equations for any $T>0$ and $|2\kappa+|\mathcal{Z}_0||<\delta$ with $\delta>0$ small enough for the right hand side to be a contraction.  The requirement that $T>1/\delta$ is only to ensure that $T$ is large enough in order to obtain the desired asymptotics. 

Recall the ratio condition {\bf (H4)}.  Under this assumption, the quadratic terms in $\bzw$ are sufficient to derive an expansion for $\bZs(T)$.  To do this, we recall that the leading order expansion for $\bZs$ can be obtained from the integral equation for $\bZs$, where we identify the dominant terms are found in the integral
\[ \bZs(\xi) =e^{\Lambda_{ss}(s,\sigma)\xi}\int_0^\xi e^{-\Lambda_{ss}(s,\sigma)\tau}\frac{\partial \gamma_{21}}{\partial \bzw}(-\kappa e^{-\nu_u^+(s)\tau},0,0,s,\sigma) \kappa^2 e^{2\nu_v^+(s,\sigma)\tau}\md\tau.\]
Of these terms, the dominant contribution comes from the quadratic terms that are independent of $\zu$ and we obtain the desired expansion.  
 \end{Proof}

\subsection{Application of Theorem~\ref{thm:shilnikov} to the manifold $W^u(\P_2)$}\label{sec:shilnikov}

Let $\kappa>0$ and  fix the sections
\[ \Sigma^{out}=\{ \zu=-\kappa\}, \quad \Sigma^{in}=\{ \bzw=\kappa\}.\]
We suppose that $\kappa$ is sufficiently small so that these sections intersect the neighborhood on which the changes of variables in Lemma~\ref{lem:cov1} and Lemma~\ref{lem:cov2} are valid and for which the existence of solutions in Theorem~\ref{thm:shilnikov} holds.

The goal is to derive an expansion for $W^u(\P_2)$ within the section $\Sigma^{out}$ so as to facilitate a comparison with the manifold $W^{ss}(\P_0)$.    Note that for fixed values of $\sigma$ and $s$, $W^u(\P_2)$ is a two dimensional manifold, so that its intersection with $\Sigma^{out}$ is one dimensional.  Recall Hypothesis {\bf (H5)}, where we assume that $W^u(\P_2)$ enters a neighborhood of $\P_1$ near the weak-stable eigendirection.  In terms of the coordinates of (\ref{eq:maindiag3}), this assumption implies that 
\begin{equation}
\begin{split}
\zu(0)&= h_u(\chi,s,\sigma),  \\
\bzus(0) &= h_{ss,u}(\chi,s,\sigma),  \\
\bzvs (0) &= h_{ss,v}(\chi,s,\sigma),
\end{split}
\label{eq:Wuenter}
\end{equation}
where $\chi$ parametrizes the intersection and we have that $h_u(0,s,\sigma)$, $ h_{ss,u}(0,s,\sigma) $ and $ h_{ss,u}(0,s,\sigma)$ are all zero.  We first match the terms in the $\zu$ component.  We find that to leading order
\[ -\kappa e^{-\nu_u^+T} +\O(e^{(-\nu_u^++\omega)T})= r(s,\sigma)\chi+\O(\chi^2),\]
where $r(s,\sigma)=\frac{\partial{h_u}}{{\partial \chi}} (0,s,\sigma)\neq0$ because the tangent space of $W^u(\P_2)$ intersects $W^s(\P_1)$ transversely, see $\bf{(H5)}$.  We then have the expansion 
\[ \chi(\rho,s,\sigma)=-\frac{\kappa}{r(s,\sigma)}e^{-\nu_u^+T}+\tilde{\chi}(T,s,\sigma),\]
see Remark~\ref{rem:nonresonance}.
Therefore, for every $T\geq \frac{1}{\delta}$ we can solve for $\chi(\rho,s,\sigma)$ and 
obtain expressions for $W^{u}(\P_2)$ within $\Sigma^{out}$.  These expressions can be given as a graph over the weak-stable direction, namely
\begin{equation}
\begin{split}
\bzw(T)&=\rho, \\
\bZs(T)&= \rho^2 \left(\gamma(s,\sigma)+\mathcal{Z}_{ss}(\rho,s,\sigma)\right).
\end{split}
\label{eq:asy2} 
\end{equation}

\begin{rmk}\label{rem:nonresonance} It is at this stage that the condition \eqref{nonresonance} on the ratio of the eigenvalues in {\bf (H4)} comes into play.  Were this condition to fail to hold, then the expansions for the strong stable components in (\ref{eq:asy}) would depend on the initial character of the manifold $W^u(\P_2)$ within $\Sigma^{in}$.  Then the particular form of the matching condition $\chi(\rho,s,\sigma)$ would be relevant and it would prove more challenging to match solutions in the following section.
\end{rmk}

\subsection{Transforming to original coordinates}
To compare the description of the manifold $W^u(\P_2)$ in (\ref{eq:asy2}) to the one for $W^{ss}(\P_0)$ we need to transform back to the original coordinates.  To do this, we first transform from $(\zu,\bzw,\bZs)$ coordinates to $(\zu,\zw,\Zs)$ coordinates.  This change of coordinates is performed in Lemma~\ref{lem:cov2} and can be inverted explicitly.  We obtain
\begin{equation}
\begin{split}
\zu&= -\kappa \\
\zw &= \frac{\rho}{1-p(-\kappa,s,\sigma)}  \\
\Zs &= \rho \frac{q(-\kappa,s,\sigma)}{1-p(-\kappa,s,\sigma)}+\rho^2\left(\gamma(s,\sigma)+\mathcal{Z}_{ss}(\rho,s,\sigma)\right) .
\end{split}
\label{eq:inZ} 
\end{equation}

Next, we need to transform this expression from the coordinates $(\zu,\zw,\Zs)$ to the coordinates $(\yu,\yw,\yus,\yvs)$.  This involves inverting the change of coordinates given in Lemma~\ref{lem:cov1}, i.e. solving the following set of implicit equations,
\begin{equation}
\begin{split}
-\kappa &= \yu-\mathcal{H}_s(\yw, \yus,\yvs)  \\
 \frac{\rho}{1-p(-\kappa,s,\sigma)} &= \yw  \\
\rho\frac{q^{(1)}(-\kappa,s,\sigma)}{1-p(-\kappa,s,\sigma)}+\rho^2\left(\gamma^{(1)}(s,\sigma)+\mathcal{Z}_{ss,u}(\rho,\sigma,s)\right) &= \yus  -\mathcal{H}_{u}(\yu) \\
\rho\frac{q^{(2)}(-\kappa,s,\sigma)}{1-p(-\kappa,s,\sigma)}+\rho^2\left(\gamma^{(2)}(s,\sigma)+\mathcal{Z}_{ss,v}(\rho,\sigma,s)\right)  &= \yvs. 
\end{split}
 \label{eq:inYprelim} 
\end{equation}
The change of coordinates can be inverted by first inputting the expressions for $\yw, \yus, $ and $\yvs$ into the first equation in (\ref{eq:inYprelim}).  This yields a scalar equation for $y_u$, 
\[ -\kappa =\yu -\mathcal{H}_s\left(\begin{array}{c}
\frac{\rho}{1-p(-\kappa,s,\sigma)} \\
\mathcal{H}_{u}(\yu)+ \rho\frac{q^{(1)}(-\kappa,s,\sigma)}{1-p(-\kappa,s,\sigma)}+\rho^2\left(\gamma^{(1)}(s,\sigma)+\mathcal{Z}_{ss,u}(\rho,\sigma,s)\right) \\
\rho\frac{q^{(2)}(-\kappa,s,\sigma)}{1-p(-\kappa,s,\sigma)}+\rho^2\left(\gamma^{(2)}(s,\sigma)+\mathcal{Z}_{ss,v}(\rho,\sigma,s)\right) 
\end{array}\right).
\] 
Applying the implicit function theorem, we obtain a solution
\[\yu=\mathcal{Y}_u(\rho,s,\sigma)=\mathcal{Y}_u^0(s)+\rho\mathcal{Y}_u^1(s,\sigma)+\rho^2\mathcal{Y}_u^2(s,\sigma)+\O(\rho^3).\]
Note that $\mathcal{Y}_u^0(s)$ is a solution of
\[ 0=\kappa+ \mathcal{Y}_u^0(s)-\mathcal{H}_s \left(0,\mathcal{H}_u\left(\mathcal{Y}_u^0(s)\right),0\right)\]
and we find an expansion in $\kappa$ of $\mathcal{Y}_u^0(s)=-\kappa+\O(\kappa^4)$.  We observe that the independence of the leading order term on $\sigma$ follows from the fact that the vector field restricted to $\yw=\yvs=0$ is indepedent of $\sigma$.
  
We then obtain an explicit representation for $\yus$ in terms of $\mathcal{Y}_u$.  For convenience we make a similar expansion,
\[ \mathcal{Y}_{uu,s}(\rho,s,\sigma)=\mathcal{H}_u(\mathcal{Y}_u(\rho,s,\sigma))=\mathcal{Y}_{uu,s}^0(s,\sigma)+\rho\mathcal{Y}_{uu,s}^1(s,\sigma)+\rho^2\mathcal{Y}_{uu,s}^2(s,\sigma)+\O(\rho^3).\]
These terms have similar expansions in $\kappa$, for example
\[ \mathcal{Y}_{uu,s}^0(s,\sigma)=\frac{\mathbf{N}_{ss,u}^{(2,0,0,0)}(s,\sigma)}{2\nu_u^+(s)-\nu_u^-(s)}\kappa^2+\O(\kappa^3). \]

To summarize, we  have found the expressions
\begin{equation}
\begin{split}
 \yu &=\mathcal{Y}_u(\rho,s,\sigma) \\
 \yw &= \frac{\rho}{1-p(-\kappa,s,\sigma)} \\
\yus &= \rho\frac{q^{(1)}(-\kappa,s,\sigma)}{1-p(-\kappa,s,\sigma)}+\rho^2\left(\gamma^{(1)}(s,\sigma)+\mathcal{Z}_{ss,u}(\rho,\sigma,s)\right) +\mathcal{Y}_{uu,s}(\rho,\sigma,s) \\
 \yvs &=  \rho\frac{q^{(2)}(-\kappa,s,\sigma)}{1-p(-\kappa,s,\sigma)}+\rho^2\left(\gamma^{(2)}(s,\sigma)+\mathcal{Z}_{ss,v}(\rho,\sigma,s)\right).  
 \end{split}
 \label{eq:inY} 
\end{equation}
Therefore, the manifold $W^u(\P_2)\cap \Sigma^{out}$ in the original variables is  
\be \left(\begin{array}{c} u_1 \\ u_2 \\ v_1 \\ v_2 \end{array}\right)= \left(\begin{array}{c} u^+ \\ 0 \\ 0 \\ 0 \end{array}\right)+T(s,\sigma) \left(\begin{array}{c}\mathcal{Y}_u(\rho,s,\sigma)  \\ 
\rho\frac{q^{(1)}(-\kappa,s,\sigma)}{1-p(-\kappa,s,\sigma)}+\rho^2\left(\gamma^{(1)}(s,\sigma)+\mathcal{Z}_{ss,u}(\rho,\sigma,s)\right) +\mathcal{Y}_{uu,s}(\rho,\sigma,s) \\
\frac{\rho}{1-p(-\kappa,s,\sigma)}  \\
\rho\frac{q^{(2)}(-\kappa,s,\sigma)}{1-p(-\kappa,s,\sigma)}+\rho^2\left(\gamma^{(2)}(s,\sigma)+\mathcal{Z}_{ss,v}(\rho,\sigma,s)\right)
  \end{array}\right). \label{eq:wuexp} \ee
For future reference, we refer to 
\be \mathcal{W}(\rho,s,\sigma):=T(s,\sigma) \left(\begin{array}{c}\mathcal{Y}_u(\rho,s,\sigma)  \\ 
\rho\frac{q^{(1)}(-\kappa,s,\sigma)}{1-p(-\kappa,s,\sigma)}+\rho^2\left(\gamma^{(1)}(s,\sigma)+\mathcal{Z}_{ss,u}(\rho,\sigma,s)\right) +\mathcal{Y}_{uu,s}(\rho,\sigma,s) \\
\frac{\rho}{1-p(-\kappa,s,\sigma)}  \\
\rho\frac{q^{(2)}(-\kappa,s,\sigma)}{1-p(-\kappa,s,\sigma)}+\rho^2\left(\gamma^{(2)}(s,\sigma)+\mathcal{Z}_{ss,v}(\rho,\sigma,s)\right)
  \end{array}\right). \label{eq:rhs} \ee

\subsection{Expansions of relevant quantities}
Before proceeding to compare $W^{ss}(\P_0)$ and $W^u(\P_2)$, we first interpret some of the terms in $\mathcal{W}$ and derive alternate expressions that will prove useful later.

\begin{lemma}\label{lem:Wexp} Recall $\mathcal{W}(\rho,s,\sigma)$ from (\ref{eq:rhs}).  We have that  $\mathcal{W}(0,s,\sigma)\subset W^u(\P_1)$. Furthermore, $\mathcal{W}(0,s^*,\sigma)$ is colinear with $\theta_1$ and 
\[ \frac{\partial \mathcal{W}}{\partial s} (0,s^*,\sigma)=\theta_s=(\theta_s^1,\theta_s^2,0,0)^T, \ \quad \frac{\partial \mathcal{W}}{\partial \sigma} (0,s^*,\sigma^*)=0. \]
\end{lemma}
\begin{Proof}
First observe that 
\[ \mathcal{W}(0,s,\sigma)=T(s,\sigma) \left(\begin{array}{c}\mathcal{Y}_u^0(s)  \\  \mathcal{Y}_{uu,s}^0(s) \\ 0 \\ 0 \end{array}\right).\]
Recalling the expression in  (\ref{eq:inZ}) we see that the limit $\rho=0$ corresponds to a value in the unstable manifold of $\P_1$.  When $s=s^*$, the unstable manifold includes the heteroclinic orbit $(U_p(\xi),U_p'(\xi),0,0)^T$, with tangent vector $\theta_1$.  
\end{Proof}

\begin{lemma} The vector 
\[ \frac{\partial W}{\partial \rho}(0,s^*,\sigma^*)=r_1\theta_1+r_2\theta_2=T(s^*,\sigma^*) \left(\begin{array}{c}\mathcal{Y}_u^1(s^*,\sigma^*)  \\ \frac{q^{(1)}(-\kappa,s^*,\sigma^*)}{1-p(-\kappa,s^*,\sigma^*)}  +\mathcal{Y}_{uu,s}^1(\sigma^*,s^*) \\ \frac{1}{1-p(-\kappa,s^*,\sigma^*)} \\    \frac{q^{(2)}(-\kappa,s^*,\sigma^*)}{1-p(-\kappa,s^*,\sigma^*)} \end{array}\right),\]
where it follows that 
\begin{eqnarray*}
r_1&=& \frac{1}{\langle \theta_1,\theta_1\rangle}\left( \mathcal{Y}_u^1(s^*,\sigma^*)  
\langle \theta_1,e_u^+\rangle+\left(\frac{q^{(1)}(-\kappa,s^*,\sigma^*)}{1-p(-\kappa,s^*,\sigma^*)}  +\mathcal{Y}_{uu,s}^1(\sigma^*,s^*)\right)\langle \theta_1,e_u^-\rangle \right. \\
&+& \left. \frac{1}{1-p(-\kappa,s^*,\sigma^*)}\langle \theta_1,e_v^+\rangle+\frac{q^{(2)}(-\kappa,s^*,\sigma^*)}{1-p(-\kappa,s^*,\sigma^*)}\langle \theta_1,e_v^-\rangle\right)
\\
r_2 &=& \frac{1}{\langle \theta_2,\theta_2\rangle}\left( \mathcal{Y}_u^1(s^*,\sigma^*)  
\langle \theta_2,e_u^+\rangle+\left(\frac{q^{(1)}(-\kappa,s^*,\sigma^*)}{1-p(-\kappa,s^*,\sigma^*)}  +\mathcal{Y}_{uu,s}^1(\sigma^*,s^*)\right)\langle \theta_2,e_u^-\rangle \right. \\
&+& \left. \frac{1}{1-p(-\kappa,s^*,\sigma^*)}\langle \theta_2,e_v^+\rangle+\frac{q^{(2)}(-\kappa,s^*,\sigma^*)}{1-p(-\kappa,s^*,\sigma^*)}\langle \theta_2,e_v^-\rangle\right),
\end{eqnarray*}
where 
\begin{eqnarray*}
\langle \theta_1,e_u^\pm\rangle &=& U_p'(\xi_0)+U_p''(\xi_0)\nu_u^\pm \\
\langle \theta_1,e_v^\pm\rangle &=& -\frac{F_{u}(\p_1)}{d_u(\nu_v^\pm)}\left(  U_p'(\xi_0)+U_p''(\xi_0)\nu_v^\pm\right) \\
\langle \theta_2,e_u^\pm\rangle &=&  a_1(\xi_0)+a_2(\xi_0)\nu_u^\pm \\
\langle \theta_2,e_v^\pm\rangle &=&  -\frac{F_{u}(\p_1)}{d_u(\nu_v^\pm)}\left(a_1(\xi_0)+a_2(\xi_0)\nu_v^\pm\right) +\phi(\xi_0)+\nu_v^\pm \phi'(\xi_0).
\end{eqnarray*}
\end{lemma}
\begin{Proof} Recall that those terms that are linear in $\rho$ originate in (\ref{eq:asy2})  and result from following the weak-unstable eigenspace along the unstable manifold of $\P_1$ to the section $\Sigma^{out}$.  The subspace $\bzus=\bzvs=0$ is invariant in (\ref{eq:maindiag2}) and therefore this vector is the weak stable tangent space of $\P_1$ tracked forward along the unstable manifold.  In Section~\ref{sec:tangent}, we calculated that this space coincides with $\mathrm{span}\{ \theta_1,\theta_2\}$ and the result therefore follows.

\end{Proof}

\begin{lemma}\label{lem:rhosigma} We have the further expansions of $\mathcal{W}(\rho,s,\sigma)$  
\[\theta_{\rho\sigma}:=\frac{\partial^2 \mathcal{W}}{\partial \rho \partial \sigma} (0,s^*,\sigma^*)= r_2\left( \beta_1\psi_1 +\beta_2\psi_2\right) ,\]
and
\[ \theta_{ \rho^2}:=\frac{1}{2}\frac{\partial^2 \mathcal{W}}{\partial \rho^2} (0,s^*,\sigma^*)=T(s^*,\sigma^*)\left(\begin{array}{c} \mathcal{Y}_u^2(s^*,\sigma^*) \\ \gamma^{(1)}(s^*,\sigma^*) +\mathcal{Y}_{uu,s}^2(s^*,\sigma^*) \\ 0 \\ \gamma^{(2)}(s^*,\sigma^*) \end{array}\right). \]
\end{lemma}

\begin{Proof}
The expression for $\theta_{\rho^2}$ follows from a calculation.  

For $\theta_{\rho\sigma}$, we recall Section~\ref{sec:tangent} where  the tangent space to the weak- unstable manifold was tracked and its dependence on $\sigma$ was ascertained; see (\ref{eq:R}) for the expression $\mathcal{R}(p_1,q_1,\sigma)$.  Using the parameterization of the subspace in terms of $r_1$ and $r_2$, we can write the subspace as 
\begin{eqnarray*} \frac{\partial \mathcal{W}}{\partial \rho} (0,s^*,\sigma)&=& r_1\theta_1+r_2\theta_2+ \frac{\langle \psi_1, \mathcal{R}(r_1U_p'(\xi_0)+r_2a_1(\xi_0),r_2\phi(\xi_0),\sigma)\rangle}{\Omega_1}  \psi_1 \\
&+& \frac{\langle \psi_2, \mathcal{R}((r_1U_p'(\xi_0)+r_2a_1(\xi_0),r_2\phi(\xi_0),\sigma)\rangle}{\Omega_2}  \psi_2,\end{eqnarray*}
where 
\[ \langle \psi_1, \mathcal{R}\rangle = r_2 (\sigma-\sigma^*)\int_{-\infty}^{\xi_0}b_2(\tau)\left(\frac{1}{(\sigma^*)^2}g(U_p(\tau),0)\phi(\tau)+\frac{s^*}{(\sigma^*)^2}\phi'(\tau) \right)\md\tau,\]
and
\[ \langle \psi_2, \mathcal{R}\rangle= r_2 (\sigma-\sigma^*)\int_{-\infty}^{\xi_0} e^{\frac{s^*}{\sigma^*}\tau}\left(\frac{1}{(\sigma^*)^2}g(U_p(\tau),0) \phi(\tau)^2+\frac{s^*}{(\sigma^*)^2} \phi(\tau)\phi'(\tau) \right)\md\tau.\]
\end{Proof}

\section{Resolving the bifurcation equation: Proof of Theorem~\ref{thm:main}}\label{sec:bif}
We now establish Theorem~\ref{thm:main}.  Recall the expression (\ref{eq:wssexp}) that describes the manifold $W^{ss}(\P_0)$ near the section $\Sigma^{out}$.  Similarly, we have expansion (\ref{eq:wuexp}) that describes $W^u(\P_2)$ within the section $\Sigma^{out}$.  Equating these expressions we obtain an implicit bifurcation equation
\begin{eqnarray*} 0=\mathcal{F}(\rho,\eta_1,\eta_2,s,\sigma;\xi_0,\kappa)&:=&   \Delta_f(\xi_0)+\eta_1\theta_1+\eta_2\theta_2+(s-s^*)\Gamma_0\psi_1+h_1(\eta_1,\eta_2,s,\sigma)\psi_1+h_2(\eta_1,\eta_2,s,\sigma)\psi_2\\
&&- \mathcal{W}(\rho,s,\sigma).
\end{eqnarray*}
First, we relate $\xi_0$ and $\kappa$ by imposing that $\mathcal{F}(0,0,0,0,s^*,\sigma^*;\xi_0,\kappa)=0$.  This is possible since $\Delta_f(\xi_0)$ and $\mathcal{W}(0,s^*,\sigma^*)$ both lie in the heteroclinic orbit $(U_p(\xi),U_p'(\xi),0,0)^T$.  We henceforth suppress the dependence of $\mathcal{F}$ on $\kappa$.  

Using the expansions in Lemma~\ref{lem:Wexp} through Lemma~\ref{lem:rhosigma}, we simplify $\mathcal{F}$ to 
\begin{eqnarray*} \mathcal{F}(\rho,\eta_1,\eta_2,s,\sigma)&=&   \eta_1\theta_1+\eta_2\theta_2+(s-s^*)\Gamma_0\psi_1+ h_1(\eta_1,\eta_2,s,\sigma)\psi_1+h_2(\eta_1,\eta_2,s,\sigma)\psi_2\\
&&- (s-s^*)\theta_s -\rho (r_1\theta_1+r_2\theta_2) - \rho^2 \theta_{\rho^2} -\rho(s-s^*) \theta_{\rho s} -\rho(\sigma-\sigma^*)\theta_{\rho\sigma}+\O(3).
\end{eqnarray*}
We wish to employ a Liapunov-Schmidt reduction and so we compute the partials of $\mathcal{F}$,
\[D_{\eta_1,\eta_2,\rho,s}\mathcal{F}=\left(\begin{array}{cccc} \theta_1 & \theta_2 & -r_1\theta_1-r_2\theta_2 & \Gamma_0 \psi_1-\theta_s \end{array}\right).\]
The Jacobian has rank three, so we project onto the range by projecting onto the vectors $\theta_1,\theta_2$ and $\psi_1$.  We obtain
\begin{eqnarray*} 0&=&\eta_1\langle \theta_1,\theta_1\rangle- (s-s^*) \langle \theta_1,\theta_s\rangle-\rho r_1 \langle \theta_1,\theta_1\rangle  \\
&&- \rho^2 \langle \theta_1, \theta_{\rho^2}\rangle -\rho(s-\sigma)\langle \theta_1, \theta_{\rho s} \rangle -\rho(\sigma-\sigma^*)\langle \theta_1, \theta_{\rho\sigma}\rangle+\O(3),
\end{eqnarray*}
\begin{eqnarray*} 0&=&\eta_2\langle \theta_2,\theta_2\rangle- (s-s^*) \langle \theta_2,\theta_s\rangle-\rho r_2 \langle \theta_2,\theta_2\rangle  \\
&&- \rho^2 \langle \theta_2, \theta_{\rho^2}\rangle -\rho(s-\sigma)\langle \theta_2, \theta_{\rho s} \rangle -\rho(\sigma-\sigma^*)\langle \theta_2, \theta_{\rho\sigma}\rangle+\O(3),
\end{eqnarray*}
\begin{eqnarray*} 0&=&(s-s^*)\Gamma_0\langle \psi_1,\psi_1\rangle+h_1(\eta_1,\eta_2,s,\sigma) \langle \psi_1,\psi_1\rangle- (s-s^*) \langle \psi_1,\theta_s\rangle\\
&&- \rho^2 \langle \psi_1, \theta_{\rho^2}\rangle -\rho(s-s^*)\langle \psi_1, \theta_{\rho s} \rangle -\rho(\sigma-\sigma^*)\langle \psi_1, \theta_{\rho\sigma}\rangle+\O(3).
\end{eqnarray*}
This constitutes an implicit set of equations which we write as $\mathcal{G}(\eta_1,\eta_2,s,\rho,\sigma)=0$.  Now, a simple computation leads to
\[ D_{\eta_1,\eta_2,s}\mathcal{G}(0,s^*,\sigma^*)=\left(\begin{array}{ccc} \langle \theta_1,\theta_1\rangle & 0 & \langle \theta_1,\theta_s\rangle \\ 0 &  \langle \theta_2,\theta_2\rangle & \langle \theta_2,\theta_s\rangle \\ 0 & 0  & \Omega_1\Gamma_0-  \langle \psi_1,\theta_s\rangle\end{array} \right). \]
At the same time, we compute
\[ D_{\rho}\mathcal{G}(0,s^*,\sigma^*)=\left(\begin{array}{c} -r_1 \langle \theta_1,\theta_1\rangle  \\  -r_2 \langle \theta_2,\theta_2 \rangle \\ 0 \end{array}\right).\] 
Therefore, the implicit function theorem ensures a solution $\mathcal{G}(\eta_1(\rho,\sigma),\eta_2(\rho,\sigma),s(\rho,\sigma))=0$ with 
\begin{eqnarray*}
\eta_1(\rho,\sigma)&=&r_1 \rho +\mathrm{g}_1(\rho,\sigma), \\
\eta_2(\rho,\sigma)&=& r_2\rho+\mathrm{g}_2(\rho,\sigma), \\
s(\rho,\sigma)-s^*&=& \mathrm{G}_s(\rho,\sigma)=\rho^2\frac{1}{\Gamma}\left( \langle \psi_1, \theta_{\rho^2}\rangle-r_1^2\Omega_1\frac{\partial^2 h_1}{\partial \eta_1^2}-r_2^2\Omega_1 \frac{\partial^2 h_1}{\partial \eta_2^2} \right)  \\
&+& \rho(\sigma-\sigma^*)\frac{1}{\Gamma}\left( \langle \psi_1, \theta_{\rho\sigma}\rangle-r_2 \Omega_1\frac{\partial^2 h_1}{\partial \eta_2\partial \sigma} \right) 
+\O(3),
\end{eqnarray*}
where the functions $\mathrm{g}_1$, $\mathrm{g}_2$ and $\mathrm{G}_s$ are all quadratic order or higher and 
\[ \Gamma=\Omega_1 \Gamma_0-  \langle \psi_1,\theta_s\rangle= \int_{-\infty}^\infty e^{s^*\tau}(U_p'(\tau))^2\md\tau.\]
We then consider the implicit equation
\begin{eqnarray*} 0=\mathcal{H}(\rho,\sigma)&:=&\langle \psi_2, \mathcal{F}\left(\rho,r_1\rho+\mathrm{g}_1(\rho,\sigma),r_2\rho+\mathrm{g}_2(\rho,\sigma),s^*+\mathrm{G}_s(\rho,\sigma),\sigma\right)\rangle \\
&=& h_2\left(r_1\rho+\mathrm{g}_1(\rho,\sigma),r_2\rho+\mathrm{g}_2(\rho,\sigma),\mathrm{G}_s(\rho,\sigma),\sigma\right)\langle \psi_2,\psi_2\rangle-\langle \psi_2,\mathcal{W}(\rho,s^*+\mathrm{G}_s(\rho,\sigma),\sigma)\rangle. 
\end{eqnarray*}
Note that $\mathcal{H}(\rho,\sigma)=\rho\tilde{\mathcal{H}}(\rho,\sigma)$.  We therefore expand, focusing on quadratic terms in $\mathcal{H}$,
\[ \mathcal{H}(\rho,\sigma)= h_2^{(2)}(r_1\rho,r_2\rho,0,\sigma)-\rho^2 \langle \psi_2,\theta_{\rho^2}\rangle -\rho(\sigma-\sigma^*)\langle \psi_2,\theta_{\rho\sigma}\rangle. \]
There are three non-zero terms in $h_2^{(2)}$ that contribute to the quadratic term -- namely  the terms $\eta_2^2$, $\eta_1\eta_2$ and $\eta_2\sigma$.  After factoring, we find the solution
\[ \rho=M_\rho (\sigma-\sigma^*)+\mathrm{G}_\sigma(\sigma),\]
where 
\[ M_\rho=\frac{\langle \psi_2,\theta_{\rho\sigma}\rangle-r_2\Omega_2\frac{\partial^2 h_2}{\partial\eta_2\partial\sigma} }{ r_2^2\Omega_2 \frac{\partial^2 h_2}{\partial\eta_2^2}+r_1r_2\Omega_2 \frac{\partial^2 h_2}{\partial\eta_1\partial\eta_2}-\langle \psi_2,\theta_{\rho^2}\rangle},\] 
and $\mathrm{G}_\sigma(\sigma)$ collects higher-order terms. We require $\rho$ to be positive to ensure positivity of the solution.  Therefore, the sign of $M_\rho$ dictates whether the bifurcation to locked fronts is sub or super critical.  With this solution, we can then determine whether the front is sped up or slowed down by inputting this into $\mathrm{G}_s(\rho,\sigma)$.

\paragraph{ Simplification of the term $M_\rho$.}
We now make several simplifications.  First, note that by Lemma~\ref{lem:rhosigma} the numerator simplifies with
\[
\frac{1}{r_2}\langle \psi_2,\theta_{\rho\sigma}\rangle-\Omega_2\frac{\partial^2 h_2}{\partial\eta_2\partial\sigma} =\int_{-\infty}^\infty e^{\frac{s^*}{\sigma^*}\xi}\left(\frac{g(U_p(\xi),0)}{(\sigma^*)^2} (\phi(\xi))^2 +\frac{s^*}{(\sigma^*)^2} \phi'(\xi)\phi(\xi) \right)\md\xi.
\]
We then use the identity $g(U_p(\xi),0)\phi(\xi)+s^*\phi'(\xi)=-\sigma^*\phi''(\xi)$ and integrate by parts
\begin{eqnarray*}\frac{1}{r_2}\langle \psi_2,\theta_{\rho\sigma}\rangle-\Omega_2\frac{\partial^2 h_2}{\partial\eta_2\partial\sigma} &=& -\frac{1}{\sigma^*}\int_{-\infty}^\infty e^{\frac{s^*}{\sigma^*}\xi}\phi(\xi)\phi''(\xi)\md \xi, \\
&=&\frac{1}{\sigma^*} \int_{-\infty}^\infty \phi'(\xi)\left( \phi'(\xi)e^{\frac{s^*}{\sigma^*}\xi}+\frac{s^*}{\sigma^*}\phi(\xi)e^{\frac{s^*}{\sigma^*}\xi}\right) \md\xi,  \\
&=& \frac{1}{\sigma^*} \int_{-\infty}^\infty \phi'(\xi)\phi(\xi)e^{\frac{s^*}{\sigma^*}\xi}\left( \frac{\phi'(\xi)}{\phi(\xi)}+\frac{s^*}{\sigma^*}\right) \md\xi.
\end{eqnarray*}
Now, we note that the term inside the parenthesis is positive, since for any $\xi$ we have that $g(U_p(\xi),0)>0$ from {\bf (H3)},  and therefore
\[ Z_{22}(\xi)=\frac{\phi'(\xi)}{\phi(\xi)}>-\frac{s^*}{2\sigma^*}-\frac{s^*}{2\sigma^*}\sqrt{(s^*)^2 -4g(U_p(\xi),0)}>-\frac{s^*}{\sigma^*}.\]
We finally find that 
\[
\mathrm{sign}\left(\frac{1}{r_2}\langle \psi_2,\theta_{\rho\sigma}\rangle-\Omega_2\frac{\partial^2 h_2}{\partial\eta_2\partial\sigma} \right) = \mathrm{sign}(\phi') =-1,
\]
since $\phi'<0$. And thus the sign of $M_\rho$ is determined by the opposite sign of its denominator:
\[ \mathrm{sign}{M_\rho}=- \mathrm{sign}  \left( r_2 \Omega_2 \frac{\partial^2 h_2}{\partial\eta_2^2}+r_1 \Omega_2 \frac{\partial^2 h_2}{\partial\eta_1\partial\eta_2}-\frac{1}{r_2}\langle \psi_2,\theta_{\rho^2}\rangle \right),\]
where we recall the following expressions for each term
\begin{align*}
 \Omega_2 \frac{\partial^2 h_2}{\partial\eta_2^2}&=\int_{\xi_0}^\infty e^{\frac{s^*}{\sigma^*}\xi}\left( \frac{G_{uv}(U_p(\xi),0)}{\sigma^*} a_1(\xi) \phi(\xi)^2 +\frac{G_{vv}(U_p(\xi),0)}{2\sigma^*} \phi^3(\xi) \right)\md\xi,\\
 \Omega_2 \frac{\partial^2 h_2}{\partial\eta_1\partial\eta_2}&=\left(\tilde\phi''(\xi_0)\tilde\phi(\xi_0)-(\tilde\phi'(\xi_0))^2\right),\\
 \langle \psi_2,\theta_{\rho^2}\rangle &= e^{\frac{s^*}{\sigma^*}\xi_0}\gamma^{(2)}(s^*,\sigma^*)\left(\nu_v^-(s^*,\sigma^*)\phi(\xi_0)-\phi'(\xi_0) \right),
\end{align*}
with from Lemma~\ref{lemexpNssv},
\[ \gamma^{(2)}(s^*,\sigma^*)=\frac{1}{\sigma(\nu_v^-(s,\sigma)-\nu_v^+(s,\sigma))(2\nu_v^+(s,\sigma)-\nu_v^-(s,\sigma))}\left( \frac{F_v(\p_1)}{d_u(\nu_v^-(s,\sigma))}G_{uv}(\p_1)-\frac{G_{vv}(\p_1)}{2}\right).\]

\paragraph{ Expansion of $s-s^*$.}
With an expansion for $\rho$ as a function of $\sigma-\sigma^*$, we finally obtain an expansion for $s-s^*$ as a function of $\sigma-\sigma^*$.  Let
\[ s-s^*=M_s (\sigma-\sigma^*)^2 +\O(3),\]
where 
\[ M_s=\frac{M_\rho}{\Gamma}\left(M_\rho \left( \langle \psi_1, \theta_{\rho^2}\rangle-r_1^2\Omega_1\frac{\partial^2 h_1}{\partial \eta_1^2}-r_2^2\Omega_1 \frac{\partial^2 h_1}{\partial \eta_2^2} \right) + r_2\int_{-\infty}^\infty b_2(\xi)\left(\frac{g(U_p(\xi),0)}{(\sigma^*)^2} \phi(\xi) +\frac{s^*}{(\sigma^*)^2} \phi'(\xi)\right)\md\xi \right). \]

\section*{Acknowledgements} The authors are grateful to Jim Nolen for suggesting this class of equations for study.   GF received support from the ANR project NONLOCAL ANR-14-CE25-0013.  MH received partial support from the National Science Foundation through grant NSF-DMS-1516155.     

\begin{appendix}

\section{Expansions of $h_{1,2}$}\label{sec:hexp}
We return to derive expressions for those terms in the quadratic expansions of $h_{1}$ and $h_2$ from Lemma~\ref{lem:h1h2} that are required for the resolution of the bifurcation equation.  To simplify the presentation, we recall some of the notations that were used in Section~\ref{sec:ssb}. The maps $h_{1,2}$ are determined by projecting equation \eqref{eq:defQs} onto $\psi_{1,2}$ to obtain the expressions
\begin{equation}
h_{1,2}(\eta_1,\eta_2,s,\sigma) = -\frac{1}{\Omega_{1,2}} \int_{\xi_0}^\infty \left\langle \psi_{1,2}(\xi),N(Q^*(\xi,\eta_1,\eta_2,s,\sigma),\xi,s,\sigma) \right\rangle \md\xi ,
\label{eq:defh12}
\end{equation}
where $\Omega_{1,2}=\langle \psi_{1,2}(\xi_0),\psi_{1,2}(\xi_0) \rangle$ and $Q^*(\cdot,\eta_1,\eta_2,s,\sigma)$ is the fixed point solution of the operator $T$ introduced in Lemma~\ref{lem:SScontraction} (see equation \eqref{eq:T}). As shown in the proof of Lemma~\ref{lem:h1h2}, the maps $h_{1,2}$ are at least quadratic or of higher order in all their arguments, and the associated quadratic expansions of $h_{1,2}$ can be obtained by collecting the quadratic expansions of the following quantities:
\begin{equation}
\tilde{h}_{1,2}(\eta_1,\eta_2,s,\sigma) := -\frac{1}{\Omega_{1,2}} \int_{\xi_0}^\infty \left\langle \psi_{1,2}(\xi),N(\eta_1\theta_1(\xi)+\eta_2\theta_2(\xi)+(s-s^*)\theta_s(\xi),\xi,s,\sigma) \right\rangle \md\xi ,
\label{eq:defth12}
\end{equation}
where we approximated $Q^*(\xi,\eta_1,\eta_2,s,\sigma)$ by $Q^0(\xi)=\eta_1\theta_1(\xi)+\eta_2\theta_2(\xi)+(s-s^*)\theta_s(\xi)$. The definition of the nonlinear term $N(z,\xi,s,\sigma)$ is 
\[N(z,\xi,s,\sigma)=\left(\begin{array}{c} 0 \\ N_p(z,\xi,s,\sigma) \\ 0 \\ N_q(z,\xi,s,\sigma) \end{array}\right), \quad z=(p_1,p_2,q_1,q_2)^T,\]
with quadratic expansions of $N_{p,q}$ denoted $N_{p,q}^{(2)}$  given by
\begin{align*}
N_p^{(2)}(z,\xi,s,\sigma) =& -(s-s^*)p_2-\frac{F_{uu}(U_p(\xi),0)}{2}p_1^2-F_{uv}(U_p(\xi),0)p_1q_1-\frac{F_{vv}(U_p(\xi),0)}{2}q_1^2, \\
N_q^{(2)}(z,\xi,s,\sigma) =& -\frac{1}{\sigma^*}(s-s^*)q_2+\frac{s^*}{(\sigma^*)^2}(\sigma-\sigma^*)q_2 -\frac{G_{uv}(U_p(\xi),0)}{\sigma^*}p_1q_1 -  \frac{G_{vv}(U_p(\xi),0)}{2\sigma^*}q_1^2\\
& +\frac{g(U_p(\xi),0)}{(\sigma^*)^2}q_1(\sigma-\sigma^*).
\end{align*}
To continue, we need expansions for $N_{p,q}^{(2)}$ in terms of $\eta_1$, $\eta_2$, $(s-s^*)$ and $(\sigma-\sigma^*)$.  To accomplish this, we recall that we have
\begin{eqnarray*}
p_1(\xi) &=& \eta_1U_p'(\xi)+\eta_2a_1(\xi)+(s-s^*)\theta_s^1(\xi), \\
p_2(\xi) &=& \eta_1U_p''(\xi)+\eta_2a_2(\xi)+(s-s^*)\theta_s^2(\xi), \\
q_1(\xi)&=& \eta_2\phi(\xi), \\
q_2(\xi) &=& \eta_2\phi'(\xi).
\end{eqnarray*}
To simplify the presentation, we will use the following notation
\begin{equation*}
N_{p,q}^{(2)}(z,\xi,s,\sigma)= \sum_{i+j+k+l=2}\eta_1^{i}\eta_2^{j}(s-s^*)^k(\sigma-\sigma^*)^l \mathbf{N}^{(i,j,k,l)}_{p,q}(\xi).
\end{equation*}
We then obtain the expressions:
\begin{equation*}
\begin{array}{rccl}
\mathcal{O}(\eta_1^2): & \mathbf{N}^{(2,0,0,0)}_{p}(\xi)&=&-\frac{F_{uu}(U_p(\xi),0)}{2} \left(U_p'(\xi)\right)^2,\\
\mathcal{O}(\eta_1\eta_2): & \mathbf{N}^{(1,1,0,0)}_{p}(\xi)&=&-\left(F_{uu}(U_p(\xi),0)U_p'(\xi)a_1(\xi)+F_{uv}(U_p(\xi),0)U_p'(\xi)\phi(\xi) \right),\\
\mathcal{O}(\eta_2^2): &\mathbf{N}^{(0,2,0,0)}_{p}(\xi)&=&-\left( \frac{F_{uu}(U_p(\xi),0)}{2} a_1^2(\xi) +F_{uv}(U_p(\xi),0)a_1(\xi)\phi(\xi) +\frac{F_{vv}(U_p(\xi),0)}{2}\phi^2(\xi)\right)\\
\mathcal{O}(\eta_1|s-s^*|): & \mathbf{N}^{(1,0,1,0)}_{p}(\xi)&=&-\left(U_p''(\xi)+F_{uu}(U_p(\xi),0)U_p'(\xi)\theta_s^1(\xi)\right),\\
 \mathcal{O}(\eta_2|s-s^*|): & \mathbf{N}^{(0,1,1,0)}_{p}(\xi)&=& -\left(a_2(\xi)+F_{uv}(U_p(\xi),0)\phi(\xi)\theta_s^1(\xi)+F_{uu}(U_p(\xi),0)a_1(\xi)\theta_s^1(\xi) \right),\\
\mathcal{O}(|s-s^*|^2): &\mathbf{N}^{(0,0,2,0)}_{p}(\xi)&=&-\left(\theta_s^2(\xi)+\frac{F_{uu}(U_p(\xi),0)}{2} (\theta_s^1(\xi))^2\right),
\end{array}
\end{equation*}
all other quadratic terms in the expansion being equal to zero. Regarding $N^{(2)}_q$, we get
\begin{equation*}
\begin{array}{rccl}
\mathcal{O}(\eta_1\eta_2): & \mathbf{N}^{(1,1,0,0)}_{q}(\xi)&=&-\frac{G_{uv}(U_p(\xi),0)}{\sigma^*}U_p'(\xi)\phi(\xi),\\
\mathcal{O}(\eta_2^2):  &\mathbf{N}^{(0,2,0,0)}_{q}(\xi)&=&- \left( \frac{G_{uv}(U_p(\xi),0)}{\sigma^*} a_1(\xi) \phi(\xi) +\frac{G_{vv}(U_p(\xi),0)}{2\sigma^*} \phi^2(\xi) \right),   \\
\mathcal{O}(\eta_2|s-s^*|):  &\mathbf{N}^{(0,1,1,0)}_{q}(\xi)&=&-\left( \frac{1}{\sigma^*} \phi'(\xi)+\frac{G_{uv}(U_p(\xi),0)}{\sigma^*}\theta_s^1(\xi)\phi(\xi) \right),\\
\mathcal{O}(\eta_2|\sigma-\sigma^*|): &\mathbf{N}^{(0,1,0,1)}_{q}(\xi)&=&\frac{g(U_p(\xi),0)}{(\sigma^*)^2} \phi(\xi) +\frac{s^*}{(\sigma^*)^2} \phi'(\xi),
\end{array}
\end{equation*}
all other quadratic terms in the expansion being equal to zero.

As a consequence, we can now collect all quadratic terms in the expansions of the maps $h_{1,2}$ by identification. Namely, if one sets
\begin{equation*}
h_{1,2}(\eta_1,\eta_2,s,\sigma) :=\sum_{\substack{i,j,k,l\geq0\\ i+j+k+l\geq 2}}\eta_1^{i}\eta_2^{j}(s-s^*)^k(\sigma-\sigma^*)^l \mathbf{h}^{(i,j,k,l)}_{1,2},
\end{equation*}
then using equation \eqref{eq:defth12}, we get the following relations for the quadratic terms. For all $i,j,k,l\geq0$ with $i+j+k+l=2$ we have
\begin{align*}
\mathbf{h}^{(i,j,k,l)}_{1}&=-\frac{1}{\Omega_{1}} \int_{\xi_0}^\infty \left(e^{s^*\xi}U_p'(\xi)\mathbf{N}^{(i,j,k,l)}_{p}(\xi) + b_2(\xi)\mathbf{N}^{(i,j,k,l)}_{q}(\xi) \right) \md\xi, \\
\mathbf{h}^{(i,j,k,l)}_{2}&=-\frac{1}{\Omega_{2}} \int_{\xi_0}^\infty e^{\frac{s^*}{\sigma^*}\xi}\phi(\xi) \mathbf{N}^{(i,j,k,l)}_{q}(\xi) \md\xi.
\end{align*}

We have the following Lemma which summarizes the previous computations.

\begin{lemma}\label{lem:relevantexp}
The nonlinear maps $h_1(\eta_1,\eta_2,s,\sigma)$ and $h_2(\eta_1,\eta_2,s,\sigma)$ from Lemma~\ref{lem:h1h2} admit the following quadratic expansions. For all $i,j,k,l\geq0$ with $i+j+k+l=2$ we have for $h_1(\eta_1,\eta_2,s,\sigma)$:
\begin{align*}
\mathcal{O}(\eta_1^2): ~ \mathbf{h}^{(2,0,0,0)}_{1}&=\frac{1}{\Omega_{1}} \int_{\xi_0}^\infty e^{s^*\xi} \frac{F_{uu}(U_p(\xi),0)}{2} \left(U_p'(\xi)\right)^3\md\xi,\\
\mathcal{O}(\eta_1\eta_2): ~ \mathbf{h}^{(1,1,0,0)}_{1}&=\frac{1}{\Omega_{1}} \int_{\xi_0}^\infty e^{s^*\xi} U'_p(\xi)\left(F_{uu}(U_p(\xi),0)U_p'(\xi)a_1(\xi)+F_{uv}(U_p(\xi),0)U_p'(\xi)\phi(\xi) \right)\md\xi\\
&\qquad+\frac{1}{\Omega_{1}} \int_{\xi_0}^\infty b_2(\xi)\frac{G_{uv}(U_p(\xi),0)}{\sigma^*}U_p'(\xi)\phi(\xi) \md\xi ,\\
\mathcal{O}(\eta_2^2): ~\mathbf{h}^{(0,2,0,0)}_{1}&=\frac{1}{\Omega_{1}} \int_{\xi_0}^\infty e^{s^*\xi} U'_p(\xi) \left( \frac{F_{uu}(U_p(\xi),0)}{2} a_1^2(\xi) +F_{uv}(U_p(\xi),0)a_1(\xi)\phi(\xi)\right)\md\xi\\
&\qquad + \frac{1}{\Omega_{1}} \int_{\xi_0}^\infty b_2(\xi)\left( \frac{G_{uv}(U_p(\xi),0)}{\sigma^*} a_1(\xi) \phi(\xi) +\frac{G_{vv}(U_p(\xi),0)}{2\sigma^*} \phi^2(\xi) \right)\md\xi\\
&\qquad +\frac{1}{\Omega_{1}} \int_{\xi_0}^\infty e^{s^*\xi} U'_p(\xi) \frac{F_{vv}(U_p(\xi),0)}{2}\phi^2(\xi)\md\xi,
\end{align*}
\begin{align*}
\mathcal{O}(\eta_1|s-s^*|): ~ \mathbf{h}^{(1,0,1,0)}_{1}&=\frac{1}{\Omega_{1}} \int_{\xi_0}^\infty e^{s^*\xi} U'_p(\xi) \left(U_p''(\xi)+F_{uu}(U_p(\xi),0)U_p'(\xi)\theta_s^1(\xi)\right)\md\xi,\\
\mathcal{O}(\eta_2|s-s^*|): ~ \mathbf{h}^{(0,1,1,0)}_{1}&=\frac{1}{\Omega_{1}} \int_{\xi_0}^\infty e^{s^*\xi} U'_p(\xi)  \left(a_2(\xi)+F_{uv}(U_p'(\xi),0)\phi(\xi)\theta_s^1(\xi)+F_{uu}(U_p(\xi),0)a_1(\xi)\theta_s^1(\xi) \right)\md\xi\\
&\qquad + \frac{1}{\Omega_{1}} \int_{\xi_0}^\infty b_2(\xi) \left( \frac{1}{\sigma^*} \phi'(\xi)+\frac{G_{uv}(U_p(\xi),0)}{\sigma^*}\theta_s^1(\xi)\phi(\xi) \right) \md\xi,\\
\mathcal{O}(\eta_2|\sigma-\sigma^*|): ~ \mathbf{h}^{(0,1,0,1)}_{1}&=-\frac{1}{\Omega_{1}} \int_{\xi_0}^\infty b_2(\xi)\left(\frac{g(U_p(\xi),0)}{(\sigma^*)^2} \phi(\xi) +\frac{s^*}{(\sigma^*)^2} \phi'(\xi)\right)\md\xi,\\
\mathcal{O}(|s-s^*|^2): ~ \mathbf{h}^{(0,0,2,0)}_{1}&=\frac{1}{\Omega_{1}} \int_{\xi_0}^\infty e^{s^*\xi} U'_p(\xi) \left(\theta_s^2(\xi)+\frac{F_{uu}(U_p(\xi),0)}{2} (\theta_s^1(\xi))^2\right)\md\xi,
\end{align*}
and for $h_2(\eta_1,\eta_2,s,\sigma)$:
\begin{align*}
\mathcal{O}(\eta_1\eta_2): ~ \mathbf{h}_2^{(1,1,0,0)}&=\frac{1}{\Omega_{2}} \int_{\xi_0}^\infty e^{\frac{s^*}{\sigma^*}\xi}\frac{G_{uv}(U_p(\xi),0)}{\sigma^*}U_p'(\xi)\left(\phi(\xi)\right)^2\md\xi,\\
\mathcal{O}(\eta_2^2): ~ \mathbf{h}_2^{(0,2,0,0)}&= \frac{1}{\Omega_{2}} \int_{\xi_0}^\infty e^{\frac{s^*}{\sigma^*}\xi}\left( \frac{G_{uv}(U_p(\xi),0)}{\sigma^*} a_1(\xi) \phi(\xi)^2 +\frac{G_{vv}(U_p(\xi),0)}{2\sigma^*} \phi^3(\xi) \right)\md\xi,   \\
\mathcal{O}(\eta_2|s-s^*|):  ~ \mathbf{h}_2^{(0,1,1,0)}&=\frac{1}{\Omega_{2}} \int_{\xi_0}^\infty e^{\frac{s^*}{\sigma^*}\xi}\left( \frac{1}{\sigma^*} \phi'(\xi)\phi(\xi)+\frac{G_{uv}(U_p(\xi),0)}{\sigma^*}\theta_s^1(\xi)(\phi(\xi))^2 \right)\md\xi,\\
\mathcal{O}(\eta_2|\sigma-\sigma^*|): ~ \mathbf{h}_2^{(0,1,0,1)}&=-\frac{1}{\Omega_{2}} \int_{\xi_0}^\infty e^{\frac{s^*}{\sigma^*}\xi}\left(\frac{g(U_p(\xi),0)}{(\sigma^*)^2} (\phi(\xi))^2 +\frac{s^*}{(\sigma^*)^2} \phi'(\xi)\phi(\xi) \right)\md\xi.
\end{align*}
All stated integrals converge in the limit $\xi_0\to -\infty$.  
\end{lemma}

\begin{Proof}  Asymptotic exponential decay rates for the relevant quantities are collected in Table~\ref{table}.  We focus on the convergence of the integrands as $\xi_0\to -\infty$.  Recall Hypothesis {\bf (H4)} and the assumed ordering of the eigenvalues
\[ \nu_v^-(s^*,\sigma^*)<\nu_u^-(s^*)<\nu_v^+(s^*,\sigma^*)<0<\nu_u^+(s^*),\]
as well as the condition on the ratio of the eigenvalues $\nu_u^-(s^*)<2\nu_v^+(s^*,\sigma^*)$.

We now proceed through the terms in the quadratic expansions of $h_{1,2}$ and show that each of the integrands converge exponentially as $\xi\to -\infty$.  The condition on the ratio of the eigenvalues is key for the convergence of the integrals listed -- in particular those that are quadratic in $\eta_{1,2}$.  

\begin{table}
\centering
  \begin{tabular}{ | c | c  | c |}
    \hline
    Term  & Exponential rate as $\xi\to-\infty$  \\ \hline
    $\phi(\xi)$, $a_1(\xi)$, $a_2(\xi)$  &  $\nu_v^+(s^*,\sigma^*)=-\frac{s^*}{2\sigma^*}+\frac{1}{2\sigma^*}\sqrt{(s^*)^2-4\sigma^* g(\p_1)}$ \\ 
$U_p'(\xi)$, $U_p''(\xi)$  ,$\theta_s^{1}(\xi)$, $ \theta_s^{2}(\xi)$ &  $\nu_u^+(s^*)=-\frac{s^*}{2}+\frac{1}{2}\sqrt{(s^*)^2-4F_u(\p_1)} $ \\
$b_2(\xi) $  & $ -\nu_u^-(s^*)=\frac{s^*}{2}+\frac{1}{2}\sqrt{(s^*)^2-4F_u(\p_1)}$ \\ 
   \hline 
  \end{tabular}
\caption{Asymptotic exponential decay rates of the terms arising in Lemma~\ref{lem:relevantexp}.  These expressions are derived from  (\ref{eq:a1})  for $a_1$ and $a_2$, (\ref{eq:b2}) for $b_2$ and (\ref{eq:thetas}) for $\theta_s^1$ and $\theta_s^2$.}
\label{table}
\end{table}

\begin{itemize}
\item For $\mathbf{h}^{(2,0,0,0)}_{1}$, the asymptotic exponential rate of the integrand is $s^*+3\nu_u^+>0$ and the integral converges as $\xi\to -\infty$.  
\item For $\mathbf{h}^{(1,1,0,0)}_{1}$, the asymptotic exponential rate of the first term in the expansions is 
\[s^*+2\nu_u^+(s^*)+\nu_v^+(s^*,\sigma^*)> s^*+2\nu_u^+(s^*)+\frac{\nu_u^-(s^*)}{2}=\frac{1}{2}s^* +\frac{3}{4}\nu_u^+(s^*)>0.\]
The second term has exponential rate
\[ -\nu_u^-(s^*)+\nu_u^+(s^*) +\nu_v^+(s^*,\sigma^*)>\nu_u^+(s^*)-\nu_v^+(s^*,\sigma^*)>0.\]
\item For $\mathbf{h}^{(0,2,0,0)}_{1}$, the asymptotic exponential rate of the first term in the expansions is $s^*+\nu_u^+(s^*)+2\nu_v^+(s^*,\sigma^*)>s^*+\nu_u^+(s^*)+\nu_u^-(s^*)=0$ and those terms converge.  For the second integral, the rate is $-\nu_u^-(s^*)+2\nu_v^+(s^*,\sigma^*)>0$ and the final integral has exponential rate $s^*+\nu_u^+(s^*)+2\nu_v^+(s^*,\sigma^*)=-\nu_u^-(s^*)+2\nu_v^+(s^*,\sigma^*)>0$.
\item For $\mathbf{h}^{(1,0,1,0)}_{1}$ all exponential rates are positive and the integral therefore converges as $\xi\to -\infty$.  
\item For  the first integral in $\mathbf{h}^{(0,1,1,0)}_{1}$, the term $e^{s^*\xi} U_p'(\xi)a_2(\xi)$ as asymptotic exponential rate $s^*+\nu_u^+(s^*)+\nu_v^+(s^*,\sigma^*)>0$ and therefore converges.  All other terms in the first integral possess stronger decay rates and therefore also converge.  The exponential rate of the first term in the second integral is $-\nu_u^-(s^*)+\nu_v^+(s^*,\sigma^*)>-\nu_v^+(s^*,\sigma^*)>0$.  The second term has stronger decay and therefore the second integral also converges.  
\item For $\mathbf{h}^{(0,1,0,1)}_{1}$, the asymptotic exponential rate is again $-\nu_u^-(s^*)+\nu_v^+(s^*,\sigma^*)>0$ and the integral converges.  
\item For $\mathbf{h}^{(0,0,2,0)}_{1}$, all exponential rates are positive and the integral converges.  
\item For $\mathbf{h}^{(1,1,0,0)}_{2}$, the asymptotic exponential rate is 
\[ \frac{s^*}{\sigma^*}+\nu_u^+(s^*)+2\nu_v^+(s^*,\sigma^*)>\frac{s^*}{\sigma^*}+\nu_u^+(s^*)+\nu_v^+(s^*,\sigma^*)+\nu_v^-(s^*,\sigma^*)=\nu_u^+(s^*)>0,\]
and the integral converges.  
\item For $\mathbf{h}^{(0,2,0,0)}_{2}$, the asymptotic exponential rate is 
\[ \frac{s^*}{\sigma^*}+3\nu_v^+(s^*,\sigma^*)>\frac{s^*}{\sigma^*}+\nu_v^+(s^*,\sigma^*)+\nu_v^-(s^*,\sigma^*)=0, \]
and the integral converges. 
\item For $\mathbf{h}^{(0,1,1,0)}_{2}$, the asymptotic rate of the first term in the integral is
\[
\frac{s^*}{\sigma^*}+2\nu_v^+(s^*,\sigma^*)>\frac{s^*}{\sigma^*}+\nu_v^+(s^*,\sigma^*)+\nu_v^-(s^*,\sigma^*)=0,
\]
while the term gives 
\[ \frac{s^*}{\sigma^*}+\nu_u^+(s^*)+2\nu_v^+(s^*,\sigma^*)>\frac{s^*}{\sigma^*}+\nu_u^+(s^*)+\nu_v^+(s^*,\sigma^*)+\nu_v^-(s^*,\sigma^*)=\nu_u^+(s^*)>0,\]
and the integral converge.  A similar argument implies the convergence of $\mathbf{h}^{(0,1,0,1)}_{2}$.
\end{itemize}
\end{Proof}

\begin{lemma}\label{lem:h2iszero} We have that 
\[ \mathbf{h}_2^{(1,1,0,0)}=\frac{1}{\Omega_{2}}\left(\tilde{\phi}''(\xi_0)\tilde\phi(\xi_0)-(\tilde{\phi}'(\xi_0))^2\right),\]
and 
\[\mathbf{h}_2^{(1,1,0,0)}\sim \gamma_{11} e^{(\nu_v^+(s^*,\sigma^*)-\nu_v^-(s^*,\sigma^*)+\nu_u^+(s^*))\xi_0},\]
as $\xi_0\to -\infty$ where 
\[ \mathrm{sign}(\gamma_{11}) =\mathrm{sign}\left( g_u(\p_1)\right).\]
\end{lemma}
\begin{Proof}
Recall from Lemma~\ref{lem:relevantexp} that 
\[ \mathbf{h}_2^{(1,1,0,0)}=\frac{1}{\Omega_{2}} \int_{\xi_0}^\infty e^{\frac{s^*}{\sigma^*}\xi}\frac{G_{uv}(U_p(\xi),0)}{\sigma^*}U_p'(\xi)\left(\phi(\xi)\right)^2\md\xi.\]
Observe that 
\[ G_{uv}(U_p(\xi),0)U_p'(\xi)=\frac{\md}{\md \xi} g(U_p(\xi),0).\]
After also recalling that $\phi(\xi)=e^{-\frac{s^*}{2\sigma^*}\xi}\tilde{\phi}(\xi)$ we are able to transform the integral as follows and obtain the desired result
\begin{eqnarray*} \mathbf{h}_2^{(1,1,0,0)}&=& \frac{1}{\sigma^*\Omega_{2}} \int_{\xi_0}^\infty \left(\frac{\md}{\md \xi} g(U_p(\xi),0)\right)\tilde{\phi}^2(\xi)\md\xi \\
&=&  \frac{1}{\sigma^*\Omega_{2}}\left[ g(U_p(\xi),0)\tilde{\phi}^2(\xi)\right]_{\xi=\xi_0}^\infty - \frac{2}{\sigma^*\Omega_{2}} \int_{\xi_0}^\infty g(U_p(\xi),0)\tilde{\phi}(\xi)\tilde{\phi}'(\xi)\md \xi \\
&=&  -\frac{1}{\sigma^*\Omega_{2}}g(U_p(\xi_0),0)\tilde{\phi}^2(\xi_0) - \frac{2}{\sigma^*\Omega_{2}} \int_{\xi_0}^\infty \left(-\sigma^*\tilde{\phi}''(\xi)+\frac{(s^*)^2}{4\sigma^*}\phi(\xi)\right)\tilde{\phi}'(\xi)\md \xi \\
&=&  -\frac{1}{\sigma^*\Omega_{2}}\left( g(U_p(\xi_0),0)\tilde{\phi}^2(\xi_0) +\sigma^*\left(\tilde{\phi}'(\xi_0)\right)^2-\frac{(s^*)^2}{4\sigma^*}\left(\tilde{\phi}(\xi_0)\right)^2\right) \\
&=& \frac{1}{\Omega_{2}}\left(\tilde{\phi}''(\xi_0)\tilde\phi(\xi_0)-(\tilde{\phi}'(\xi_0))^2\right).
\end{eqnarray*}
To determine the asymptotics of the final form, we expand the second order system defining $\tilde{\phi}(\xi)$ into a system,
\begin{eqnarray*}
\tilde{\phi}'&=& \tilde{\psi} \\
\tilde{\psi}'&=& \frac{(s^*)^2}{4(\sigma^*)^2} \tilde{\phi}-\frac{g(U_p(\xi),0)}{\sigma^*} \tilde{\phi}
\end{eqnarray*}
We then diagonalize and expand $g(U_p,0)=g(\p_1)+g_u(\p_1)U_p +\O(2)$, arriving at the following system that is relevant for the determination of the asymptotic decay rates, 
\begin{eqnarray*}
\tilde{\phi}_{ws}'&=& \tilde{\nu}_v^+(s^*,\sigma^*)\tilde\phi_{ws}+\frac{g_u(\p_1)}{\sigma^*(\tilde{\nu}_v^-(s^*,\sigma^*)-\tilde{\nu}_v^+(s^*,\sigma^*))} \left(U_p-u^+\right)\tilde \phi_{ws} +\O(2), \\
U_p'&=& \nu_u^+(s^*) (U_p-u^+)+\O(2),
\end{eqnarray*}
where $\tilde{\nu}_v^\pm = \frac{1}{2\sigma^*}\sqrt{ (s^*)^2 -4\sigma^*g(\p_1)}$. 
Then 
\[ U_p(\xi)\sim u^+-c_u e^{\nu_u^+(s^*) \xi},\]
from which we determine that 
\[  \left(\tilde{\phi}''(\xi_0)\tilde\phi(\xi_0)-(\tilde{\phi}'(\xi_0))^2\right)=-C^2 \frac{g_u(\p_1)\nu_u^+(s^*)}{\sigma^*(\tilde{\nu}_v^-(s^*,\sigma^*)-\tilde{\nu}_v^+(s^*,\sigma^*))}e^{\left( 2\tilde{\nu}_v^+(s^*,\sigma^*)+\nu_u^+(s^*)\right)\xi_0 }.\]
Therefore $\gamma_{11}$ is the constant multiplying the exponential and the exponential decay rate is obtained by noting that $2\tilde{\nu}_v^+=\nu_v^+-\nu_v^-$. 
\end{Proof}

\section{Expressions for $\langle \psi_1, \mathcal{R}(p_1,q_1,\sigma)\rangle$ and $\langle \psi_2, \mathcal{R}(p_1,q_1,\sigma)\rangle$}

\begin{lemma}\label{lemexppsiR}
We have the following expressions for the projections of $\mathcal{R}(p_1,q_1,\sigma)$, defined in \eqref{eq:R}, onto $\psi_1$ and $\psi_2$:
\begin{subequations}
\begin{align} 
\langle \psi_1, \mathcal{R}(p_1,q_1,\sigma)\rangle &= \frac{q_1 (\sigma-\sigma^*)}{\phi(\xi_0)} \int_{-\infty}^{\xi_0}b_2(\tau)\left(\frac{1}{(\sigma^*)^2}g(U_p(\tau),0)\phi(\tau)+\frac{s^*}{(\sigma^*)^2}\phi'(\tau)\right)\md\tau,\\
\langle \psi_2, \mathcal{R}(p_1,q_1,\sigma)\rangle &= \frac{q_1 (\sigma-\sigma^*)}{\phi(\xi_0)}\int_{-\infty}^{\xi_0} e^{\frac{s^*}{\sigma^*}\tau}\left(\frac{1}{(\sigma^*)^2}g(U_p(\tau),0) \phi(\tau)^2+\frac{s^*}{(\sigma^*)^2} \phi(\tau)\phi'(\tau) \right)\md\tau.
\end{align}
\label{eq:exppsiR}
\end{subequations}
\end{lemma}

\begin{Proof}
We first prove the second equality of \eqref{eq:exppsiR} on $\mathcal{R}(p_1,q_1,\sigma)$. From the definition of $\psi_{2}$ in \eqref{eq:psi12}, we have that
\begin{align*}
\langle \psi_2, \mathcal{R}(p_1,q_1,\sigma)\rangle &= q_1 e^{\frac{s^*}{\sigma^*}\xi_0} \left(-\phi'(\xi_0)+\phi(\xi_0)Z_{22}(\xi_0)+\phi(\xi_0)E_5(\xi_0)h_5(\sigma)\right),\\
&= \frac{q_1 (\sigma-\sigma^*)}{\phi(\xi_0)}\int_{-\infty}^{\xi_0} e^{\frac{s^*}{\sigma^*}\tau}\left(\frac{1}{(\sigma^*)^2}g(U_p(\tau),0) \phi(\tau)^2+\frac{s^*}{(\sigma^*)^2} \phi(\tau)\phi'(\tau) \right)\md\tau,
\end{align*}
where we used the facts that $Z_{22}(\xi_0)=\phi'(\xi_0)/\phi(\xi_0)$ and $E_5(\xi_0)=\phi(\xi_0)^2 e^{\frac{s^*}{\sigma^*}\xi_0}$, together with \eqref{eq:h5sigma}.

We now turn our attention to the first equality of \eqref{eq:exppsiR}. Using this time the definition of $\psi_1$, we get that
\begin{align*}
\langle \psi_1, \mathcal{R}(p_1,q_1,\sigma)\rangle &= p_1\left(-e^{s^*\xi_0}U_p''(\xi_0)+e^{s^*\xi_0}U_p'(\xi_0)Z_{11}(\xi_0) \right)\\
&~~+q_1\left( e^{s^*\xi_0}U_p'(\xi_0)Z_{12}(\xi_0)+b_1(\xi_0)+Z_{22}(\xi_0)b_2(\xi_0) \right)\\
&~~+q_1\left( e^{s^*\xi_0}U_p'(\xi_0)D_4(\xi_0)h_4(\sigma)+b_2(\xi_0)E_5(\xi_0)h_5(\sigma)\right).
\end{align*}
As $Z_{11}(\xi_0)=U_p''(\xi_0)/U_p'(\xi_0)$ the first term in the factor of $p_1$ vanishes. We are going to show that the second term also vanishes. From the definition of $Z_{12}$, we have
\bqs
e^{s^*\xi_0}U_p'(\xi_0)Z_{12}(\xi_0)=\frac{1}{\phi(\xi_0)}\int_{\xi_0}^\infty e^{s^*\tau} U_p'(\tau) F_v(U_p(\tau),0)\phi(\tau)\md \tau.
\eqs
By definition, $b_1$ and $b_2$ satisfy 
\begin{eqnarray*}
b_1'(\xi)&=&\frac{g(U_p(\xi),0)}{\sigma^*} b_2(\xi)+F_v(U_p(\xi),0) e^{s^*\xi}U_p'(\xi) \\
b_2'(\xi)&=& -b_1(\xi)+\frac{s^*}{\sigma^*}b_2(\xi).
\end{eqnarray*}
Multiplying the first equation by $\phi(\xi)$ and integrating we get that
\begin{align*}
\int_{\xi_0}^\infty e^{s^*\tau} U_p'(\tau) F_v(U_p(\tau),0)\phi(\tau)\md \tau &= \int_{\xi_0}^\infty \phi(\tau)b_1'(\tau)\md \tau - \int_{\xi_0}^\infty \phi(\tau)\frac{g(U_p(\tau),0)}{\sigma^*} b_2(\tau)\md\tau \\
&= -\phi(\xi_0)b_1(\xi_0)-\int_{\xi_0}^\infty \phi'(\tau)b_1(\tau)\md \tau - \int_{\xi_0}^\infty \phi(\tau)\frac{g(U_p(\tau),0)}{\sigma^*} b_2(\tau)\md\tau.
\end{align*}
On the other hand, we have
\begin{align*}
-\int_{\xi_0}^\infty \phi'(\tau)b_1(\tau)\md \tau&=\int_{\xi_0}^\infty \phi'(\tau)b_2'(\tau)\md \tau-\frac{s^*}{\sigma^*}\int_{\xi_0}^\infty \phi'(\tau)b_2(\tau)\md \tau\\
&= -\phi'(\xi_0)b_2(\xi_0)-\int_{\xi_0}^\infty \phi''(\tau)b_2(\tau)\md \tau-\frac{s^*}{\sigma^*}\int_{\xi_0}^\infty \phi'(\tau)b_2(\tau)\md \tau.
\end{align*}
Combining all the terms, and using the fact that $\mathcal{L}_v\phi=0$, we obtain that
\bqs
\int_{\xi_0}^\infty e^{s^*\tau} U_p'(\tau) F_v(U_p(\tau),0)\phi(\tau)\md \tau=-\phi(\xi_0)b_1(\xi_0)-\phi'(\xi_0)b_2(\xi_0)
\eqs
from which we deduce that
\bqs
e^{s^*\xi_0}U_p'(\xi_0)Z_{12}(\xi_0)+b_1(\xi_0)+Z_{22}(\xi_0)b_2(\xi_0)=0,
\eqs
as $Z_{22}(\xi_0)=\phi'(\xi_0)/\phi(\xi_0)$. So far, we have thus obtained
\bqs
\langle \psi_1, \mathcal{R}(p_1,q_1,\sigma)\rangle = q_1\left( e^{s^*\xi_0}U_p'(\xi_0)D_4(\xi_0)h_4(\sigma)+b_2(\xi_0)E_5(\xi_0)h_5(\sigma)\right),
\eqs
and each term simplifies to
\begin{align*}
e^{s^*\xi_0}U_p'(\xi_0)D_4(\xi_0)h_4(\sigma)&=\frac{(\sigma-\sigma^*)}{\phi(\xi_0)}\int_{-\infty}^{\xi_0} E_4(\tau) \left(\frac{1}{(\sigma^*)^2}g(U_p(\tau),0) +\frac{s^*}{(\sigma^*)^2} Z_{22}(\tau) \right)\md\tau,\\
b_2(\xi_0)E_5(\xi_0)h_5(\sigma)&=\frac{\tilde{c}_1 (\sigma-\sigma^*)}{\phi(\xi_0)}\int_{-\infty}^{\xi_0} E_5(\tau) \left(\frac{1}{(\sigma^*)^2}g(U_p(\tau),0) +\frac{s^*}{(\sigma^*)^2} Z_{22}(\tau) \right)\md\tau,
\end{align*}
where we used the fact that $b_2(\xi_0)=\tilde{c}_1e^{\frac{s^*}{\sigma^*}\xi_0}\phi(\xi_0)$ from \eqref{eq:b2} and $E_5(\xi_0)=e^{\frac{s^*}{\sigma^*}\xi_0}\phi(\xi_0)^2$. As a consequence, we have
\bqs
\langle \psi_1, \mathcal{R}(p_1,q_1,\sigma)\rangle=\frac{(\sigma-\sigma^*)}{\phi(\xi_0)}\int_{-\infty}^{\xi_0} (E_4(\tau)+\tilde{c}_1E_5(\tau)) \left(\frac{1}{(\sigma^*)^2}g(U_p(\tau),0) +\frac{s^*}{(\sigma^*)^2} Z_{22}(\tau) \right)\md\tau.
\eqs
To conclude, we are going to show that
\bqs
E_4(\xi)+\tilde{c}_1E_5(\xi)=b_2(\xi)\phi(\xi).
\eqs
We recall that by definition, $E_4(\xi)$ satisfies
\bqs
E_4'(\xi)=Z_{12}(\xi)U'_p(\xi)\phi(\xi)e^{s^*\xi}+\left(\frac{s^*}{\sigma^*} + 2Z_{22}(\xi)\right)E_4(\xi), \quad E_4(\xi_0)=0.
\eqs
As a consequence, $E_4(\xi)$ can be written as 
\bqs
E_4(\xi)=\left(\int_{\xi_0}^\xi \frac{e^{-\frac{s^*}{\sigma^*}\zeta}}{\phi(\zeta)^2} \left[ \int_{\zeta}^\infty e^{s^*\tau} U_p'(\tau) F_v(U_p(\tau),0)\phi(\tau)\md \tau\right]\md \zeta \right)E_5(\xi).
\eqs
Integrating by parts, we obtain 
\begin{align*}
\int_{\xi_0}^\xi \frac{e^{-\frac{s^*}{\sigma^*}\zeta}}{\phi(\zeta)^2} \left[ \int_{\zeta}^\infty e^{s^*\tau} U_p'(\tau) F_v(U_p(\tau),0)\phi(\tau)\md \tau\right]\md \zeta
&= \left(\int_{\xi_0}^\xi \frac{e^{-\frac{s^*}{\sigma^*}\zeta}}{\phi(\zeta)^2}\md\zeta\right)\left(\int_{\xi}^\infty e^{s^*\tau} U_p'(\tau) F_v(U_p(\tau),0)\phi(\tau)\md \tau\right)\\
&~~+ \int_{\xi_0}^\xi \left(\int_{\xi_0}^\tau \frac{e^{-\frac{s^*}{\sigma^*}\zeta}}{\phi(\zeta)^2}\md\zeta \right)e^{s^*\tau} U_p'(\tau) F_v(U_p(\tau),0)\phi(\tau)\md \tau.
\end{align*}
Using the fact that $E_5(\xi)=B_1(\xi)\phi(\xi)$ and equation \eqref{eq:b2}, we obtain that
\bqs
E_4(\xi)=b_2(\xi)\phi(\xi)-\tilde{c}_1B_1(\xi)\phi(\xi)=b_2(\xi)\phi(\xi)-\tilde{c}_1E_5(\xi).
\eqs
As a conclusion, we get
\bqs
\langle \psi_1, \mathcal{R}(p_1,q_1,\sigma)\rangle=\frac{(\sigma-\sigma^*)}{\phi(\xi_0)}\int_{-\infty}^{\xi_0} b_2(\tau)\phi(\tau) \left(\frac{1}{(\sigma^*)^2}g(U_p(\tau),0) +\frac{s^*}{(\sigma^*)^2} Z_{22}(\tau) \right)\md\tau,
\eqs
which proves the lemma.
\end{Proof}

\section{Expressions for $\mathbf{N}_u^{(0,0,2,0)}(s,\sigma)$ and $\mathbf{N}_{ss,u}^{(2,0,0,0)}(s,\sigma)$}

\begin{lemma}\label{lemexpN}
The coefficients $\mathbf{N}_u^{(0,0,2,0)}(s,\sigma)$ and $\mathbf{N}_{ss,u}^{(2,0,0,0)}(s,\sigma)$ appearing in the expansions of $\mathcal{N}_u$ and $\mathcal{N}_{ss,u}$ defined in equation \eqref{nonlinearN} depend only on $s$ and have the expressions:
\begin{subequations}
\begin{align}
\mathbf{N}_u^{(0,0,2,0)}(s,\sigma)&=\frac{F_{uu}(\p_1)}{2(\nu_u^-(s)-\nu_u^+(s))},\\
\mathbf{N}_{ss,u}^{(2,0,0,0)}(s,\sigma)&=-\frac{F_{uu}(\p_1)}{2(\nu_u^-(s)-\nu_u^+(s))}.
\end{align}
\label{eq:expN}
\end{subequations}
\end{lemma}

\begin{Proof}
Let us recall that we have the change of variables 
\[ \left(\begin{array}{c} u_1-u^+ \\ u_2 \\ v_1 \\ v_2 \end{array}\right)=T(s,\sigma)\left(\begin{array}{c} \yu \\ \yus \\ \yw \\ \yvs \end{array}\right),\]
where $T(s,\sigma)$ is defined in \eqref{matrixT}. We set $U:=(u_1-u^+,u_2,v_1,v_2)^{T}$ and $Y:=(y^u,y^{ss,u},y^{ws},y^{ss,v})^T$. Let us also remark that in the original coordinates, the quadratic terms in $(u_1-u^+,u_2,v_1,v_2)^{T}$ of the nonlinear part are given by
\bqs
N_2(U):=\left(
\begin{array}{c}
0 \\
-\frac{F_{uu}(\p_1)}{2}(u_1-u^+)^2-F_{uv}(\p_1)(u_1-u^+)v_1-\frac{F_{vv}(\p_1)}{2}v_1^2\\
0 \\
-\frac{G_{uv}(\p_1)}{\sigma}(u_1-u^+)v_1-\frac{G_{vv}(\p_1)}{2\sigma}v_1^2
\end{array}
 \right),
\eqs
where we have used the fact that $G_{uu}(\p_1)=0$. Then, we note that with our change of variables both $v_1$ and $v_2$ in the new coordinates do not depend in $y^u$ and $y^{ss,u}$. As consequence, if one keeps only the quadratic terms in $y^u$ and $y^{ss,u}$ in the expression of $N_2$, expressed in the new coordinates, we get
\bqs
N_2(T(s,\sigma)Y)=\left(
\begin{array}{c}
0 \\
-\frac{F_{uu}(\p_1)}{2}(y^u)^2-\frac{F_{uu}(\p_1)}{2}(y^{ss,u})^2-F_{uu}(\p_1)y^uy^{ss,u}\\
0 \\
0
\end{array}
 \right)+\mathcal{O}(2).
\eqs
To conclude, it is enough to remark that 
\bqs
\left(
\begin{array}{c}
\mathcal{N}_u(Y,s,\sigma) \\
\mathcal{N}_{ss,u}(Y,s,\sigma) \\
\mathcal{N}_{ws}(Y,s,\sigma) \\
\mathcal{N}_{ss,v}(Y,s,\sigma) 
\end{array}
\right) = T(s,\sigma)^{-1} N_2(T(s,\sigma)Y)+ \mathcal{O}(3),
\eqs
and that the matrix $T(s,\sigma)$ is block triangular so that
\bqs
T(s,\sigma) =  \left(\begin{matrix} T_{11}(s) & T_{12}(s,\sigma) \\ 0 & T_{22}(s,\sigma) \end{matrix} \right) \text{ and } T(s,\sigma)^{-1} =  \left(\begin{matrix} T_{11}^{-1}(s) & -T_{11}^{-1}(s)T_{1,2}(s,\sigma)T_{22}^{-1}(s,\sigma) \\ 0 & T_{22}^{-1}(s,\sigma) \end{matrix} \right).
\eqs
Finally, a direct computation shows that 
\bqs
T_{11}^{-1}(s)=\frac{1}{\nu_u^-(s)-\nu_u^+(s)}\left(\begin{matrix} \nu_u^-(s) & -1 \\ -\nu_u^+(s) & 1 \end{matrix} \right),
\eqs
which in turns implies that the quadratic terms in $y^u$ and $y^{ss,u}$ in the expression of $\mathcal{N}_u(Y,s,\sigma)$ are
\bqs
\mathcal{N}_u(Y,s,\sigma) = \frac{1}{\nu_u^-(s)-\nu_u^+(s)} \left( \frac{F_{uu}(\p_1)}{2}(y^u)^2+\frac{F_{uu}(\p_1)}{2}(y^{ss,u})^2+F_{uu}(\p_1)y^uy^{ss,u} \right)+\mathcal{O}(2),
\eqs
and similarly for $\mathcal{N}_{ss,u}(Y,s,\sigma)$
\bqs
\mathcal{N}_{ss,u}(Y,s,\sigma) = -\frac{1}{\nu_u^-(s)-\nu_u^+(s)} \left( \frac{F_{uu}(\p_1)}{2}(y^u)^2+\frac{F_{uu}(\p_1)}{2}(y^{ss,u})^2+F_{uu}(\p_1)y^uy^{ss,u} \right)+\mathcal{O}(2),
\eqs
which concludes the proof.
\end{Proof}

\section{Expression for $\mathbf{N}_{ss,v}^{(0,2,0,0)}(s,\sigma)$}

\begin{lemma}\label{lemexpNssv}
The coefficient $\mathbf{N}_{ss,v}^{(0,2,0,0)}(s,\sigma)$ appearing in the expansion of $\mathcal{N}_{ss,v}$ defined in equation \eqref{nonlinearN} has the following expression:
\begin{equation}
\mathbf{N}_{ss,v}^{(0,2,0,0)}(s,\sigma)=\frac{1}{\sigma(\nu_v^-(s,\sigma)-\nu_v^+(s,\sigma))}\left( \frac{F_v(\p_1)}{d_u(\nu_v^-(s,\sigma))}G_{uv}(\p_1)-\frac{G_{vv}(\p_1)}{2}\right).
\label{eq:expNssv}
\end{equation}
\end{lemma}

\begin{Proof}
The proof is similar to the proof of Lemma~\ref{lemexpN}. One only needs to keep track of the quadratic terms $y^{ws}$ in the nonlinear part of the system and notice that
\bqs
N_2(T(s,\sigma)Y)=\left(
\begin{array}{c}
0 \\
-\frac{F_{uu}(\p_1)}{2}\left(\frac{F_v(\p_1)}{d_u(\nu_v^-(s,\sigma))}\right)^2(y^{ws})^2+F_{uv}(\p_1)\frac{F_v(\p_1)}{d_u(\nu_v^-(s,\sigma))}(y^{ws})^2-\frac{F_{vv}(\p_1)}{2}(y^{ws})^2
\\
0 \\
\frac{F_v(\p_1)}{d_u(\nu_v^-(s,\sigma))}\frac{G_{uv}(\p_1)}{\sigma}(y^{ws})^2-\frac{G_{vv}(\p_1)}{2\sigma}(y^{ws})^2
\end{array}
 \right)+\mathcal{O}(2).
\eqs
This implies that
\[
\mathcal{N}_{ss,v}(Y,s,\sigma)=\frac{1}{\sigma(\nu_v^-(s,\sigma)-\nu_v^+(s,\sigma))}\left( \frac{F_v(\p_1)}{d_u(\nu_v^-(s,\sigma))}G_{uv}(\p_1)-\frac{G_{vv}(\p_1)}{2}\right)(y^{ws})^2+\mathcal{O}(2),
\]
where we have used the explicit form of the inverse of $T_{22}^{-1}(s,\sigma)$.
\end{Proof}

\section{Quadratic expansions of $p(\zu,s,\sigma)$ and $q(\zu,s,\sigma)$}

In the following Lemma we will use the notations
\[ \gamma_{ij}(\zu,0,0,s,\sigma)=\gamma_{ij}^{(1)}(s,\sigma)\zu+\gamma_{ij}^{(2)}(s,\sigma)(\zu)^2+\O\left((\zu)^3\right), \quad i,j\in\{1,2\}\]
 together with
\[ \mathcal{M}_u(\zu,0,0,s,\sigma)=\mathcal{M}_u^{(2)}(s,\sigma)(\zu)^2+\O\left((\zu)^3\right),\]
where $\gamma_{ij}$ and $\mathcal{M}_u$ are defined in equation \eqref{eq:maindiag2}, Lemma~\ref{lem:cov1}.

\begin{lemma}\label{lem:expqzu}
The quadratic expansions for the maps $p(\zu,s,\sigma)$ and $q(\zu,s,\sigma)$ defined in equations \eqref{eq:bartrans} from Lemma~\ref{lem:cov2} are:
\begin{align*}
p(\zu,s,\sigma)&= {\cal P}_1(s,\sigma) \zu + {\cal P}_2(s,\sigma)(\zu)^2 +\O\left((\zu)^3\right),\\
q(\zu,s,\sigma) &= {\cal Q}_1(s,\sigma) \zu +{\cal Q}_2 (s,\sigma)(\zu)^2 +\O\left((\zu)^3\right),
\end{align*}
with 
\begin{align*}
{\cal P}_1(s,\sigma) &= \frac{\gamma_{11}^{(1)}(s,\sigma)}{\nu_u^+(s)}, \\
{\cal Q}_1(s,\sigma) &= - \left( \Lambda_{ss}(s,\sigma)-(\nu_v^+(s,\sigma)+\nu_u^+(s))\mathrm{I}\right)^{-1}  \gamma_{21}^{(1)}(s,\sigma),
\end{align*}
and
\begin{align*}
{\cal P}_2(s,\sigma) &=\frac{1}{2\nu_u^+(s)}\left[-{\cal P}_1(s,\sigma) \left(\gamma_{11}^{(1)}(s,\sigma)+\mathcal{M}_u^{(2)}(s,\sigma) \right)+\gamma_{11}^{(2)} (s,\sigma)+ \gamma_{12}^{(1)}(s,\sigma){\cal Q}_1(s,\sigma)\right],\\
{\cal Q}_2(s,\sigma) &= \left( \Lambda_{ss}(s,\sigma)-(\nu_v^+(s,\sigma)+2\nu_u^+(s))\mathrm{I}\right)^{-1}\left( -\gamma_{21}^{(2)}(s,\sigma)+(\mathcal{M}_u^{(2)}(s,\sigma)-\gamma_{22}^{(1)}(s,\sigma)+\gamma_{11}^{(1)}(s,\sigma)){\cal Q}_1(s,\sigma) \right).
\end{align*}
\end{lemma}

\end{appendix}

\bibliographystyle{abbrv}
\bibliography{lockedbib}

\end{document}